\documentclass[twoside,11pt]{article}

\usepackage{amsmath,amsfonts}
\usepackage{amssymb,latexsym}
\usepackage{amsxtra}
\usepackage{upref}
\usepackage{exscale}
\usepackage[mathscr]{eucal}

\newcommand{\be}{\begin{equation}}
\newcommand{\ee}{\end{equation}}
\newcommand{\beq}{\begin{eqnarray}}
\newcommand{\eeq}{\end{eqnarray}}
\newcommand{\no}{\nonumber}
\newcommand{\bea}{\begin{array}}
\newcommand{\eea}{\end{array}}
\newcommand{\lb}{\label}
\newcommand{\mcal}{\mathcal}
\newcommand{\mscr}{\mathscr}
\newcommand{\mfrak}{\mathfrak}
\newcommand{\ve}{\varepsilon}

\newcommand{\ts}{\textstyle}
\newcommand{\pp}{\partial}
\newcommand{\im}{\imath}
\newcommand{\ppr}{^{\boldsymbol{\prime}}}

\newcommand{\pdag}{^{\dagger}}
\newcommand{\wt}{\widetilde}
\newcommand{\ovv}{\overline}
\newcommand{\trs}{\raisebox{-4pt}{$\mbox{\large tr}\atop {\scriptstyle s,s^{\boldsymbol{\prime}}=\uparrow,\downarrow}$}}
\newcommand{\TRS}{\raisebox{-4pt}{$\mbox{\large Tr}\atop {\scriptstyle a,s;b,s^{\boldsymbol{\prime}}}$}}
\newcommand{\sums}{\sum_{s=\uparrow,\downarrow}}
\newcommand{\sumss}{\sum_{s,s^{\boldsymbol{\prime}}=\uparrow,\downarrow}}
\newcommand{\ph}{\phantom}

\newcommand{\scz}{\scriptsize}

\numberwithin{equation}{section}
\allowdisplaybreaks[4]

\begin{document}

\begin{center}
{\large\bf Coherent state path integral and nonlinear sigma model for a condensate} \vspace*{0.1cm}\\
{\large\bf composed of fermions with precise, discrete steps in the time development and} \vspace*{0.1cm}\\
{\large\bf with transformations from Euclidean path integrations to spherical field variables} \vspace*{0.3cm}\\
{\bf Bernhard Mieck}\footnote{e-mail: "bjmeppstein@arcor.de";\newline freelance activity during 2007-2009;
current location : Zum Kohlwaldfeld 16, D-65817 Eppstein, Germany.}
\end{center}

\begin{abstract}
A coherent state path integral is considered for fermions with precise, discrete time separation
so that non-hermitian actions are obtained; the complex-conjugated, anti-commuting fields
\(\psi_{\vec{x},s}^{*}(t_{p}+\Delta t_{p})\) always follow a discrete time step '\(\Delta t_{p}\)'
later on the non-equilibrium time contour than the corresponding fields \(\psi_{\vec{x},s}(t_{p})\)
according to the chosen normal ordering within the original, second quantized Hamilton operator. We describe
details of the derivation for a nonlinear sigma model of a pair condensate composed of fermions
concerning the precise, discrete time step order which is usually abbreviated by the appealing
(but in fact non-existent) hermitian form of the fields. The non-hermitian kind of actions with
additional time shift '\(\Delta t_{p}\)' in \(\psi_{\vec{x},s}^{*}(t_{p}+\Delta t_{p})\)
(relative to \(\psi_{\vec{x},s}(t_{p})\)) is necessary in order to avoid infinities which arise from
the simultaneous action of field operators with their hermitian conjugates caused by the defining
anti-commutators. This problem is ubiquitous in quantum many-body physics and already occurs in
the original Dyson equation. However, one has only to include a few amendments concerning the
precise, discrete time step development of coherent state path integrals,
compared to previous, abridged approaches, so that one can accomplish the
exact treatment and derivation of a spontaneous symmetry breaking (SSB) with a coset decomposition
for coset matrices of pair condensates and sub-algebra elements of density parts. The
involved Hubbard-Stratonovich transformation (HST) to self-energies is also stated more
precisely regarding the separation to exact, discrete time steps with a few amendments for
the matrix operations of hermitian conjugation and transposition of dyadic product related density
matrices and hermitian self-energies. Finally, we define and give details for the transformation
of path integrals from the Euclidean base manifold with Cartesian coordinates to spherical field
variables so that inherent, rotational symmetries can be used to simplify non-perturbative calculations
of path integrals. The involved metric tensor of spatial coordinates, considered as indices
in addition to the spin indices, is therefore given as an extension of the metric tensor of internal
degrees of freedom so that the supplementary invariant integration measure is achieved from the inverse
square root of the determinant of the metric tensor with spatial indices.\newline

\vspace*{0.1cm}

\end{abstract}

\newpage

\tableofcontents

\newpage

\section{Introduction} \lb{s1}

\subsection{The problem of discrete time steps in coherent state path integrals} \lb{s11}

Coherent states are a convenient tool to transform the time development of second quantized Hamilton operators to path
integrals for quantum many particle problems. In general coherent states are determined by a coset decomposition of the
underlying dynamical group where transformations with the maximal, commuting Cartan sub-algebra only change the reference
state (as a vacuum or ground state) by redundant phases; on the contrary the cosets generate from the chosen reference
state new states always being distinct up to redundant phase transformations by the subgroup elements \cite{gil1}. In this paper
we restrict to coherent states for the Heisenberg algebra of anti-commuting field operators in a spontaneous
symmetry breaking (SSB,\cite{gold,nambu}) for pair condensates composed of fermionic constituents \cite{neg,klau}.
Previous articles of this problem
with SSB in a coherent state path integral contain abbreviated, simplified relations and equations concerning the
problem of precise, subsequent time steps of the time development and concerning the problem of
Hubbard-Stratonovich transformations (HST,\cite{st}) to self-energies for coset decompositions \cite{physica6,pop1,pop2}.
Therefore, we briefly describe
in sections \ref{s2} to \ref{s3} various details of these transformations with appropriate time steps in a coherent state
path integral on the non-equilibrium time contour. Despite of widespread use in many particle physics,
these details of 'correct time step limits' of second quantized field operators are usually omitted for brevity, but are ubiquitous and have to be considered as soon as path integrations
of coherent state fields are performed for quantum effects beyond
classical approximations of the actions in the exponentials.

Although it is straightforward to obtain coherent state path integrals from insertion of overcomplete sets of coherent
states for discrete, subsequently separated steps of the time development, there arises the problem of a suitable, discrete
time-labelling of coherent state fields which substitute the field operators of second quantization in many particle systems.
Apart from the derivation of the pair condensate of fermions in sections to \ref{s2} to \ref{s3}, we briefly illustrate a
simpler problem for the second quantized Hamilton operator (\ref{s1_2}) with Fermi field operators \(\hat{\psi}_{\alpha}\),
\(\hat{\psi}_{\beta}\pdag\) (\ref{s1_1}) which are indexed by the first Greek letters \(\alpha,\,\beta,\,\gamma,\,\ldots\)
for various states as e.\ g.\ momentum space, coordinate space states or random states of disorder, etc.\ . Aside from
a one-particle operator \(\hat{h}_{\alpha\beta}\), a real, symmetric interaction potential \(\hat{V}_{\alpha\beta}\) is
introduced for nontrivial contributions in a many particle problem
\beq \lb{s1_1}
\delta_{\alpha\beta} &=&
\hat{\psi}_{\alpha}\;\hat{\psi}_{\beta}\pdag+
\hat{\psi}_{\beta}\pdag\;\hat{\psi}_{\alpha} =
\big\{\hat{\psi}_{\alpha}\;\boldsymbol{\Large,}\;\hat{\psi}_{\beta}\pdag\big\}_{\boldsymbol{+}}
\;;\hspace*{0.6cm}
\hat{\psi}_{\alpha}\;\hat{\psi}_{\beta}+\hat{\psi}_{\beta}\;\hat{\psi}_{\alpha} =
\big\{\hat{\psi}_{\alpha}\;\boldsymbol{\Large,}\;\hat{\psi}_{\beta}\big\}_{\boldsymbol{+}}=0\;;   \\ \lb{s1_2}
\hat{H}(\hat{\psi}_{\alpha}\pdag,\hat{\psi}_{\beta})
&=&\sum_{\alpha,\beta}\hat{\psi}_{\alpha}\pdag\;\hat{h}_{\alpha\beta}\;\hat{\psi}_{\beta}+\frac{1}{2}
\sum_{\alpha,\beta}\hat{\psi}_{\alpha}\pdag\;\hat{\psi}_{\beta}\pdag\;\hat{V}_{\beta\alpha}\;
\hat{\psi}_{\beta}\;\hat{\psi}_{\alpha}\;;   \\ \lb{s1_3}
\hat{h}_{\alpha\beta}\pdag &=& \hat{h}_{\alpha\beta}\;;\hspace*{0.6cm}
\hat{V}_{\alpha\beta}^{T} = \hat{V}_{\alpha\beta}\in\mathsf{R} \;.
\eeq
There exist various kinds of coherent state path integrals for the time development with a particular
second quantized Hamilton operator as the {\it normal ordering} '\(\hat{\psi}_{\alpha}\pdag\;\hat{\psi}_{\beta}\)',
the so called {\it Weyl ordering} '\((\hat{\psi}_{\alpha}\pdag\;\hat{\psi}_{\beta}-
\hat{\psi}_{\beta}\;\hat{\psi}_{\alpha}\pdag)/2\)' or the {\it anti-normal order}
'\(\hat{\psi}_{\beta}\;\hat{\psi}_{\alpha}\pdag\)' \cite{kash}; arbitrary mixtures between normal and anti-normal
order can also be chosen with the Weyl-ordering as an intermediate choice between the two mentioned extremes
of normal and anti-normal order. In this paper we restrict to a normal ordering and take for this
special case the precise, discrete steps of the time development involving the derivation of a nonlinear
sigma model with a coset decomposition.
The normal order of the prevailing Hamilton operator for the coherent state path integral avoids infinities or
ambiguities following from
\(\{\hat{\psi}_{\alpha}\;\boldsymbol{\Large,}\;\hat{\psi}_{\beta}\pdag\}_{\boldsymbol{+}}
=\delta_{\alpha\beta}\). As one considers the corresponding coherent
state path integral of (\ref{s1_1}-\ref{s1_3}) with normal ordering between two coherent
states \(|\psi(T_{\mathsf{ini}})\rangle\) and \(\langle\psi(T_{\mathsf{fin}})|\)
for the '{\sf ini}'tial and '{\sf fin}'al field configuration,
one has to notice the occurrence of coherent state fields in combinations as
'\(\psi_{\alpha}^{*}(t+\Delta t)\;\hat{h}_{\alpha\beta}\;\psi_{\beta}(t)\)' and
'\(\psi_{\alpha}^{*}(t+\Delta t)\;\psi_{\beta}^{*}(t+\Delta t)\;\frac{1}{2}\;\hat{V}_{\beta\alpha}\;
\psi_{\beta}(t)\;\psi_{\alpha}(t)\)' (\ref{s1_4}) as the
precise, discrete time steps instead of the more appealing 'hermitian' combinations
'\(\psi_{\alpha}^{*}(t)\;\hat{h}_{\alpha\beta}\;\psi_{\beta}(t)\)' and
'\(\psi_{\alpha}^{*}(t)\;\psi_{\beta}^{*}(t)\;\frac{1}{2}\;
\hat{V}_{\beta\alpha}\;\psi_{\beta}(t)\;\psi_{\alpha}(t)\)' (\ref{s1_5}) \cite{physica1}.
The latter combination (\ref{s1_5}) with equal times of \(\psi_{\alpha}^{*}(t)\), \(\psi_{\beta}(t)\) seems to be more preferable for
HST's to self-energies, coset decompositions and other transformations and approximations using an (indeed non-existent) hermitian
property of the actions under subsequent separation into discrete time steps. We apply the definition
\(|\psi(t)\rangle=\exp\{\sum_{\alpha}\psi_{\alpha}(t)\;\hat{\psi}_{\alpha}\pdag\}|0\rangle\)
(\ref{s1_6}-\ref{s1_9}) without the normalizing weight
\(\exp\{-\sum_{\alpha}\psi_{\alpha}^{*}(t)\;\psi_{\alpha}(t)\}\) for times \(T_{\mathsf{fin}} > t > T_{\mathsf{ini}}\)
so that this weight has additionally to be included
into the unit operator \(\hat{1}\) (\ref{s1_8},\ref{s1_9}) of overcomplete sets and into the definition of
appropriate initial and final configurations of coherent states
\(|\psi(T_{\mathsf{ini}})\rangle\), \(\langle\psi(T_{\mathsf{fin}})|\)
\beq\lb{s1_4}
\lefteqn{\big\langle\psi(T_{\mathsf{fin}})\big|\overleftarrow{\exp}\bigg\{-\frac{\im}{\hbar}\int_{T_{\mathsf{ini}}}^{T_{\mathsf{fin}}}
d t\;\hat{H}(\hat{\psi}_{\alpha}\pdag,\hat{\psi}_{\beta})
\bigg\}\big|\psi(T_{\mathsf{ini}})\big\rangle= } \\ \no &=&
\int\ldots d[\psi_{\alpha}^{*}(t+\Delta t),\psi_{\beta}(t+\Delta t)]\;
d[\psi_{\alpha}^{*}(t),\psi_{\beta}(t)]\;\ldots\;
e^{-\sum_{\alpha}\psi_{\alpha}^{*}(t+\Delta t)\;\psi_{\alpha}(t+\Delta t)}\;
e^{-\sum_{\alpha}\psi_{\alpha}^{*}(t)\;\psi_{\alpha}(t)}\;\ldots \times \\ \no &\times&\ldots
\big|\psi(t+\Delta t)\big\rangle\;\big\langle\psi(t+\Delta t)\big|
\underbrace{\exp\big\{-{\ts\frac{\im}{\hbar}}\Delta t\;
\overbrace{\hat{H}(\hat{\psi}_{\alpha}\pdag,\hat{\psi}_{\beta})}^{\mbox{\scz normal ordering !}}\big\}}_{
\approx 1-\frac{\im}{\hbar}\Delta t\:
\hat{H}(\hat{\psi}_{\alpha}\pdag,\hat{\psi}_{\beta})}\big|\psi(t)\big\rangle\;
\big\langle\psi(t)\big|\;\ldots\; = \\ \no &=&
\int d[\psi_{\alpha}^{*}(t),\psi_{\beta}(t)]\;\;
\exp\bigg\{-\frac{\im}{\hbar}\int_{T_{\mathsf{ini}}}^{T_{\mathsf{fin}}}d t\sum_{\alpha}\psi_{\alpha}^{*}(t+\Delta t)\;(-\im\hbar)
\frac{\psi_{\alpha}(t+\Delta t)-\psi_{\alpha}(t)}{\Delta t}\bigg\}\;\times \\ \no &\times&
\exp\bigg\{-\frac{\im}{\hbar}\int_{T_{\mathsf{ini}}}^{T_{\mathsf{fin}}}d t\sum_{\alpha,\beta}
\bigg(\psi_{\alpha}^{*}(t+\Delta t)\Big(\hat{h}_{\alpha\beta}-\im\:\ve_{+}\;\delta_{\alpha\beta}\Big)\psi_{\beta}(t)+  \\ \no &+&\frac{1}{2}
\psi_{\alpha}^{*}(t+\Delta t)\;\psi_{\beta}^{*}(t+\Delta t)\;\hat{V}_{\beta\alpha}\;\psi_{\beta}(t)\;\psi_{\alpha}(t)\bigg)\bigg\}\;;\hspace*{0.6cm}(\ve_{+}>0)\;;
\eeq
\beq\no
&&\mbox{convenient, 'lax' hermitian form :} \\  \lb{s1_5}
\lefteqn{\big\langle\psi(T_{\mathsf{fin}})\big|\overleftarrow{\exp}\bigg\{-\frac{\im}{\hbar}\int_{T_{\mathsf{ini}}}^{T_{\mathsf{fin}}}
d t\;\hat{H}(\hat{\psi}_{\alpha}\pdag,\hat{\psi}_{\beta})
\bigg\}\big|\psi(T_{\mathsf{ini}})\big\rangle= \int d[\psi_{\alpha}^{*}(t),\psi_{\beta}(t)]\;\;\times} \\ \no &\times&
\exp\bigg\{-\frac{\im}{\hbar}\int_{T_{\mathsf{ini}}}^{T_{\mathsf{fin}}}d t\sum_{\alpha,\beta}
\bigg(\psi_{\alpha}^{*}(t)\;\bigg(\Big(-\im\hbar\,
\frac{\pp}{\pp t}-\im\:\ve_{+}\Big)\delta_{\alpha\beta}+\hat{h}_{\alpha\beta}\bigg)\;
\psi_{\beta}(t)+ \frac{1}{2}\;\psi_{\alpha}^{*}(t)\;
\psi_{\beta}^{*}(t)\;\hat{V}_{\beta\alpha}\;\psi_{\beta}(t)\;\psi_{\alpha}(t)\bigg)\bigg\}\;;
\eeq
\beq \lb{s1_6}
|\psi(t)\rangle &=&\exp\Big\{\sum_{\alpha}\psi_{\alpha}(t)\;\hat{\psi}_{\alpha}\pdag\Big\}|0\rangle\;; \hspace*{0.65cm}
\hat{\psi}_{\alpha}\big|\psi(t)\big\rangle = \psi_{\alpha}(t)\big|\psi(t)\big\rangle \;;  \\  \lb{s1_7}
\langle\psi(t)| &=&\langle0|\exp\Big\{\sum_{\alpha}\hat{\psi}_{\alpha}\;\psi_{\alpha}^{*}(t)\Big\}\;; \hspace*{0.6cm}
\big\langle\psi(t)\big|\hat{\psi}_{\alpha}\pdag =\big\langle\psi(t)\big|\psi_{\alpha}^{*}(t) \;; \\  \lb{s1_8}
\hat{1}&=&\int d[\psi_{\alpha}^{*}(t),\psi_{\alpha}(t)]\;\;
\exp\Big\{-\sum_{\alpha}\psi_{\alpha}^{*}(t)\;\psi_{\alpha}(t)\Big\}\;\;
\big|\psi(t)\big\rangle\;\big\langle\psi(t)\big| \;;  \\   \lb{s1_9}
\mbox{Insert unit operator}&'\hat{1}'&\mbox{(\ref{s1_8}) with fields }\;
\psi_{\alpha}(t),\;\psi_{\alpha}^{*}(t)\mbox{ for times between }
(T_{\mathsf{fin}} > t > T_{\mathsf{ini}}) !\;.
\eeq
The problem of taking the proper, discrete time steps (\ref{s1_4}) in coherent state path integrals can be circumvented by a pragmatic point of view
which can be summarized as follows (and will also be applied to the more sophisticated problem for a pair condensate creation
from HST's, coset decompositions and a gradient expansion in sections \ref{s2} to \ref{s3})
\begin{itemize}
\item As far as classical approximations are only taken from the variations \(\delta/\delta \psi_{\alpha}^{*}(t+\Delta t)\),
\(\delta/\delta \psi_{\alpha}^{*}(t)\) of the actions in (\ref{s1_4}), (\ref{s1_5}),
the precise, discrete time steps '\(t+\Delta t\)' or '\(t\)'
do not matter if the chosen intervals '\(\Delta t\)' are sufficiently small for the maximal energy range \(\Delta\!E_{max}\)
of the physical problem (\(\Delta t\cdot\Delta\!E_{max}\ll 1\cdot\hbar\))
\beq\no
\frac{\delta}{\delta \psi_{\alpha}^{*}(t+\Delta t)}\Big(\mbox{Eq. (\ref{s1_4})}\Big)&\Longrightarrow&  \\   \lb{s1_10}
\im\hbar\frac{\psi_{\alpha}(t+\Delta t)-\psi_{\alpha}(t)}{\Delta t} &=&\sum_{\beta}\bigg(\hat{h}_{\alpha\beta}\;\psi_{\beta}(t)+
\psi_{\beta}^{*}(t+\Delta t)\;\hat{V}_{\beta\alpha}\;\psi_{\beta}(t)\;\psi_{\alpha}(t)\bigg)\;;  \\ \no
\frac{\delta}{\delta \psi_{\alpha}^{*}(t)}\Big(\mbox{Eq. (\ref{s1_5})}\Big)&\Longrightarrow&  \\   \lb{s1_11}
\im\hbar\frac{\psi_{\alpha}(t)}{\pp t} &=&\sum_{\beta}\bigg(\hat{h}_{\alpha\beta}\;\psi_{\beta}(t)+
\psi_{\beta}^{*}(t)\;\hat{V}_{\beta\alpha}\;\psi_{\beta}(t)\;\psi_{\alpha}(t)\bigg)\;.
\eeq
Hence, one should achieve physical results independent of the chosen discrete, time separation in the classical equations for
sufficiently small time intervals \(\Delta t\) adequate to the considered energy range of the classical many body problem.

\item However, as one tries to include quantum effects (e.\ g.\ quadratic quantum fluctuations around classical field solutions of the
actions), the precise form (\ref{s1_4}) of discrete time steps has to be used in the path integrations over coherent states. This is obvious
because \(\psi_{\alpha}^{*}(t+\Delta t)\) and \(\psi_{\alpha}^{*}(t)\) are unrelated, completely independent, different
integration variables in the coherent state path integral so that the path integration of the laxed form (\ref{s1_5})
'\(\psi_{\alpha}^{*}(t)\;\hat{h}_{\alpha\beta}\;\psi_{\beta}(t)+\psi_{\alpha}^{*}(t)\;\psi_{\beta}^{*}(t)\;
\frac{1}{2}\;\hat{V}_{\beta\alpha}\;
\psi_{\beta}(t)\;\psi_{\alpha}(t)\)' fails to give correct physical results and the precise, discrete time steps (\ref{s1_4})
'\(\psi_{\alpha}^{*}(t+\Delta t)\;\hat{h}_{\alpha\beta}\;\psi_{\beta}(t)+\psi_{\alpha}^{*}(t+\Delta t)\;
\psi_{\beta}^{*}(t+\Delta t)\;\frac{1}{2}\;
\hat{V}_{\beta\alpha}\;\psi_{\beta}(t)\;\psi_{\alpha}(t)\)' should be taken into account instead.
We have to point out that it is a very convenient standard form in literature to abbreviate the coherent state path integrals by their
(non-existent) hermitian versions (\ref{s1_5}) of the actions in order to simplify the appearance of equations, but has always to keep in mind
the precise, discrete separations of time steps (\ref{s1_4}) as soon as a path integration of coherent state fields is performed beyond
classical approximations from variations of the actions.
\end{itemize}

Despite of the non-hermitian property of the interaction potential
'\(\psi_{\alpha}^{*}(t+\Delta t)\;\psi_{\beta}^{*}(t+\Delta t)\;\frac{1}{2}\;\hat{V}_{\beta\alpha}\;
\psi_{\beta}(t)\;\psi_{\alpha}(t)\)' in (\ref{s1_4}),
it is possible to transform to a {\it hermitian} self-energy density \(\hat{\sigma}_{\alpha\beta}(t)\)
by a small modification of the HST for the case with the lax, hermitian form of interaction. Apart from the imaginary increment
\(-\im\,\ve_{+}\;\delta_{\alpha\beta}\) of \(\hat{h}_{\alpha\beta}\) for appropriate convergence properties of
Green functions, we include a suitable imaginary increment in the real, symmetric interaction
potential \(\hat{V}_{\alpha\beta}\) for a convergent Gaussian integration of the self-energy \(\hat{\sigma}_{\alpha\beta}(t)\)
\beq \lb{s1_12}
\hat{\sigma}_{\alpha\beta}\pdag(t) &=& \hat{\sigma}_{\alpha\beta}(t)\;;\hspace*{0.6cm}
\hat{V}_{\beta\alpha}\in\mathsf{R}\;;\hspace*{0.3cm}\hat{V}_{\alpha\beta}=\hat{V}_{\beta\alpha}\;;\hspace*{0.6cm}
\hat{\ve}_{+}=\delta_{\alpha\beta}\;\ve_{+}\;; \;\;\;\;(\ve_{+}>0)\;;  \\   \lb{s1_13}
1 &:=& \int d[\hat{\sigma}_{\alpha\beta}(t)]\;\;
\exp\bigg\{-\frac{\im}{2\:\hbar}\int_{T_{\mathsf{ini}}}^{T_{\mathsf{fin}}}d t\sum_{\alpha\beta}
\frac{1}{\hat{V}_{\alpha\beta}+\im\;\hat{\ve}_{+}}\;\times   \\  \no  &\times&
\Big(\hat{\sigma}_{\alpha\beta}(t)-\hat{V}_{\alpha\beta}\;\psi_{\alpha}(t)\;\psi_{\beta}^{*}(t+\Delta t)\Big)\;
\;\Big(\hat{\sigma}_{\beta\alpha}(t)-\hat{V}_{\beta\alpha}\;\psi_{\beta}(t)\;\psi_{\alpha}^{*}(t+\Delta t)\Big)\bigg\}  \;;
\\ \no \Longrightarrow &&   \\  \lb{s1_14}
\lefteqn{\exp\bigg\{-\frac{\im}{2\:\hbar}\int_{T_{\mathsf{ini}}}^{T_{\mathsf{fin}}}d t\sum_{\alpha\beta}
\psi_{\alpha}^{*}(t+\Delta t)\;\psi_{\beta}^{*}(t+\Delta t)\;
\hat{V}_{\beta\alpha}\;\psi_{\beta}(t)\;\psi_{\alpha}(t)\bigg\}=  }   \\ \no  &=&\lim_{\ve_{+}\rightarrow 0_{+}}
\int d[\hat{\sigma}_{\alpha\beta}(t)]\;\;
\exp\bigg\{-\frac{\im}{2\:\hbar}\int_{T_{\mathsf{ini}}}^{T_{\mathsf{fin}}}d t\sum_{\alpha\beta}
\frac{\hat{\sigma}_{\alpha\beta}(t)\;\;\hat{\sigma}_{\beta\alpha}(t)}{\hat{V}_{\alpha\beta}+\im\;\hat{\ve}_{+}}\bigg\}
\;\times   \\ \no &\times& \exp\bigg\{-\frac{\im}{\hbar}\int_{T_{\mathsf{ini}}}^{T_{\mathsf{fin}}}d t\sum_{\alpha\beta}
\psi_{\beta}^{*}(t+\Delta t)\;\hat{\sigma}_{\beta\alpha}(t)\;\psi_{\alpha}(t)\;
\underbrace{\frac{\hat{V}_{\alpha\beta}}{\hat{V}_{\alpha\beta}+\im\;\hat{\ve}_{+}}}_{:=1}\bigg\}_{\mbox{.}}
\eeq
Note the precise, discrete time steps '\(\psi_{\beta}^{*}(t+\Delta t)\;\hat{\sigma}_{\beta\alpha}(t)\;
\psi_{\alpha}(t)\)' in relation (\ref{s1_14}) which coincides with '\(\psi_{\alpha}^{*}(t+\Delta t)\;\hat{h}_{\alpha\beta}\;
\psi_{\beta}(t)\)' in (\ref{s1_4}) along the first lower subdiagonal of matrix elements of the discrete time indices.
Therefore, we define the matrix \(\hat{M}_{\beta\alpha}(t\ppr,t)\) (\ref{s1_15},\ref{s1_16})
for the states '\(\beta,\alpha\)' with the particular,
discrete time steps '\(t\ppr,t\)' specified by Kronecker deltas '\(\delta_{t\ppr,t}\simeq\delta(t\ppr-t)\)' and
'\(\delta_{t\ppr,t+\Delta t}\simeq\delta(t\ppr-t-\Delta t)\)'
which replace the corresponding delta functions; however, we omit the complete labeling with time step indices
for brevity (\(t,t\ppr=(t_{0}=T_{\mathsf{ini}}),\,(t_{1}=T_{\mathsf{ini}}+\Delta t),\,
\ldots,\,(t_{N-1}=T_{\mathsf{fin}}-\Delta t),\,(t_{N}=T_{\mathsf{fin}})\,\))
\beq \lb{s1_15}
\hspace*{-0.6cm}\Delta t{\cdot}\hat{M}_{\beta\alpha}(t\ppr,t)&=& \delta_{\beta\alpha}
\underbrace{\frac{\delta_{t\ppr\boldsymbol{\large,}t}-
\delta_{t\ppr\boldsymbol{\large,}t+\Delta t}}{\Delta t}}_{\simeq\pp/\pp t=-(\im/\hbar)\:\hat{E}}+
\underbrace{\frac{\im}{\hbar}\Big(\hat{h}_{\beta\alpha}-\im\;\hat{\ve}_{+}+
\hat{\sigma}_{\beta\alpha}(t)\Big)}_{\hat{m}_{\beta\alpha}(t)}\;\delta_{t\ppr\boldsymbol{\large,}t+\Delta t}  \;;    \\  \lb{s1_16}
\hspace*{-0.6cm}\Delta t{\cdot}\hat{M}_{\beta\alpha}(t\ppr,t)&=&\left(\bea{ccccccc}
\hat{1}_{\beta\alpha} &  \\ \hat{m}_{\beta\alpha}(t_{0})-\hat{1}_{\beta\alpha}  & \hat{1}_{\beta\alpha} & \\ &
\hat{m}_{\beta\alpha}(t_{1})-\hat{1}_{\beta\alpha} &  \\ &&&
\hspace*{-0.4cm}\ddots \hspace*{-0.4cm}& \\ &&&& \hat{1}_{\beta\alpha} \\
&&&& \hat{m}_{\beta\alpha}(t_{N-2})-\hat{1}_{\beta\alpha}  & \hat{1}_{\beta\alpha} & \\ &&&&&
\hat{m}_{\beta\alpha}(t_{N-1})-\hat{1}_{\beta\alpha} & \hat{1}_{\beta\alpha} \eea\right)_{\mbox{.}}
\eeq
After substitution of the HST (\ref{s1_12}-\ref{s1_14}) and the matrix
\(\hat{M}_{\beta\alpha}(t\ppr,t)\) (\ref{s1_15},\ref{s1_16}) into (\ref{s1_4}),
we acquire the path integral (\ref{s1_17},\ref{s1_18}) with bilinear coherent state fields which are to be removed by
Gaussian integration with convergence generating imaginary factor '\(-\im\:\ve_{+}\:\delta_{\beta\alpha}\)'
added to one particle operator \(\hat{h}_{\beta\alpha}\)
\beq  \lb{s1_17}
\lefteqn{\big\langle\psi(T_{\mathsf{fin}})\big|
\overleftarrow{\exp}\bigg\{-\frac{\im}{\hbar}\int_{T_{\mathsf{ini}}}^{T_{\mathsf{fin}}}
d t\;\hat{H}(\hat{\psi}_{\alpha}\pdag,\hat{\psi}_{\beta})
\bigg\}\big|\psi(T_{\mathsf{ini}})\big\rangle= } \\ \no &=&
\int d[\hat{\sigma}_{\alpha\beta}(t)]\;\;
\exp\bigg\{-\frac{\im}{2\:\hbar}\int_{T_{\mathsf{ini}}}^{T_{\mathsf{fin}}}d t\sum_{\alpha\beta}
\frac{\hat{\sigma}_{\alpha\beta}(t)\;\;\hat{\sigma}_{\beta\alpha}(t)}{\hat{V}_{\alpha\beta}+\im\;\hat{\ve}_{+}}\bigg\}
\;\times   \\ \no &\times& \int d[\psi_{\alpha}^{*}(t),\psi_{\beta}(t)]\;\;
\exp\bigg\{-\int_{T_{\mathsf{ini}}}^{T_{\mathsf{fin}}}d t\;d t\ppr\sum_{\alpha,\beta}
\psi_{\beta}^{*}(t\ppr)\;\hat{M}_{\beta\alpha}(t\ppr,t)\;\psi_{\alpha}(t)\bigg\} =  \\ \no &=&
\int d[\hat{\sigma}_{\alpha\beta}(t)]\;\;
\exp\bigg\{-\frac{\im}{2\:\hbar}\int_{T_{\mathsf{ini}}}^{T_{\mathsf{fin}}}d t\sum_{\alpha\beta}
\frac{\hat{\sigma}_{\alpha\beta}(t)\;\;\hat{\sigma}_{\beta\alpha}(t)}{\hat{V}_{\alpha\beta}+\im\;\hat{\ve}_{+}}\bigg\}
\;\mbox{det}\Big[\hat{M}_{\beta\alpha}(t\ppr,t)\Big]  \;;  \\   \lb{s1_18}
\mbox{det}\Big[\hat{M}_{\beta\alpha}(t\ppr,t)\Big] &=&
\mbox{det}\bigg[\delta_{\beta\alpha}\frac{\delta_{t\ppr\boldsymbol{\large,}t}-
\delta_{t\ppr\boldsymbol{\large,}t+\Delta t}}{\Delta t}+\frac{\im}{\hbar}
\Big(\hat{h}_{\beta\alpha}-\im\:\hat{\ve}_{+}+\hat{\sigma}_{\beta\alpha}(t)\Big)\;
\delta_{t\ppr\boldsymbol{\large,}t+\Delta t}\bigg]_{\mbox{.}}
\eeq
In this manner one has mapped the original coherent state path integral (\ref{s1_4}) of Grassmann fields
with precise, discrete time steps to a path integral (\ref{s1_17},\ref{s1_18}) with the hermitian self-energy integration
variables \(\hat{\sigma}_{\beta\alpha}(t)\) (\ref{s1_12}-\ref{s1_14})
by using the matrix \(\hat{M}_{\beta\alpha}(t\ppr,t)\) (\ref{s1_15},\ref{s1_16})
with one-particle operator \(\hat{h}_{\beta\alpha}\). If one only applies a lax kind of matrix
as \(\underline{\hat{M}}_{\beta\alpha}(t\ppr,t)\) (\ref{s1_19}) {\it with missing} infinitesimal time
shifts \(\Delta t\) as already given in relation (\ref{s1_5})
\beq  \lb{s1_19}
\lefteqn{
\underline{\hat{M}}_{\beta\alpha}(t\ppr,t) =
\delta_{\beta\alpha}\frac{\delta_{t\ppr\boldsymbol{\large,}t}-
\delta_{t\ppr\boldsymbol{\large,}t+\Delta t}}{\Delta t}+\underbrace{\frac{\im}{\hbar}
\Big(\hat{h}_{\beta\alpha}-\im\:\ve_{+}+\hat{\sigma}_{\beta\alpha}(t)\Big)}_{\hat{m}_{\beta\alpha}(t)}\;
\delta_{t\ppr\boldsymbol{\large,}t}    =  } \\ \no &=&
\left(\bea{ccccccc}
\hat{1}_{\beta\alpha}+\hat{m}_{\beta\alpha}(t_{0}) &  \\ -\hat{1}_{\beta\alpha}  &
\hat{1}_{\beta\alpha} +\hat{m}_{\beta\alpha}(t_{1})& \\  & -\hat{1}_{\beta\alpha}  &     \\  &&&
\hspace*{-0.4cm}\ddots \hspace*{-0.4cm}& \\ &&&& \hat{1}_{\beta\alpha} +\hat{m}_{\beta\alpha}(t_{N-2}) \\
&&&& -\hat{1}_{\beta\alpha}  & \hat{1}_{\beta\alpha}+\hat{m}_{\beta\alpha}(t_{N-1}) & \\ &&&&&
-\hat{1}_{\beta\alpha} & \hat{1}_{\beta\alpha}+\hat{m}_{\beta\alpha}(t_{N}) \eea\right)_{\mbox{,}}
\eeq
one has obviously made a mistake because the precise location of the self-energy as integration variable
enters into the computation of the fermion determinant and path integral. This problem of the shift
of \(\hat{m}_{\beta\alpha}(t)\) to its first lower sub-diagonal is also inherent in the original
Dyson equation which utilizes the corresponding relation (\ref{s1_20}) with the self-energy matrix
\(\hat{\Sigma}_{\beta\alpha}(t\ppr,t)\) (\ref{s1_22}),
the total Green function \(\hat{G}_{\beta\alpha}(t\ppr,t)\)
and the free Green function \(\hat{g}_{\beta\alpha}(t\ppr,t)\) (\ref{s1_21})
instead of our path integral (\ref{s1_17},\ref{s1_18})
\beq \lb{s1_20}
\hat{M}_{\beta\alpha}(t\ppr,t) &=&\hat{G}_{\beta\alpha}^{-1}(t\ppr,t)=\hat{g}_{\beta\alpha}^{-1}(t\ppr,t)+
\hat{\Sigma}_{\beta\alpha}(t\ppr,t) \;;  \\    \lb{s1_21}
\hat{g}_{\beta\alpha}^{-1}(t\ppr,t) &=&\delta_{\beta\alpha}\;
\frac{\delta_{t\ppr\boldsymbol{\large,}t}-\delta_{t\ppr\boldsymbol{\large,}t+\Delta t}}{\Delta t}+
\frac{\im}{\hbar}\big(\hat{h}_{\beta\alpha}-\im\;\ve_{+}\big)\;
\underbrace{\delta_{t\ppr\boldsymbol{\large,}t+\Delta t}}_{\propto\;\delta(t\ppr-t-\Delta t)} \;;  \\   \lb{s1_22}
\hat{\Sigma}_{\beta\alpha}(t\ppr,t) &=&\frac{\im}{\hbar}\;\hat{\sigma}_{\beta\alpha}(t)\;
\underbrace{\delta_{t\ppr\boldsymbol{\large,}t+\Delta t}}_{\propto\;\delta(t\ppr-t-\Delta t)} \;.
\eeq
The exact form of the Dyson equation includes an infinitesimal shift \(\Delta t\) in the delta functions
\(\delta(t\ppr-t-\Delta t)\) in order to avoid the action of a field operator
\(\hat{\psi}_{\alpha}(t)\) with its hermitian conjugate
\(\hat{\psi}_{\alpha}\pdag(t)\) at the same time point. (This essential time shift
\(\Delta t\) is usually omitted for convenience in standard representations of the Dyson equation !).
As one starts to solve the Dyson equation (e.\ g.\ by iteration) without these time shifts \(\Delta t\)
in the delta functions, one fails to achieve correct, meaningful physical results. Therefore, the removal
of these infinitesimal, inconspicuous time shifts \(\Delta t\) proves to be erroneous at the
quantum level where one has to perform correct path integrations of the prevailing independent
integration variables. However, we again emphasize that these shifts and limits with time interval \(\Delta t\)
are assumed to be present in the physical literature of many particle physics even though they are usually omitted
in the presentation of equations for brevity and clarity of mathematical relations.

In the remainder of this paper we briefly describe by various amendments the precise location
of proper time shifts \(\Delta t_{p=\pm}\) on the non-equilibrium time contour; these brief
amendments of the necessary, precise time shifts \(\Delta t_{p=\pm}\) at the quantum level
are performed for the more sophisticated problem with a spontaneous symmetry breaking to a
pair condensate composed of fermionic constituents where various HST's and coset decompositions
are involved in the derivation for coset matrices \(\hat{T}(\vec{x},t_{p})\) consisting of
anomalous terms. We distinguish between the two cases of a spatially short-ranged
\(V_{0}\,\delta_{\vec{x},\vec{x}\ppr}\) and long-ranged \(V_{|\vec{x}-\vec{x}\ppr|}\) interaction
potential in sections \ref{s22}-\ref{s24} and \ref{s31},\ref{s32}, respectively. The short-ranged
interaction only yields local self-energies after the HST whereas the general, long-ranged
interaction case results into spatially nonlocal self-energies whose dimensions are extended
by the total number \(\mcal{N}_{x}\) (\ref{s1_23}) of spatial grid points
for a D-dimensional spherical volume with discrete intervals \(\Delta x\) in each dimension
and radial length \(L\) (\(\Omega_{D}:=\)surface angle, \(\Omega_{D=2}=2\pi\),
\(\Omega_{D=3}=4\pi\))
\be \lb{s1_23}
\mcal{N}_{x}=\frac{\Omega_{D}}{D}\bigg(\frac{L}{\Delta x}\bigg)^{D}\;.
\ee
We emphasize that the total number of considered space points \(\mcal{N}_{x}\) (\ref{s1_23})
can be applied as the relevant parameter for saddle point approximations and gradient expansions
in particular for the bulk of an ordered (or even disordered) system where the (coupled) field variables
of the self-energy at the various, different space points can be taken as equivalent in the
case of an approximate translational space symmetry.

The parameter \(\mcal{N}_{x}\) (\ref{s1_23}) with the maximum possible energy
\(\hbar\,\Omega=\hbar\,(1/|\Delta t_{p}|\,)\), due to discrete time steps, appears in
the transformation of the determinant of an operator \(\breve{\mscr{O}}\) to its
exponential-trace-logarithm form because this transformation requires space and time integrals
for the extraction of actions from the trace-logarithm term. The trace does not only comprise
internal degrees of freedom as angular momentum or spin \(s=\uparrow,\downarrow\), but has
also to incorporate the discrete, spatial grid and involved discrete time points
\beq \lb{s1_24}
\mbox{DET}\big\{\breve{\mscr{O}}\big\} &=&\exp\Big\{\mbox{TR}\ln\breve{\mscr{O}}\Big\} =
\exp\bigg\{\int_{C}\frac{d t_{p}}{\hbar}\eta_{p}\sum_{\vec{x}}\mcal{N}\mbox{Tr}\ln\breve{\mscr{O}}\bigg\} \;;
\\ \lb{s1_25} \mbox{TR}\Big[\ldots\Big] &\stackrel{\wedge}{=}& \mbox{complete trace with summation over discrete
space and time points} \\ \no && \mbox{{\it without infinitesimal volume elements}} \;; \\ \lb{s1_26}
\mbox{Tr}\Big[\ldots\Big] &\stackrel{\wedge}{=}& \mbox{trace over internal degrees of freedom} \;; \\ \lb{s1_27}
\sum_{\vec{x}}\ldots &\stackrel{\wedge}{=}& \int_{|\vec{x}|<L}\Big(d^{D}\!x/S^{D}\Big)\;\ldots\;\;\;;
(\mbox{space integral normalized by spherical volume }\;S^{D})\;;  \\ \lb{s1_28}
\int_{C}d t_{p}\ldots &\stackrel{\wedge}{=}& \mbox{time contour integral }\;;\;\;\;
\eta_{p}=\mbox{time contour metric}\;;\;\;(\mbox{cf. following Eqs. (\ref{s2_8},\ref{s2_9}) })\;; \\ \lb{s1_29}
\mcal{N} &\stackrel{\wedge}{=}& \big(\hbar\Omega\big)\cdot\mcal{N}_{x}\;\;\;;\;\;\;
\Omega=1/|\Delta t_{p}|\;\;\;.
\eeq
In the final sections \ref{s41}, \ref{s42} we describe how to transform from
'path integration field variables' on an underlying Euclidean, 'flat' spatial grid to
spherical coordinates (for \(D=2,3\)) in order to take into account rotational symmetries
of the actions. This transformation is performed under the assumption that the coherent state
path integrals in discrete form can be regarded as ordinary integrals of complex variables
in multiple dimensions following from the \(\mcal{N}_{x}\) (\ref{s1_23})
discrete grid points. Therefore, we start out from following metric of \((dS)^{2}\) (\ref{s1_30})
with complex (,anti-commuting) fields \(\psi_{\vec{x},s}^{*}(t_{p})\), \(\psi_{\vec{x},s}(t_{p})\)
where the summations over the spatial grid points are included apart from internal degrees
of freedom (e.g. the spin summation in following sections)
\be \lb{s1_30}
\big(dS\big)^{2} =\sum_{\vec{x}}\sum_{s=\uparrow,\downarrow}
d\psi_{\vec{x},s}^{*}(t_{p})\;\;d\psi_{\vec{x},s}(t_{p})\;.
\ee
Since the inverse square root of the metric tensor of \((dS)^{2}\) (\ref{s1_30})
determines the invariant integration measure, one obtains the corresponding integration measure
in spherical coordinates by transforming with the spherical basis functions and by
regrouping in \((dS)^{2}\) the transformed fields with spherical coordinate 'indices'.

\section{Coherent state path integral for a pair condensate composed of fermions} \lb{s2}

\subsection{Grassmann fields of fermionic operators and the non-equilibrium time contour} \lb{s21}

It is our main destination to supplement the various, essential time shifts of matrix elements
analogous to \(\hat{M}_{\beta\alpha}(t\ppr,t)\) (\ref{s1_15},\ref{s1_16})
in order to derive a pair condensate in terms of
coset matrices \(\hat{T}(\vec{x},t_{p})\) after HST's, coset decompositions and also a gradient
expansion. We consider fermionic field operators \(\hat{\psi}_{\vec{x},s}\),
\(\hat{\psi}_{\vec{x},s}\pdag\) (\ref{s2_1},\ref{s2_2}) on a D-dimensional, Cartesian space grid, specified
by the D-dimensional vectors \(\vec{x},\,\vec{x}\ppr,\,\vec{x}_{1},\,\vec{x}_{1}\ppr,\,\ldots,\,
\vec{x}_{2},\,\vec{x}_{2}\ppr,\,\ldots\) , and include the spin $1/2$ angular momentum
\(s,\,s\ppr,\,\ldots=\uparrow,\,\downarrow\) as internal degree of freedom
\beq  \lb{s2_1}
\hat{\psi}_{\vec{x},s}\;;\;\hat{\psi}_{\vec{x}\ppr,s\ppr}\pdag &:&
s,\,s\ppr,\,s_{1},\,s_{1}\ppr,\,\ldots\,,\,s_{2},\,s_{2}\ppr,\,\ldots=\uparrow\;\mbox{ or }\;
\downarrow \;;   \\   \lb{s2_2}
\big\{\hat{\psi}_{\vec{x},s}\;\boldsymbol{\large,}\;
\hat{\psi}_{\vec{x}\ppr,s\ppr}\pdag\big\}_{\boldsymbol+}  &=&\delta_{\vec{x},\vec{x}\ppr}\;
\delta_{ss\ppr}\;;\hspace*{0.6cm}\big\{\hat{\psi}_{\vec{x},s}\;\boldsymbol{\large,}\;
\hat{\psi}_{\vec{x}\ppr,s\ppr}\big\}_{\boldsymbol+} = 0 \;.
\eeq
Aside from the real two-body potential \(V_{|\vec{x}\ppr-\vec{x}|}\), the second quantized Hamilton
operator \(\hat{H}(\hat{\psi}\pdag,\hat{\psi},t)\) (\ref{s2_3}) consists of the
spin independent one-particle operator \(\hat{h}(\vec{x})\) (\ref{s2_4}) with kinetic energy and external potential
\(u(\vec{x})\) which is shifted by the zero-temperature, chemical potential \(\mu_{0}\) as a reference
energy. Furthermore, the second quantized Hamilton operator
\(\hat{H}(\hat{\psi}\pdag,\hat{\psi},t)\) (\ref{s2_3}) contains symmetry breaking
source fields, as the anti-commuting fields \(j_{\psi;s}(\vec{x},t)\), \(j_{\psi;s}\pdag(\vec{x},t)\) (\ref{s2_5})
for a coherent, macroscopic field \(\langle\psi_{\vec{x},s}(t)\rangle\) of fermions and the complex,
even-valued, anti-symmetric field matrices \(\hat{j}_{\psi\psi;ss\ppr}(\vec{x},t)\),
\(\hat{j}_{\psi\psi;ss\ppr}\pdag(\vec{x},t)\) (\ref{s2_6}) for creating anomalous pair condensates as
\(\langle\psi_{\vec{x},s}(t)\;\psi_{\vec{x}\ppr,s\ppr}(t)\rangle\) and their hermitian conjugates
\(\langle\psi_{\vec{x},s}\pdag(t)\;\psi_{\vec{x}\ppr,s\ppr}\pdag(t)\rangle\)
\beq \lb{s2_3}
\hat{H}(\hat{\psi}\pdag,\hat{\psi},t) &=&
\sum_{\vec{x}}\sums \hat{\psi}_{\vec{x},s}\pdag\;\;
\hat{h}(\vec{x})\;\; \hat{\psi}_{\vec{x},s}+
\sum_{\vec{x},\vec{x}\ppr}\sumss
\hat{\psi}_{\vec{x}\ppr,s\ppr}\pdag\;\hat{\psi}_{\vec{x},s}\pdag\;V_{|\vec{x}\ppr-\vec{x}|}\;
\hat{\psi}_{\vec{x},s}\;\hat{\psi}_{\vec{x}\ppr,s\ppr}  +  \\
\no &+& \sum_{\vec{x}}\sums
\Big(j_{\psi;s}\pdag(\vec{x},t)\;\hat{\psi}_{\vec{x},s} +
\hat{\psi}_{\vec{x},s}\pdag\;j_{\psi;s}(\vec{x},t)\Big) +
\\ \no &-&\frac{1}{2}
\sum_{\vec{x}}\trs\bigg[\hat{j}_{\psi\psi;s\ppr s}\pdag(\vec{x},t)\;
\hat{\psi}_{\vec{x},s}\;\hat{\psi}_{\vec{x},s\ppr}+
\hat{\psi}_{\vec{x},s}\pdag\;\hat{\psi}_{\vec{x},s\ppr}\pdag\;
\hat{j}_{\psi\psi;s\ppr s}(\vec{x},t)\bigg]\;;  \\    \lb{s2_4}
\hat{h}(\vec{x})&=&\frac{\vec{p}^{\;2}}{2m}+u(\vec{x})-\mu_{0} \;;   \\    \lb{s2_5}
j_{\psi;s}(\vec{x},t)\;,\;j_{\psi;s}\pdag(\vec{x},t)&\in& \mscr{C}_{odd}\;\;; \hspace*{0.3cm}
\hat{j}_{\psi\psi;ss\ppr}(\vec{x},t)\;\in\;\mathsf{C}_{even}\;\;;\;\;\;
\hat{j}_{\psi\psi;ss\ppr}^{T}(\vec{x},t)=-\hat{j}_{\psi\psi;ss\ppr}(\vec{x},t)\;; \\  \lb{s2_6}
\hat{j}_{\psi\psi;ss\ppr}(\vec{x},t)&=&\big(\hat{\tau}_{2}\big)_{ss\ppr}\;j_{\psi\psi}(\vec{x},t)\;;\;\;
j_{\psi\psi}(\vec{x},t)=\big|j_{\psi\psi}(\vec{x},t)\big|\;\;\exp\{\im\:\gamma(\vec{x},t)\}\;.
\eeq
We point out the particular definition (\ref{s2_7}) for the complex conjugation of a product of anti-commuting fields
\(\xi_{1}\,\ldots\,\xi_{n}\) so that
the '{\it real}' term \(\psi_{\vec{x},s}^{*}(t)\;\psi_{\vec{x},s}(t)\) or \(\xi_{j}^{*}\xi_{j}\)
of coherent state fields is reproduced under the defined complex conjugated transformation
\be \lb{s2_7}
\left(\xi_{1}\ldots\xi_{n}\right)^{*}=\xi_{n}^{*}\ldots\xi_{1}^{*}\;;\hspace*{0.3cm}
\xi_{j}^{**}=\xi_{j}\;;\hspace*{0.3cm}\big(\xi_{j}^{*}\xi_{j}\big)^{*}=\xi_{j}^{*}\xi_{j}^{**}=\xi_{j}^{*}\xi_{j}\;.
\ee
In comparison to section \ref{s11}, we introduce a forward and an additional backward time development (\ref{s2_8}) for the coherent
state matrix elements of the propagator of the second quantized Hamilton operator
\(\hat{H}(\hat{\psi}\pdag,\hat{\psi},t)\) (\ref{s2_3}); this guarantees the generating functional to
be normalized to unity. The '{\sf ini}'tial ('{\sf fin}'al) coherent state field configurations are created from the vacuum by the source
fields in the last two lines of \(\hat{H}(\hat{\psi}\pdag,\hat{\psi},t)\) (\ref{s2_3}). The
forward '$+$' and backward '$-$' time development is combined by the non-equilibrium time contour
'\(t_{p=\pm}\)','\(t_{q=\pm}\ppr\)' (\ref{s2_8}) where one has to distinguish between time variables
'\(t_{+}\)', '\(t_{-}\)' , '\(t_{+}\ppr\)', '\(t_{-}\ppr\)'
(indices '\(p,\,q=\pm\)') for forward and backward propagation by using a time contour metric \(\eta_{p=\pm}=p=\pm1\)
\beq   \lb{s2_8}
\int_{C}d t_{p}\ldots &=&\int_{T_{\mathsf{ini}}}^{T_{\mathsf{fin}}}
dt_{+}\ldots+ \int_{T_{\mathsf{fin}}}^{T_{\mathsf{ini}}}d t_{-}\ldots= \int_{T_{\mathsf{ini}}}^{T_{\mathsf{fin}}}d t_{+}\ldots-
\int_{T_{\mathsf{ini}}}^{T_{\mathsf{fin}}}d t_{-}\ldots \\ \no &=&
\sum_{p=\pm}\int_{T_{\mathsf{ini}}}^{T_{\mathsf{fin}}}\big|d t_{p}\big|\;\;\eta_{p}\;\;\ldots\;\;;\hspace*{0.6cm}
\eta_{p=\pm}=p=\pm1\;;  \\ \lb{s2_9}  \Delta t_{p} &=&\big|\Delta t_{p}\big|\;\;\eta_{p}\;\;\;.
\eeq
Since we aim on the derivation of pair condensates with coset matrices, we have to perform an anomalous doubling of fields
\(\psi_{\vec{x},s}(t_{p})\) with their complex conjugates \(\psi_{\vec{x},s}^{*}(t_{p})\) to the composed field
\(\Psi_{\vec{x},s}^{a}(t_{p})\) (\ref{s2_10}) where the first letters \(a,\,b,\,\ldots\) of the Latin alphabet specify one out of the
two possible components of \(\Psi_{\vec{x},s}^{a}(t_{p})\) (\ref{s2_10})
(and similar for its anomalous doubled, hermitian-conjugated form \(\Psi_{\vec{x},s}^{\dag a}(t_{p})\) (\ref{s2_11}))
\beq
\no \mbox{(1)}&:& \mbox{'{\it equal time}', anomalous-doubled field :}  \\  \lb{s2_10}
\Psi_{\vec{x},s}^{a(=1/2)}(t_{p})&=&\left(\bea{c}\psi_{\vec{x},s}(t_{p}) \\ \psi_{\vec{x},s}^{*}(t_{p})\eea\right)^{a}=
\Big(\underbrace{\psi_{\vec{x},s}(t_{p})}_{a=1}\;;\;\underbrace{\psi_{\vec{x},s}^{*}(t_{p})}_{a=2}\Big)^{T}\;;  \\
\no \mbox{(2)} &:& \mbox{'hermitian-conjugation' '$\pdag$' of '{\it equal time}', anomalous-doubled field :} \\  \lb{s2_11}
\Psi_{\vec{x},s}^{\dag a(=1/2)}(t_{p})&=&
\Big(\underbrace{\psi_{\vec{x},s}^{*}(t_{p})}_{a=1}\;;\;\underbrace{\psi_{\vec{x},s}(t_{p})}_{a=2}\Big)\;.
\eeq
According to the analogous normal ordering of \(\hat{H}(\hat{\psi}\pdag,\hat{\psi},t)\) (\ref{s2_3})
for the 'contour time' development '\(\mscr{T}_{p}\)' similar to (\ref{s1_4})
\beq \lb{s2_12}
\big\langle0\big|\mscr{T}_{p}\overleftarrow{\exp}\bigg\{-\frac{\im}{\hbar}\int_{C}dt_{p}\;\;
\hat{H}(\hat{\psi}\pdag,\hat{\psi},t)\bigg\}\big|0\big\rangle &\equiv& 1 \;, \\  \no
\mbox{ordering operator on the time contour} & : & \mscr{T}_{p} \;\;\;,
\eeq
one has to add an infinitesimal, non-equilibrium time shift \(\Delta t_{p}=|\Delta t_{p}|\;\eta_{p}\) in the complex
conjugated second part of \(\Psi_{\vec{x},s}^{a=2}(t_{p})=\psi_{\vec{x},s}^{*}
(t_{p})\rightarrow\psi_{\vec{x},s}^{*}(t_{p}+\Delta t_{p})\); however, the 'true' hermitian conjugation of
\(\Psi_{\vec{x},s}^{a(=1/2)}(t_{p})\) (\ref{s2_10}) just yields 'equal time' fields in \(\Psi_{\vec{x},s}^{\dag a(=1/2)}(t_{p})\) (\ref{s2_11})
or under consideration of the just mentioned time shift \(\psi_{\vec{x},s}^{*}(t_{p})\rightarrow
\psi_{\vec{x},s}^{*}(t_{p}+\Delta t_{p})\) in \(\Psi_{\vec{x},s}^{a=2}(t_{p})\) (\ref{s2_10}) the incorrect, hermitian conjugated
case where the complex conjugated field \(\psi_{\vec{x},s}^{*}(t_{p})\) in the first part \((a=1)\) acts before its second
part \(\psi_{\vec{x},s}(t_{p}+\Delta t_{p})\) \((a=2)\) contrary to the normal ordering of the second quantized field
operators in \(\hat{H}(\hat{\psi}\pdag,\hat{\psi},t)\) (\ref{s2_3}) for the
non-equilibrium time development. Therefore, we have to define an additional hermitian conjugation
\(\breve{\Psi}_{\vec{x},s}^{\sharp a(=1/2)}(t_{p})\) (\ref{s2_14}) of a slightly modified, anomalous doubled field
\(\breve{\Psi}_{\vec{x},s}^{a(=1/2)}(t_{p})\) (\ref{s2_13}) which is adapted to the normal ordering of
\(\hat{H}(\hat{\psi}\pdag,\hat{\psi},t)\) (\ref{s2_3}) with the field creation operators
\(\hat{\psi}_{\vec{x},s}\pdag\) left to the annihilation operators
\(\hat{\psi}_{\vec{x},s}\), thereby always acting the time interval \(\Delta t_{p}\) later on the time contour
\footnote{The fields \(\breve{\Psi}_{\vec{x},s}^{a(=1/2)}(t_{p})\) or \(\breve{\Psi}_{\vec{x},s}^{\sharp a(=1/2)}(t_{p})\), which
include the additional time shifts '\(\Delta t_{p}\)' in the complex parts \(\psi_{\vec{x},s}^{*}(t_{p}+\Delta t_{p})\)
relative to \(\psi_{\vec{x},s}(t_{p})\), are marked by the symbol '\(\boldsymbol{\breve{\ph{\Psi}}}\)'
above the Greek, capital letter of the field and also above an analogous matrix.}
\beq
\no \mbox{(1)}&:& \mbox{'{\it time shifted}' $\Delta t_{p}$, anomalous-doubled field '$\boldsymbol{\breve{\ph{\Psi}}}$' :}  \\  \lb{s2_13}
\breve{\Psi}_{\vec{x},s}^{a(=1/2)}(t_{p})&=&\left(\bea{c}\psi_{\vec{x},s}(t_{p}) \\
\psi_{\vec{x},s}^{*}(t_{p}+\Delta t_{p})\eea\right)^{a}=
\Big(\underbrace{\psi_{\vec{x},s}(t_{p})}_{a=1}\;;\;\underbrace{\psi_{\vec{x},s}^{*}(t_{p}+\Delta t_{p})}_{a=2}\Big)^{T}\;;  \\
\no \mbox{(2)} &:& \mbox{'hermitian-conjugation' '$^{\sharp}$' with '{\it time shift correction}' \(\Delta t_{p}\)} \\ \no &&
\mbox{in the resulting complex part :}    \\  \lb{s2_14}
\breve{\Psi}_{\vec{x},s}^{a(=1/2)}(t_{p}) &\stackrel{'\sharp'}{\Longrightarrow}&
\breve{\Psi}_{\vec{x},s}^{\sharp a(=1/2)}(t_{p}) = \Big(\underbrace{\psi_{\vec{x},s}^{*}(t_{p}+\Delta t_{p})}_{a=1}\;;\;
\underbrace{\psi_{\vec{x},s}(t_{p})}_{a=2}\Big)\;.
\eeq
In later steps to the derivation of the pair condensates, we have to take dyadic products (\ref{s2_15}) and (\ref{s2_16})
of the anomalous doubled fields \(\Psi_{\vec{x},s}^{a(=1/2)}(t_{p})\) (\ref{s2_10}), \(\breve{\Psi}_{\vec{x},s}^{a(=1/2)}(t_{p})\) (\ref{s2_13})
with their corresponding hermitian conjugates
\(\Psi_{\vec{x},s}^{\dag a(=1/2)}(t_{p})\) (\ref{s2_11}), \(\breve{\Psi}_{\vec{x},s}^{\sharp a(=1/2)}(t_{p})\) (\ref{s2_14})
so that there arise two distinct forms of dyadic products
\beq \lb{s2_15}
\hat{\Phi}_{\vec{x},s;\vec{x}\ppr,s\ppr}^{ab}(t_{p})&=&
\Psi_{\vec{x},s}^{a}(t_{p})\otimes \Psi_{\vec{x}\ppr,s\ppr}^{\dag b}(t_{p})  \;;  \\   \lb{s2_16}
\breve{\Phi}_{\vec{x},s;\vec{x}\ppr,s\ppr}^{ab}(t_{p})&=&
\breve{\Psi}_{\vec{x},s}^{a}(t_{p})\otimes \breve{\Psi}_{\vec{x}\ppr,s\ppr}^{\sharp b}(t_{p})   \;.
\eeq
The latter dyadic product \(\breve{\Psi}_{\vec{x},s}^{a}(t_{p})\otimes \breve{\Psi}_{\vec{x}\ppr,s\ppr}^{\sharp b}(t_{p})\)
(\ref{s2_16}) is more appropriate for the correct time ordering in the anomalous doubled coherent state
path integrals and for the definition of density matrices whereas we have to apply the first, 'equal time' version
\(\Psi_{\vec{x},s}^{a}(t_{p})\otimes \Psi_{\vec{x}\ppr,s\ppr}^{\dag b}(t_{p})\) (\ref{s2_15}) for the definition of a total, hermitian
self-energy matrix. In consequence, we list in relations (\ref{s2_17},\ref{s2_18}) and (\ref{s2_19},\ref{s2_20})
the two different cases of dyadic products of anomalous
doubled fields corresponding to (\ref{s2_15}) and to (\ref{s2_16}), respectively.
The first case (\ref{s2_15},\ref{s2_17}-\ref{s2_18}) is completely hermitian by using the ordinary
hermitian conjugation '\(^{\dag}\)' for defining a hermitian self-energy whereas the second case (\ref{s2_16},\ref{s2_19}-\ref{s2_20})
takes into account the shifts with the time interval \(\Delta t_{p}\) for \(\psi_{\vec{x}\ppr,s\ppr}^{*}(t_{p}+\Delta t_{p})\) (compared to
\(\psi_{\vec{x},s}(t_{p})\)) in order to avoid incorrect, 'equal time' combinations as
\(\psi_{\vec{x}\ppr,s\ppr}^{*}(t_{p})\,\ldots\,\psi_{\vec{x},s}(t_{p})\) in the coherent state path integral.
The two different kinds of dyadic products of anomalous doubled fields can be considered as spatially nonlocal order parameters
\(\hat{\Phi}_{\vec{x},s;\vec{x}\ppr,s\ppr}^{ab}(t_{p})\) (\ref{s2_17}),
\(\breve{\Phi}_{\vec{x},s;\vec{x}\ppr,s\ppr}^{ab}(t_{p})\) (\ref{s2_19}) with
density terms on the block diagonals (\(a=b\)) and anomalous pair condensates on the off-diagonal blocks (\(a\neq b\))
\beq  \lb{s2_17}
\hat{\Phi}_{\vec{x},s;\vec{x}\ppr,s\ppr}^{ab}(t_{p})&=&
\Psi_{\vec{x},s}^{a}(t_{p})\;\otimes\;
\Psi_{\vec{x}\ppr,s\ppr}^{\dagger b}(t_{p}) = \left( \bea{c}
\psi_{\vec{x},s}(t_{p}) \\ \psi_{\vec{x},s}^{*}(t_{p}) \eea\right)^{a} \otimes
\Big(\psi_{\vec{x}\ppr,s\ppr}^{*}(t_{p})\;;\;\psi_{\vec{x}\ppr,s\ppr}(t_{p})\Big)^{b}
\\ \no &=&\left( \bea{cc}
\langle\psi_{\vec{x},s}(t_{p})\;\psi_{\vec{x}\ppr,s\ppr}^{*}(t_{p})\rangle &
\langle\psi_{\vec{x},s}(t_{p})\;\psi_{\vec{x}\ppr,s\ppr}(t_{p})\rangle \\
\langle\psi_{\vec{x},s}^{*}(t_{p})\;\psi_{\vec{x}\ppr,s\ppr}^{*}(t_{p})\rangle &
\langle\psi_{\vec{x},s}^{*}(t_{p})\;\psi_{\vec{x}\ppr,s\ppr}(t_{p})\rangle \eea\right) =
\left( \bea{cc}
\hat{\Phi}_{\vec{x},s;\vec{x}\ppr,s\ppr}^{11}(t_{p}) &
\hat{\Phi}_{\vec{x},s;\vec{x}\ppr,s\ppr}^{12}(t_{p}) \\
\hat{\Phi}_{\vec{x},s;\vec{x}\ppr,s\ppr}^{21}(t_{p}) &
\hat{\Phi}_{\vec{x},s;\vec{x}\ppr,s\ppr}^{22}(t_{p})
\eea\right)_{\mbox{;}}
\eeq
\be  \lb{s2_18}
\bea{rclrcl}
\hat{\Phi}_{\vec{x},s;\vec{x}\ppr,s\ppr}^{aa,\dagger}(t_{p}) &=&
\hat{\Phi}_{\vec{x},s;\vec{x}\ppr,s\ppr}^{aa}(t_{p})  & \hspace*{0.6cm}
\hat{\Phi}_{\vec{x},s;\vec{x}\ppr,s\ppr}^{22}(t_{p}) &=&
-\hat{\Phi}_{\vec{x},s;\vec{x}\ppr,s\ppr}^{11,T}(t_{p})  \\
\hat{\Phi}_{\vec{x},s;\vec{x}\ppr,s\ppr}^{21}(t_{p}) &=&
\hat{\Phi}_{\vec{x},s;\vec{x}\ppr,s\ppr}^{12,\dagger}(t_{p})  & \hspace*{0.6cm}
\hat{\Phi}_{\vec{x},s;\vec{x}\ppr,s\ppr}^{12,T}(t_{p}) &=&
-\hat{\Phi}_{\vec{x},s;\vec{x}\ppr,s\ppr}^{12}(t_{p})
\eea_{\mbox{;}}
\ee
\beq  \lb{s2_19}
\breve{\Phi}_{\vec{x},s;\vec{x}\ppr,s\ppr}^{ab}(t_{p})&=&
\breve{\Psi}_{\vec{x},s}^{a}(t_{p})\;\otimes\;
\breve{\Psi}_{\vec{x}\ppr,s\ppr}^{\sharp b}(t_{p}) = \left( \bea{c}
\psi_{\vec{x},s}(t_{p}) \\ \psi_{\vec{x},s}^{*}(t_{p}+\Delta t_{p}) \eea\right)^{a} \otimes
\Big(\psi_{\vec{x}\ppr,s\ppr}^{*}(t_{p}+\Delta t_{p})\;;\;\psi_{\vec{x}\ppr,s\ppr}(t_{p})\Big)^{b}
\\ \no &=&\left( \bea{cc}
\langle\psi_{\vec{x},s}(t_{p})\;\psi_{\vec{x}\ppr,s\ppr}^{*}(t_{p}+\Delta t_{p})\rangle &
\langle\psi_{\vec{x},s}(t_{p})\;\psi_{\vec{x}\ppr,s\ppr}(t_{p})\rangle \\
\langle\psi_{\vec{x},s}^{*}(t_{p}+\Delta t_{p})\;\psi_{\vec{x}\ppr,s\ppr}^{*}(t_{p}+\Delta t_{p})\rangle &
\langle\psi_{\vec{x},s}^{*}(t_{p}+\Delta t_{p})\;\psi_{\vec{x}\ppr,s\ppr}(t_{p})\rangle \eea\right)  \\ \no &=&
\left( \bea{cc}
\breve{\Phi}_{\vec{x},s;\vec{x}\ppr,s\ppr}^{11}(t_{p}) &
\breve{\Phi}_{\vec{x},s;\vec{x}\ppr,s\ppr}^{12}(t_{p}) \\
\breve{\Phi}_{\vec{x},s;\vec{x}\ppr,s\ppr}^{21}(t_{p}) &
\breve{\Phi}_{\vec{x},s;\vec{x}\ppr,s\ppr}^{22}(t_{p})
\eea\right)_{\mbox{;}}
\eeq
\be   \lb{s2_20}
\bea{rclrcl}
\breve{\Phi}_{\vec{x},s;\vec{x}\ppr,s\ppr}^{aa,\sharp}(t_{p}) &=&
\breve{\Phi}_{\vec{x},s;\vec{x}\ppr,s\ppr}^{aa}(t_{p})  & \hspace*{0.6cm}
\breve{\Phi}_{\vec{x},s;\vec{x}\ppr,s\ppr}^{22}(t_{p}) &=&
-\breve{\Phi}_{\vec{x},s;\vec{x}\ppr,s\ppr}^{11,T}(t_{p})  \\
\breve{\Phi}_{\vec{x},s;\vec{x}\ppr,s\ppr}^{21}(t_{p}) &=&
\breve{\Phi}_{\vec{x},s;\vec{x}\ppr,s\ppr}^{12,\sharp}(t_{p})  & \hspace*{0.6cm}
\breve{\Phi}_{\vec{x},s;\vec{x}\ppr,s\ppr}^{12,T}(t_{p}) &=&
-\breve{\Phi}_{\vec{x},s;\vec{x}\ppr,s\ppr}^{12}(t_{p})
\eea_{\mbox{.}}
\ee
Apart from the ordinary hermitian conjugation and transposition relations (\ref{s2_18}) between the various
block parts of \(\hat{\Phi}_{\vec{x},s;\vec{x}\ppr,s\ppr}^{ab}(t_{p})\), we extend to the combined transposition,
trace and hermitian conjugation (\ref{s2_21}-\ref{s2_23})
of all four involved blocks of \(\hat{\Phi}_{\vec{x},s;\vec{x}\ppr,s\ppr}^{ab}(t_{p})\) in its entity
\beq \lb{s2_21}
\Big(\hat{\Phi}_{\vec{x},s;\vec{x}\ppr,s\ppr}^{ab}(t_{p})\Big)^{T}&=&\hspace*{-0.3cm}\left( \bea{cc}
\hat{\Phi}_{\vec{x},s;\vec{x}\ppr,s\ppr}^{11}(t_{p}) & \hat{\Phi}_{\vec{x},s;\vec{x}\ppr,s\ppr}^{12}(t_{p}) \\
\hat{\Phi}_{\vec{x},s;\vec{x}\ppr,s\ppr}^{21}(t_{p}) &
\hat{\Phi}_{\vec{x},s;\vec{x}\ppr,s\ppr}^{22}(t_{p}) \eea\right)^{T}\hspace*{-0.3cm}= \left( \bea{cc}
\big(\hat{\Phi}_{\vec{x},s;\vec{x}\ppr,s\ppr}^{11}(t_{p})\big)^{T} &
\big(\hat{\Phi}_{\vec{x},s;\vec{x}\ppr,s\ppr}^{21}(t_{p})\big)^{T} \\
\big(\hat{\Phi}_{\vec{x},s;\vec{x}\ppr,s\ppr}^{12}(t_{p})\big)^{T} &
\big(\hat{\Phi}_{\vec{x},s;\vec{x}\ppr,s\ppr}^{22}(t_{p})\big)^{T}
\eea\right)_{\mbox{;}} \\    \lb{s2_22}
\TRS\Big[\hat{\Phi}_{\vec{x},s;\vec{x}\ppr,s\ppr}^{ab}(t_{p})\Big]&=&
\trs\Big[\hat{\Phi}_{\vec{x},s;\vec{x}\ppr,s\ppr}^{11}(t_{p})\Big]+
\trs\Big[\hat{\Phi}_{\vec{x},s;\vec{x}\ppr,s\ppr}^{22}(t_{p})\Big]  \;; \\  \lb{s2_23}
\Big(\hat{\Phi}_{\vec{x},s;\vec{x}\ppr,s\ppr}^{ab}(t_{p})\Big)\pdag&=&\left( \bea{cc}
\hat{\Phi}_{\vec{x},s;\vec{x}\ppr,s\ppr}^{11}(t_{p}) &
\hat{\Phi}_{\vec{x},s;\vec{x}\ppr,s\ppr}^{12}(t_{p}) \\
\hat{\Phi}_{\vec{x},s;\vec{x}\ppr,s\ppr}^{21}(t_{p}) &
\hat{\Phi}_{\vec{x},s;\vec{x}\ppr,s\ppr}^{22}(t_{p}) \eea\right)\pdag= \left( \bea{cc}
\big(\hat{\Phi}_{\vec{x},s;\vec{x}\ppr,s\ppr}^{11}(t_{p})\big)\pdag &
\big(\hat{\Phi}_{\vec{x},s;\vec{x}\ppr,s\ppr}^{21}(t_{p})\big)\pdag \\
\big(\hat{\Phi}_{\vec{x},s;\vec{x}\ppr,s\ppr}^{12}(t_{p})\big)\pdag &
\big(\hat{\Phi}_{\vec{x},s;\vec{x}\ppr,s\ppr}^{22}(t_{p})\big)\pdag
\eea\right)_{\mbox{.}}
\eeq
In a similar manner we have defined an order parameter \(\breve{\Phi}_{\vec{x},s;\vec{x}\ppr,s\ppr}^{ab}(t_{p})\)
in (\ref{s2_19},\ref{s2_20}) appropriate for anomalous doubled density
matrices and take into account the proper time ordering of the normal ordered Hamilton operator
\(\hat{H}(\hat{\psi}\pdag,\hat{\psi},t)\) (\ref{s2_3}) with inclusion of
time shifts \(\Delta t_{p}\). Therefore, one has to adapt the ordinary hermitian conjugation
'\(\pdag\)' (\ref{s2_10},\ref{s2_11}) to the introduced hermitian conjugation '\(^{\sharp}\)' (\ref{s2_13},\ref{s2_14})
which comprises an additional time shift correction \(\Delta t_{p}\) in such a manner
that the resulting complex parts \(\psi_{\vec{x},s}^{*}(t_{p}+\Delta t_{p})\) follow a
particular time step \(\Delta t_{p}\) later from \(\psi_{\vec{x},s}(t_{p})\) on the non-equilibrium time contour \(t_{p}\).
In analogy to relations (\ref{s2_21}-\ref{s2_23}), we specify transposition, traces and the time shifted, hermitian conjugation
'\(^{\sharp}\)' in relations (\ref{s2_24}-\ref{s2_26}) for the total dyadic product
'\(\breve{\Psi}_{\vec{x},s}^{a}(t_{p})\otimes \breve{\Psi}_{\vec{x}\ppr,s\ppr}^{\sharp b}(t_{p})\)' (\ref{s2_16})
related order parameter \(\breve{\Phi}_{\vec{x},s;\vec{x}\ppr,s\ppr}^{ab}(t_{p})\) (\ref{s2_19}). As already mentioned,
this kind of order order parameter with the appropriate time shifts \(\Delta t_{p}\) has to be used for
the density matrices in the time development on the non-equilibrium time contour
\beq \lb{s2_24} \hspace*{-0.3cm}
\Big(\breve{\Phi}_{\vec{x},s;\vec{x}\ppr,s\ppr}^{ab}(t_{p})\Big)^{T}&=&\hspace*{-0.3cm}\left( \bea{cc}
\breve{\Phi}_{\vec{x},s;\vec{x}\ppr,s\ppr}^{11}(t_{p}) & \breve{\Phi}_{\vec{x},s;\vec{x}\ppr,s\ppr}^{12}(t_{p}) \\
\breve{\Phi}_{\vec{x},s;\vec{x}\ppr,s\ppr}^{21}(t_{p}) &
\breve{\Phi}_{\vec{x},s;\vec{x}\ppr,s\ppr}^{22}(t_{p}) \eea\right)^{T}\hspace*{-0.3cm}= \left( \bea{cc}
\big(\breve{\Phi}_{\vec{x},s;\vec{x}\ppr,s\ppr}^{11}(t_{p})\big)^{T} &
\big(\breve{\Phi}_{\vec{x},s;\vec{x}\ppr,s\ppr}^{21}(t_{p})\big)^{T} \\
\big(\breve{\Phi}_{\vec{x},s;\vec{x}\ppr,s\ppr}^{12}(t_{p})\big)^{T} &
\big(\breve{\Phi}_{\vec{x},s;\vec{x}\ppr,s\ppr}^{22}(t_{p})\big)^{T}
\eea\right) \;;   \\    \lb{s2_25}
\TRS\Big[\breve{\Phi}_{\vec{x},s;\vec{x}\ppr,s\ppr}^{ab}(t_{p})\Big]&=&
\trs\Big[\breve{\Phi}_{\vec{x},s;\vec{x}\ppr,s\ppr}^{11}(t_{p})\Big]+
\trs\Big[\breve{\Phi}_{\vec{x},s;\vec{x}\ppr,s\ppr}^{22}(t_{p})\Big] \;;  \\  \lb{s2_26}
\Big(\breve{\Phi}_{\vec{x},s;\vec{x}\ppr,s\ppr}^{ab}(t_{p})\Big)^{\sharp}&=&\left( \bea{cc}
\breve{\Phi}_{\vec{x},s;\vec{x}\ppr,s\ppr}^{11}(t_{p}) &
\breve{\Phi}_{\vec{x},s;\vec{x}\ppr,s\ppr}^{12}(t_{p}) \\
\breve{\Phi}_{\vec{x},s;\vec{x}\ppr,s\ppr}^{21}(t_{p}) &
\breve{\Phi}_{\vec{x},s;\vec{x}\ppr,s\ppr}^{22}(t_{p}) \eea\right)^{\sharp}= \left( \bea{cc}
\big(\breve{\Phi}_{\vec{x},s;\vec{x}\ppr,s\ppr}^{11}(t_{p})\big)^{\sharp} &
\big(\breve{\Phi}_{\vec{x},s;\vec{x}\ppr,s\ppr}^{21}(t_{p})\big)^{\sharp} \\
\big(\breve{\Phi}_{\vec{x},s;\vec{x}\ppr,s\ppr}^{12}(t_{p})\big)^{\sharp} &
\big(\breve{\Phi}_{\vec{x},s;\vec{x}\ppr,s\ppr}^{22}(t_{p})\big)^{\sharp}
\eea\right)_{\mbox{.}}
\eeq
Using the above definitions and notations, we can state the following path integral (\ref{s2_27}) of coherent states with
the precise, appropriate subsequent steps of the time development on the time contour. Note the described, essential time
shifts \(\Delta t_{p}\) in the complex parts \(\psi_{\vec{x},s}^{*}(t_{p}+\Delta t_{p})\) and the additional,
anomalous doubled source matrix \(\hat{\mscr{J}}_{\vec{x}\ppr,s\ppr;\vec{x},s}^{ba}(t_{q}\ppr,t_{p})\) for generating
bilinear or higher even order correlation functions with \(\breve{\Psi}_{\vec{x}\ppr,s\ppr}^{\sharp b}(t_{q}\ppr)\;
\breve{\Psi}_{\vec{x},s}^{a}(t_{p})\). The source fields \(j_{\psi;s}(\vec{x},t_{p})\) (\ref{s2_5}),
\(\hat{j}_{\psi\psi;ss\ppr}(\vec{x},t_{p})\) (\ref{s2_6}), which are extended to a dependence on the time contour for
generating observables, are set to equivalent values on the two branches of the time contour at the final end after
complete transformations (\ref{s2_29},\ref{s2_30}) so that these eventually act as condensate seeds for a macroscopic wavefunction
\(\langle \psi_{\vec{x},s}(t_{p})\rangle\) and for a non-vanishing pair condensate correlation function
\(\langle\psi_{\vec{x},s}(t_{p})\;\psi_{\vec{x}\ppr,s\ppr}(t_{q}\ppr)\rangle\)
\beq \lb{s2_27} \hspace*{-0.9cm}
Z[\hat{\mscr{J}},j_{\psi},\hat{j}_{\psi\psi}] &=&
\int d[\psi_{\vec{x}\ppr,s\ppr}^{*}(t_{p}), \psi_{\vec{x},s}(t_{p})] \;\;
\exp\bigg\{-\frac{\im}{\hbar}\int_{C}d t_{p}\sum_{\vec{x}}\sums
\psi_{\vec{x},s}^{*}(t_{p}+\Delta t_{p})\;\;\hat{H}_{p}(\vec{x},t_{p})\;\;
\psi_{\vec{x},s}(t_{p})\bigg\}     \\ \no &\times& \exp\bigg\{-\frac{\im}{\hbar}
\int_{C}dt_{p}\sum_{\vec{x},\vec{x}\ppr}\sumss
\psi_{\vec{x}\ppr,s\ppr}^{*}(t_{p}+\Delta t_{p})\;\psi_{\vec{x},s}^{*}(t_{p}+\Delta t_{p})\;
V_{|\vec{x}\ppr-\vec{x}|}\;\psi_{\vec{x},s}(t_{p})\;\psi_{\vec{x}\ppr,s\ppr}(t_{p})
\bigg\} \\ \no &\times &
\exp\bigg\{-\frac{\im}{\hbar}\int_{C}d t_{p}\sum_{\vec{x}}\sums
\Big(j_{\psi;s}^{*}(\vec{x},t_{p})\;\psi_{\vec{x},s}(t_{p})+
\psi_{\vec{x},s}^{*}(t_{p}+\Delta t_{p})\;j_{\psi;s}(\vec{x},t_{p})\Big)\bigg\}
\\ \no &\times&
\exp\bigg\{-\frac{\im}{2\hbar}\int_{C}d t_{p} \sum_{\vec{x}}\trs\bigg[
\hat{j}_{\psi\psi;s\ppr s}\pdag(\vec{x},t_{p})\;
\psi_{\vec{x},s\ppr}(t_{p})\;
\psi_{\vec{x},s}(t_{p})+  \\ \no &+&
\psi_{\vec{x},s\ppr}^{*}(t_{p}+\Delta t_{p})\;\psi_{\vec{x},s}^{*}(t_{p}+\Delta t_{p})\;
\hat{j}_{\psi\psi;s\ppr s}(\vec{x},t_{p})\bigg]\bigg\}   \\ \no &\times&
\exp\bigg\{-\frac{\im}{2\hbar}\int_{C}d t_{p}\; d t_{q}\ppr
\sum_{\vec{x},\vec{x}\ppr}\sumss \breve{\Psi}_{\vec{x}\ppr,s\ppr}^{\sharp b}(t_{q}\ppr)\;
\hat{\mscr{J}}_{\vec{x}\ppr,s\ppr;\vec{x},s}^{ba}(t_{q}\ppr;
t_{p})\;\breve{\Psi}_{\vec{x},s}^{a}(t_{p})\bigg\} \;;  \\  \no
\hat{H}_{p}(\vec{x},t_{p}) &=& -\hat{E}_{p}-\im\;\hat{\ve}_{p}+\hat{h}(\vec{x}) =
-\im\hbar\frac{\pp}{\pp t_{p}}-\im\;\hat{\ve}_{p}+
\frac{\vec{p}^{\;2}}{2m}+u(\vec{x})-\mu_{0}\;;
\eeq
\beq \lb{s2_28}
\lefteqn{\hspace*{-1.9cm}\int_{C}d t_{p}\sum_{\vec{x};s=\uparrow,\downarrow}
\psi_{\vec{x},s}^{*}(t_{p}+\Delta t_{p})\;\hat{H}_{p}(\vec{x},t_{p})\;\psi_{\vec{x},s}(t_{p}) = } \\ \no
&\hspace*{-3.3cm}=&\hspace*{-2.0cm} \int_{C}d t_{p}\sum_{\vec{x};s=\uparrow,\downarrow}\bigg[
-\im\hbar\;\psi_{\vec{x},s}^{*}(t_{p}+\Delta t_{p})\;
\frac{\psi_{\vec{x},s}(t_{p}+\Delta t_{p})-\psi_{\vec{x},s}(t_{p})}{\Delta t_{p}}+
\psi_{\vec{x},s}^{*}(t_{p}+\Delta t_{p})\;\big(\hat{h}(\vec{x})-\im\;\hat{\ve}_{p}\big)\;
\psi_{\vec{x},s}(t_{p})\bigg]  \\ \no &\hspace*{-3.3cm}=&\hspace*{-2.0cm} \int_{C}d t_{p}\;dt_{q}\ppr
\sum_{\vec{x};s=\uparrow,\downarrow}\;\;
\psi_{\vec{x},s}^{*}(t_{q}\ppr)\;\delta_{qp}\;\eta_{q}\;\bigg(-\im\hbar\;
\frac{\delta_{t_{q}\ppr\boldsymbol{,}t_{p}}-
\delta_{t_{q}\ppr\boldsymbol{,}t_{p}+\Delta t_{p}}}{\Delta t_{p}}+
\big(\hat{h}(\vec{x})-\im\;\hat{\ve}_{p}\big)\;
\delta_{t_{q}\ppr\boldsymbol{,}t_{p}+\Delta t_{p}}\bigg)\;\psi_{\vec{x},s}(t_{p})  \\ \no
&\hspace*{-3.3cm}=&\hspace*{-2.0cm} \int_{C}dt_{p}\;dt_{q}\ppr\sum_{\vec{x};s=\uparrow,\downarrow}\;\;
\psi_{\vec{x},s}^{*}(t_{q}\ppr+\Delta t_{q}\ppr)\;\delta_{qp}\;\eta_{q}\;\bigg(-\im\hbar\;
\frac{\delta_{t_{q}\ppr\boldsymbol{,}t_{p}-\Delta t_{p}}-
\delta_{t_{q}\ppr\boldsymbol{,}t_{p}}}{\Delta t_{p}}+
\big(\hat{h}(\vec{x})-\im\;\hat{\ve}_{p}\big)\;
\delta_{t_{q}\ppr\boldsymbol{,}t_{p}}\bigg)\;\psi_{\vec{x},s}(t_{p})\;;  \\  \lb{s2_29}
j_{\psi;s}(\vec{x},t_{p}) &:=&j_{\psi;s}(\vec{x},t)\;\;;\;\;\;\;
\mbox{'condensate seed' for } \langle\psi_{\vec{x},s}(t_{p})\rangle \;; \\  \lb{s2_30}
\hat{j}_{\psi\psi;ss\ppr}(\vec{x},t_{p})&:=&
\hat{j}_{\psi\psi;ss\ppr}(\vec{x},t)\;\;;\mbox{'condensate seed' for }
\langle\psi_{\vec{x},s\ppr}(t_{p})\;\;\psi_{\vec{x},s}(t_{p})\rangle\;.
\eeq

\subsection{Self-energy with anomalous terms and
coset decomposition into density and pair condensate parts} \lb{s22}

In order to achieve a HST for anomalous doubled, quartic fields of the interaction, we consider
following transformation from the ordinary, 'appropriate' time shifted density (\ref{s2_31}) to its
anomalous doubled density form with metric \(\hat{S}_{4\times 4}\) (\ref{s2_32}). One has to apply
the fields \(\breve{\Psi}_{\vec{x},s}^{a}(t_{p})\) with
the hermitian conjugation '\(^{\sharp}\)' (\ref{s2_13},\ref{s2_14})
instead of the 'equal time' fields  \(\Psi_{\vec{x},s}^{a}(t_{p})\) with 'equal time' hermitian
conjugation '\(^{\dag}\)' (\ref{s2_10},\ref{s2_11})
so that the additional time shift \(\Delta t_{p}\) is always preserved
in the complex parts \(\psi_{\vec{x},s}^{*}(t_{p}+\Delta t_{p})\) relative to \(\psi_{\vec{x},s}(t_{p})\)
\beq \lb{s2_31}
\psi_{\vec{x},s}^{*,T}(t_{p}+\Delta t_{p})\; \psi_{\vec{x},s}(t_{p})  &=&
\frac{1}{2}\Big( \psi_{\vec{x},s}^{*,T}(t_{p}+\Delta t_{p})\;\psi_{\vec{x},s}(t_{p})-
\psi_{\vec{x},s}^{T}(t_{p})\;\psi_{\vec{x},s}^{*}(t_{p}+\Delta t_{p}) \Big) \\ \no &=&
\frac{1}{2}\;\breve{\Psi}_{\vec{x},s}^{\sharp a}(t_{p})\;\hat{S}\;
\breve{\Psi}_{\vec{x},s}^{a}(t_{p}) \;;  \\  \lb{s2_32}
\hat{S}_{4\times 4}&=&\Big\{\underbrace{\hat{1}_{2\times 2}}_{a=1}\;;\;
\underbrace{-\hat{1}_{2\times 2}}_{a=2}\Big\}\;;\;\;\;
\hat{1}_{2\times 2}=\hat{1}_{ss\ppr}\;;\;\;\;s,s\ppr=\uparrow,\downarrow \;.
\eeq
Straightforward application of (\ref{s2_31},\ref{s2_32}) for dyadic products (\ref{s2_16}) transforms
the quartic interaction part to its anomalous doubled kind (\ref{s2_33}) with anomalous doubled density matrix
\(\breve{R}_{\vec{x},s;\vec{x}\ppr,s\ppr}^{ab}(t_{p})\) (\ref{s2_34},\ref{s2_35})
having the appropriate, subsequent time shifts
in all the complex parts \(\psi_{\vec{x},s}^{*}(t_{p}+\Delta t_{p})\) (relative to \(\psi_{\vec{x},s}(t_{p})\))
\beq \lb{s2_33}
\lefteqn{\sum_{\vec{x},\vec{x}\ppr}\sumss
\psi_{\vec{x}\ppr,s\ppr}^{*,T}(t_{p}+\Delta t_{p})\;\psi_{\vec{x},s}^{*,T}(t_{p}+\Delta t_{p})\;
V_{|\vec{x}-\vec{x}\ppr|}\;\psi_{\vec{x},s}(t_{p})\;\psi_{\vec{x}\ppr,s\ppr}(t_{p})=} \\
\no &=& \frac{1}{4}\sum_{\vec{x},\vec{x}\ppr}\sumss
\bigg(\breve{\Psi}_{\vec{x},s}^{\sharp a}(t_{p})\;\hat{S}\;\breve{\Psi}_{\vec{x},s}^{a}(t_{p})\;\;
\breve{\Psi}_{\vec{x}\ppr,s\ppr}^{\sharp b}(t_{p})\;\hat{S}\;
\breve{\Psi}_{\vec{x}\ppr,s\ppr}^{b}(t_{p})\bigg)\;\; V_{|\vec{x}-\vec{x}\ppr|}= \\
\no &=&-\frac{1}{4}\sum_{\vec{x},\vec{x}\ppr}
\TRS\bigg[\hat{S}\underbrace{\left(\breve{\Psi}_{\vec{x},s}^{a}(t_{p})\otimes
\breve{\Psi}_{\vec{x}\ppr,s\ppr}^{\sharp b}(t_{p})\right)}_{
\breve{R}_{\vec{x},s;\vec{x}\ppr,s\ppr}^{ab}(t_{p})}\hat{S}
\underbrace{\left(\breve{\Psi}_{\vec{x}\ppr,s\ppr}^{b}(t_{p})\otimes
\breve{\Psi}_{\vec{x},s}^{\sharp a}(t_{p})\right)}_{
\breve{R}_{\vec{x}\ppr,s\ppr;\vec{x},s}^{ba}(t_{p})}\bigg]\;V_{|\vec{x}-\vec{x}\ppr|}= \\ \no
&=&-\frac{1}{4}\sum_{\vec{x},\vec{x}\ppr}
\TRS\left[\hat{S}\;\breve{R}_{\vec{x},s;\vec{x}\ppr,s\ppr}^{ab}(t_{p})\;\hat{S}\;
\breve{R}_{\vec{x}\ppr,s\ppr;\vec{x},s}^{ba}(t_{p})\right]\;V_{|\vec{x}-\vec{x}\ppr|}\;\;\;;  \\  \lb{s2_34}
\breve{R}_{\vec{x},s;\vec{x}\ppr,s\ppr}^{ab}(t_{p}) &=&
\left(\bea{cc} \breve{R}_{\vec{x},s;\vec{x}\ppr,s\ppr}^{11}(t_{p}) &
\breve{R}_{\vec{x},s;\vec{x}\ppr,s\ppr}^{12}(t_{p})  \\
\breve{R}_{\vec{x},s;\vec{x}\ppr,s\ppr}^{21}(t_{p}) &
\breve{R}_{\vec{x},s;\vec{x}\ppr,s\ppr}^{22}(t_{p}) \eea\right)^{ab}  \;;  \\ \no
\breve{R}_{\vec{x},s;\vec{x}\ppr,s\ppr}^{11}(t_{p}) &=&
\psi_{\vec{x},s}(t_{p})\;\psi_{\vec{x}\ppr,s\ppr}^{*}(t_{p}+\Delta t_{p})=
\breve{R}_{\vec{x},s;\vec{x}\ppr,s\ppr}^{11,\sharp}(t_{p}) \;;  \\ \no
\breve{R}_{\vec{x},s;\vec{x}\ppr,s\ppr}^{22}(t_{p}) &=&
\psi_{\vec{x},s}^{*}(t_{p}+\Delta t_{p})\;\psi_{\vec{x}\ppr,s\ppr}(t_{p})=
\breve{R}_{\vec{x},s;\vec{x}\ppr,s\ppr}^{22,\sharp}(t_{p}) \;;  \\ \no
\breve{R}_{\vec{x},s;\vec{x}\ppr,s\ppr}^{12}(t_{p}) &=&
\psi_{\vec{x},s}(t_{p})\;\psi_{\vec{x}\ppr,s\ppr}(t_{p})=
-\breve{R}_{\vec{x},s;\vec{x}\ppr,s\ppr}^{12,T}(t_{p}) \;; \\ \lb{s2_35}
\breve{R}_{\vec{x},s;\vec{x}\ppr,s\ppr}^{21}(t_{p}) &=&
\psi_{\vec{x},s}^{*}(t_{p}+\Delta t_{p})\;\psi_{\vec{x}\ppr,s\ppr}^{*}(t_{p}+\Delta t_{p})=
-\breve{R}_{\vec{x},s;\vec{x}\ppr,s\ppr}^{21,T}(t_{p}) \;; \\ \no
\breve{R}_{\vec{x},s;\vec{x}\ppr,s\ppr}^{22}(t_{p}) &=&
-\breve{R}_{\vec{x},s;\vec{x}\ppr,s\ppr}^{11,T}(t_{p}) \;; \hspace*{0.3cm}
\breve{R}_{\vec{x},s;\vec{x}\ppr,s\ppr}^{21}(t_{p}) =
\breve{R}_{\vec{x},s;\vec{x}\ppr,s\ppr}^{12,\sharp}(t_{p}) \;.
\eeq
The doubled density matrix \(\breve{R}_{\vec{x},s;\vec{x}\ppr,s\ppr}^{ab}(t_{p})\) (\ref{s2_34},\ref{s2_35}) with pair
condensate terms in the off-diagonal blocks (\(a\neq b\)) is constructed according to the
dyadic product (\ref{s2_16}) and order parameter \(\breve{\Phi}_{\vec{x},s;\vec{x}\ppr,s\ppr}^{ab}(t_{p})\)
(\ref{s2_19},\ref{s2_20},\ref{s2_24}-\ref{s2_26}); however, the self-energy matrix has to comply with
the hermitian, anomalous doubled order parameter \(\hat{\Phi}_{\vec{x},s;\vec{x}\ppr,s\ppr}^{ab}(t_{p})\)
(\ref{s2_17},\ref{s2_18},\ref{s2_21}-\ref{s2_23}) or dyadic product (\ref{s2_15}) with solely equal
time fields \(\Psi_{\vec{x},s}^{a}(t_{p})\) and equal time hermitian conjugation '\(\pdag\)' (\ref{s2_10},\ref{s2_11}).
In order to emphasize symmetries, we simplify to the case of a short-ranged interaction potential
\be  \lb{s2_36}
V_{|\vec{x}-\vec{x}\ppr|}\approx \delta_{\vec{x},\vec{x}\ppr}\;\;V_{0}\;\;\;,
\ee
which will be extended to the general case of arbitrary long-ranged interaction potentials
\(V_{|\vec{x}-\vec{x}\ppr|}\neq \delta_{\vec{x},\vec{x}\ppr}\;V_{0}\) in section \ref{s3}.
We therefore introduce the anomalous doubled self-energy matrix \(\wt{\Sigma}_{ss\ppr}^{ab}(\vec{x},t_{p})\)
(\ref{s2_37}) which consists of the block diagonal, hermitian density field \(\sigma_{D;ss\ppr}^{(0)}(\vec{x},t_{p})\)
and the additional self-energy \(\delta\wt{\Sigma}_{ss\ppr}^{ab}(\vec{x},t_{p})\) (\ref{s2_40}), having
hermitian density blocks \(\delta\hat{\Sigma}_{ss\ppr}^{11}(\vec{x},t_{p})\),
\(\delta\hat{\Sigma}_{ss\ppr}^{22}(\vec{x},t_{p})\)
(\ref{s2_38},\ref{s2_39},\ref{s2_42},\ref{s2_43}) as subalgebra elements and
anti-hermitian related coset parts \(\im\;\delta\hat{\Sigma}_{ss\ppr}^{12}(\vec{x},t_{p})\),
\(\im\;\delta\hat{\Sigma}_{ss\ppr}^{21}(\vec{x},t_{p})\) (\ref{s2_44}-\ref{s2_46}) in the off-diagonal blocks
(which is pointed out by the tilde '\(\wt{\ph{\Sigma}}\)' above
\(\delta\wt{\Sigma}_{ss\ppr}^{ab}(\vec{x},t_{p})\)). We take the analogous notation of Ref. \cite{pop1}
for the super-symmetric, ortho-symplectic case \(\mbox{Osp}(S,S|2L)\,/\,\mbox{U}(L|S)\otimes \mbox{U}(L|S)\),
but reduce relations to the even, fermion-fermion blocks in order to underline the appropriate, precise
steps in the time development. Thus, the coset matrices for anomalous parts are denoted by
\(\hat{T}(\vec{x},t_{p})\) and the remaining block diagonal densities are labeled by
\(\delta \hat{\Sigma}_{D;s_{1}s_{2}}^{aa}(\vec{x},t_{p})\) in the coset decomposition (\ref{s2_41})
\beq \lb{s2_37}
\wt{\Sigma}_{ss\ppr}^{ab}(\vec{x},t_{p}) &=&
\sigma_{D;ss\ppr}^{(0)ab}(\vec{x},t_{p})\;\delta_{ab}+
\delta\wt{\Sigma}_{ss\ppr}^{ab}(\vec{x},t_{p}) \\  \no &=&
\left(\bea{cc} \hat{\Sigma}_{ss\ppr}^{11}(\vec{x},t_{p})  &
\im\;\delta\hat{\Sigma}_{ss\ppr}^{12}(\vec{x},t_{p})  \\
\im\;\delta\hat{\Sigma}_{ss\ppr}^{21}(\vec{x},t_{p}) &
\hat{\Sigma}_{ss\ppr}^{22}(\vec{x},t_{p}) \eea\right)^{ab}\;; \\ \lb{s2_38}
\hat{\Sigma}_{ss\ppr}^{11}(\vec{x},t_{p}) &=&
\sigma_{D;ss\ppr}^{(0)}(\vec{x},t_{p})+
\delta\hat{\Sigma}_{ss\ppr}^{11}(\vec{x},t_{p}) \;;  \\ \lb{s2_39}
\hat{\Sigma}_{ss\ppr}^{22}(\vec{x},t_{p}) &=&
\sigma_{D;ss\ppr}^{(0)T}(\vec{x},t_{p})+
\delta\hat{\Sigma}_{ss\ppr}^{22}(\vec{x},t_{p}) \;;  \\
\sigma_{D;ss\ppr}^{(0)11}(\vec{x},t_{p}) &=& \sigma_{D;ss\ppr}^{(0)}(\vec{x},t_{p})\;;\hspace*{0.6cm}
\sigma_{D;ss\ppr}^{(0)22}(\vec{x},t_{p}) = \sigma_{D;ss\ppr}^{(0)T}(\vec{x},t_{p})\;; \\ \no
\sigma_{D;ss\ppr}^{(0)T}(\vec{x},t_{p}) &=& \sigma_{D;s\ppr s}^{(0)}(\vec{x},t_{p})\;;  \\   \lb{s2_40}
\delta\wt{\Sigma}_{ss\ppr}^{ab}(\vec{x},t_{p}) &=&
\left(\bea{cc} \delta\hat{\Sigma}_{ss\ppr}^{11}(\vec{x},t_{p})  &
\im\;\delta\hat{\Sigma}_{ss\ppr}^{12}(\vec{x},t_{p})  \\
\im\;\delta\hat{\Sigma}_{ss\ppr}^{21}(\vec{x},t_{p}) &
\delta\hat{\Sigma}_{ss\ppr}^{22}(\vec{x},t_{p}) \eea\right)^{ab}\;;
\eeq
\beq \lb{s2_41}
\lefteqn{\wt{\Sigma}_{ss\ppr}^{ab}(\vec{x},t_{p}) =
\left(\bea{cc}
\sigma_{D;ss\ppr}^{(0)}(\vec{x},t_{p})+
\delta\hat{\Sigma}_{ss\ppr}^{11}(\vec{x},t_{p})  &
\im\;\delta\hat{\Sigma}_{ss\ppr}^{12}(\vec{x},t_{p})  \\
\im\;\delta\hat{\Sigma}_{ss\ppr}^{21}(\vec{x},t_{p}) &
\sigma_{D;ss\ppr}^{(0)T}(\vec{x},t_{p})+
\delta\hat{\Sigma}_{ss\ppr}^{22}(\vec{x},t_{p}) \eea\right)^{ab} } \\ \no &=&
\sigma_{D;ss\ppr}^{(0)ab}(\vec{x},t_{p})\;\delta_{ab} +
\delta\wt{\Sigma}_{ss\ppr}^{ab}(\vec{x},t_{p}) = \sigma_{D;ss\ppr}^{(0)ab}(\vec{x},t_{p})+
\Big(\hat{T}(\vec{x},t_{p})\Big)_{ss_{1}}^{aa_{1}}\;
\delta\hat{\Sigma}_{D;s_{1}s_{2}}^{a_{1}=a_{2}}(\vec{x},t_{p})\;
\Big(\hat{T}^{-1}(\vec{x},t_{p})\Big)_{s_{2}s\ppr}^{a_{2}b}  \\  \no &=&
\sigma_{D;ss\ppr}^{(0)ab}(\vec{x},t_{p})+
\Big(\hat{T}(\vec{x},t_{p})\Big)_{ss_{1}}^{aa_{1}}\;
\hat{Q}_{s_{1}s_{2}\ppr}^{-1;a_{1}a_{1}}(\vec{x},t_{p})\;\;
\delta\hat{\Lambda}_{s_{2}\ppr}^{a_{1}=a_{2}}(\vec{x},t_{p})\;\;
\hat{Q}_{s_{2}\ppr s_{2}}^{a_{2}a_{2}}(\vec{x},t_{p})\;
\Big(\hat{T}^{-1}(\vec{x},t_{p})\Big)_{s_{2}s\ppr}^{a_{2}b}  \\ \no &=&\sigma_{D;ss\ppr}^{(0)ab}(\vec{x},t_{p})+
\Big(\hat{T}_{0}(\vec{x},t_{p})\Big)_{ss_{1}}^{aa_{1}} \;
\delta\hat{\Lambda}_{s_{1}=s_{2}}^{a_{1}=a_{2}}(\vec{x},t_{p})\;
\Big(\hat{T}_{0}^{-1}(\vec{x},t_{p})\Big)_{s_{2}s\ppr}^{a_{2}b}  \;;
\eeq
\beq \lb{s2_42}
\delta\hat{\Sigma}_{ss\ppr}^{11}(\vec{x},t_{p}) &=&
\delta\hat{\Sigma}_{ss\ppr}^{11,\dagger}(\vec{x},t_{p}) \;;\hspace*{0.6cm}
\delta\hat{\Sigma}_{ss\ppr}^{22}(\vec{x},t_{p}) =
\delta\hat{\Sigma}_{ss\ppr}^{22,\dagger}(\vec{x},t_{p}) \;; \\  \lb{s2_43}
\delta\hat{\Sigma}_{ss\ppr}^{22}(\vec{x},t_{p}) &=&-
\delta\hat{\Sigma}_{ss\ppr}^{11,T}(\vec{x},t_{p}) \;;   \\  \lb{s2_44}
\delta\hat{\Sigma}_{ss\ppr}^{12}(\vec{x},t_{p}) &=&
-\delta\hat{\Sigma}_{ss\ppr}^{12,T}(\vec{x},t_{p}) =
\im\left(\bea{cc}  0 & -\delta\hat{\Sigma}_{\uparrow\downarrow}^{12}(\vec{x},t_{p}) \\
+\delta\hat{\Sigma}_{\uparrow\downarrow}^{12}(\vec{x},t_{p})  & 0 \eea\right)_{ss\ppr}^{12}=
\big(\hat{\tau}_{2}\big)_{ss\ppr}\;\;\delta\hat{\Sigma}_{\uparrow\downarrow}^{12}(\vec{x},t_{p})\;;
\\  \lb{s2_45}  \delta\hat{\Sigma}_{ss\ppr}^{21}(\vec{x},t_{p}) &=&
\delta\hat{\Sigma}_{ss\ppr}^{12,\dagger}(\vec{x},t_{p}) \;;  \\  \lb{s2_46}
\delta\hat{\Sigma}_{ss\ppr}^{21}(\vec{x},t_{p}) &=&
-\delta\hat{\Sigma}_{ss\ppr}^{21,T}(\vec{x},t_{p}) =
\im\left(\bea{cc}  0 & -\delta\hat{\Sigma}_{\uparrow\downarrow}^{12,*}(\vec{x},t_{p}) \\
+\delta\hat{\Sigma}_{\uparrow\downarrow}^{12,*}(\vec{x},t_{p})  & 0 \eea\right)_{ss\ppr}^{21}=
\big(\hat{\tau}_{2}\big)_{ss\ppr}\;\;\delta\hat{\Sigma}_{\uparrow\downarrow}^{12,*}(\vec{x},t_{p}) ;  \\ \lb{s2_47}
\delta\wt{\Sigma}_{ss\ppr}^{ab}(\vec{x},t_{p}) &\simeq&
\left(\bea{cc}  \im\;\delta\hat{\Sigma}_{ss\ppr}^{12}(\vec{x},t_{p}) &
\delta\hat{\Sigma}_{ss\ppr}^{11}(\vec{x},t_{p})  \\
\underbrace{-\delta\hat{\Sigma}_{ss\ppr}^{11,T}(\vec{x},t_{p})}_{
\delta\hat{\Sigma}_{ss\ppr}^{22}(\vec{x},t_{p})} &
\im\;\delta\hat{\Sigma}_{ss\ppr}^{21}(\vec{x},t_{p}) \eea\right)\in \mbox{so}(4) \;; \\ \lb{s2_48}
\delta\hat{\Sigma}_{ss\ppr}^{11}(\vec{x},t_{p}) &\simeq&
\Big(\delta\hat{\Sigma}_{ss\ppr}^{22}(\vec{x},t_{p})=
-\delta\hat{\Sigma}_{ss\ppr}^{11,T}(\vec{x},t_{p}) \Big)\in\mbox{u}(2) \;.
\eeq
The block diagonal densities \(\delta\hat{\Sigma}_{D;s_{1}s_{2}}^{aa}(\vec{x},t_{p})\)
(\ref{s2_41},\ref{s2_49}-\ref{s2_52})
are further decomposed into eigenvalues \(\delta\hat{\Lambda}_{s\ppr s}^{aa}(\vec{x},t_{p})\),
\(\delta\hat{\lambda}_{s}(\vec{x},t_{p})\) (\ref{s2_53},\ref{s2_54}) and diagonalizing matrices
\(\hat{Q}_{ss\ppr}^{aa}(\vec{x},t_{p})\) (\ref{s2_55}-\ref{s2_57})
with generator \(\hat{\mscr{F}}_{D;ss\ppr}(\vec{x},t_{p})\)
and complex parameter \(\mscr{F}_{\uparrow\downarrow}(\vec{x},t_{p})\) (\ref{s2_58},\ref{s2_59}). The details of the
parameters and generators are listed in Eqs. (\ref{s2_49}) to (\ref{s2_59}) and are in complete accordance with the
given parameters and generators of the more general ortho-symplectic case \cite{pop1} which is just restricted to the
even, fermion-fermion parts in order to point out the problem of the appropriate, precise discrete
time steps in the coset decomposition
\beq  \lb{s2_49}
\delta\hat{\Sigma}_{D;ss\ppr}^{ab}(\vec{x},t_{p}) &=&\delta_{ab}\;\;
\delta\hat{\Sigma}_{D;ss\ppr}^{aa}(\vec{x},t_{p})  \;;  \\ \lb{s2_50}
\delta\hat{\Sigma}_{D;ss\ppr}^{aa}(\vec{x},t_{p}) &=&
\delta\hat{\Sigma}_{D;ss\ppr}^{aa,\dagger}(\vec{x},t_{p})  \;;  \\  \lb{s2_51}
\delta\hat{\Sigma}_{D;ss\ppr}^{22}(\vec{x},t_{p}) &=&
-\delta\hat{\Sigma}_{D;ss\ppr}^{11,T}(\vec{x},t_{p})  \;;  \\   \lb{s2_52}
\delta\hat{\Sigma}_{D;s_{1}s_{2}}^{aa}(\vec{x},t_{p}) &=&
\hat{Q}_{s_{1}s\ppr}^{-1;aa}(\vec{x},t_{p})\;\;
\delta\hat{\Lambda}_{s\ppr=s}^{aa}(\vec{x},t_{p})\;\;
\hat{Q}_{ss_{2}}^{aa}(\vec{x},t_{p})  \;; \\   \lb{s2_53}
\delta\hat{\Lambda}_{s\ppr s}^{aa}(\vec{x},t_{p}) &=& \delta_{s\ppr s}\;\;
\mbox{diag}\Big(\underbrace{\delta\hat{\lambda}_{s}(\vec{x},t_{p})}_{a=1}\;;\;
\underbrace{-\delta\hat{\lambda}_{s}(\vec{x},t_{p})}_{a=2} \Big) \;; \\   \lb{s2_54}
\delta\hat{\lambda}_{s}(\vec{x},t_{p}) &=& \Big(\delta\hat{\lambda}_{s=\uparrow}(\vec{x},t_{p})\,,\,
\delta\hat{\lambda}_{s=\downarrow}(\vec{x},t_{p})\Big)\;;   \\   \lb{s2_55}
\hat{Q}_{ss\ppr}^{ab}(\vec{x},t_{p}) &=&\delta_{ab}\;\;
\left(\bea{cc}  \hat{Q}_{ss\ppr}^{11}(\vec{x},t_{p})  &  \\ &
\hat{Q}_{ss\ppr}^{22}(\vec{x},t_{p})  \eea \right)^{ab} \;; \\  \lb{s2_56}
\hat{Q}_{ss\ppr}^{11}(\vec{x},t_{p})  &=&
\bigg(\exp\Big\{\im\;\hat{\mscr{F}}_{D;s_{1}s_{2}}(\vec{x},t_{p})\Big\}\bigg)_{ss\ppr}\;;  \\  \lb{s2_57}
\hat{Q}_{ss\ppr}^{22}(\vec{x},t_{p})  &=&
\bigg(\exp\Big\{-\im\;\hat{\mscr{F}}_{D;s_{1}s_{2}}^{T}(\vec{x},t_{p})\Big\}\bigg)_{ss\ppr}\;; \\ \lb{s2_58}
\hat{\mscr{F}}_{D;ss}(\vec{x},t_{p}) &\equiv& 0\;;\hspace*{0.4cm}
\hat{\mscr{F}}_{D;ss\ppr}(\vec{x},t_{p})=\hat{\mscr{F}}_{D;ss\ppr}\pdag(\vec{x},t_{p})=
\left(\bea{cc} 0  & \mscr{F}_{\uparrow\downarrow}(\vec{x},t_{p})  \\
\mscr{F}_{\uparrow\downarrow}^{*}(\vec{x},t_{p})  & 0  \eea\right)  \;;  \\ \lb{s2_59}
\hat{\mscr{F}}_{D;ss\ppr}(\vec{x},t_{p}) &\stackrel{\wedge}{=}&
\mbox{u}(2)\,/\,\big(\mbox{u}(1)\otimes\mbox{u}(1)\big)\;;\hspace*{0.3cm}
\Big(\mscr{F}_{\uparrow\downarrow}(\vec{x},t_{p})=
\big|\mscr{F}_{\uparrow\downarrow}(\vec{x},t_{p})\big|\;\;
\exp\{\im\:\varphi(\vec{x},t_{p})\} \;;\Big)
 \\ \no &\stackrel{\wedge}{=}&
\mscr{F}_{\uparrow\downarrow}(\vec{x},t_{p}) \;,\;\;
\big(\mscr{F}_{\uparrow\downarrow}^{*}(\vec{x},t_{p})\big)\;,\;\;
(\mbox{2 real parameters})\;.
\eeq
As one follows the described parameters in Eqs. (\ref{s2_37}-\ref{s2_59}),
one has performed a coset decomposition (\ref{s2_60},\ref{s2_61})
of a general \(\mbox{SO}(4)\) symmetry or \(\mbox{so}(4)\) self-energy generator
\(\delta\wt{\Sigma}_{ss\ppr}^{ab}(\vec{x},t_{p})\) (\ref{s2_40},\ref{s2_47})
into the unitary sub-algebra parts \(\mbox{u}(2)\) (\ref{s2_48})
for the block diagonal densities \(\delta\hat{\Sigma}_{D;ss\ppr}^{aa}(\vec{x},t_{p})\)
(\ref{s2_49}-\ref{s2_51}) and into
the coset algebra elements with matrices \(\hat{T}(\vec{x},t_{p})\)
\beq \lb{s2_60}
\overbrace{\mbox{SO}(4)}^{\delta\wt{\Sigma}_{ss\ppr}^{ab}} &\simeq&
\underbrace{
\overbrace{\mbox{SO}(4)}^{\delta\wt{\Sigma}_{ss\ppr}^{ab}}\,/\,
\overbrace{\mbox{U}(2)}^{\delta\hat{\Sigma}_{ss\ppr}^{aa}}}_{
\delta\hat{\Sigma}_{\uparrow\downarrow}^{12}\,,\,
\delta\hat{\Sigma}_{\uparrow\downarrow}^{12,*}} \;\otimes\;
\overbrace{\mbox{U}(2)}^{\delta\hat{\Sigma}_{ss\ppr}^{aa}}  \;\;\;;  \\ \lb{s2_61}
\bigg(\underbrace{\underbrace{\mbox{SO}(4)}_{\mbox{\scz 6 parameters}}\,/\,
\underbrace{\mbox{U}(2)}_{\mbox{\scz 4 parameters}}}_{\mbox{\scz 2 remaining parameters}}\bigg)
&\simeq& \delta\hat{\Sigma}_{\uparrow\downarrow}^{12}(\vec{x},t_{p}),\;
\big(\delta\hat{\Sigma}_{\uparrow\downarrow}^{12,*}(\vec{x},t_{p})\big) \;.
\eeq
The parametrization (\ref{s2_37}-\ref{s2_61}) and coset decomposition (\ref{s2_41},\ref{s2_60},\ref{s2_61})
of the self-energy matrix
\(\delta\wt{\Sigma}_{ss\ppr}^{ab}(\vec{x},t_{p})\) (\ref{s2_40}) has only 'equal time' fields and
'equal time' hermitian conjugation operations without any time shifts \(\Delta t_{p}\)
as in the anomalous doubled density matrices
\(\breve{R}_{\vec{x},s;\vec{x}\ppr,s\ppr}^{ab}(t_{p})\) (\ref{s2_34},\ref{s2_35}).
The coset matrix \(\hat{T}_{ss\ppr}^{ab}(\vec{x},t_{p})\) (\ref{s2_41},\ref{s2_62}) consists of the coset generator
\(\hat{Y}_{ss\ppr}^{a\neq b}(\vec{x},t_{p})\) (\ref{s2_63}) with further, anti-symmetric, complex-valued sub-matrices
\(\hat{X}_{ss\ppr}(\vec{x},t_{p})\), \(\hat{X}_{ss\ppr}\pdag(\vec{x},t_{p})\) (\ref{s2_64},\ref{s2_65}) in the
off-diagonal blocks. We use the anti-symmetric Pauli-matrix \((\hat{\tau}_{2})_{ss\ppr}\)
(as for quaternion eigenvalues) in order to define the complex pair condensate
fields \(f(\vec{x},t_{p})\), \(f^{*}(\vec{x},t_{p})\) with absolute value
\(|f(\vec{x},t_{p})\,|\) and phase \(\phi(\vec{x},t_{p})\). The modified coset matrix
\(\hat{T}_{0}(\vec{x},t_{p})\) (\ref{s2_66}) is introduced for convenience because this combination
of coset matrix \(\hat{T}(\vec{x},t_{p})\) and diagonalizing density matrix
\(\hat{Q}_{ss\ppr}^{aa}(\vec{x},t_{p})\) appears in the calculation of the
corresponding invariant integration measure \(\mbox{SO}(4)\,/\,\mbox{U}(2)\otimes\mbox{U}(2)\)
\beq \lb{s2_62}
\hat{T}_{ss\ppr}^{ab}(\vec{x},t_{p}) &=&
\bigg(\exp\Big\{-\hat{Y}_{s_{1}s_{2}}^{a_{1}\neq b_{1}}(\vec{x},t_{p})\Big\}\bigg)_{ss\ppr}^{ab}
\;;  \\ \lb{s2_63}  \hat{Y}_{ss\ppr}^{a\neq b}(\vec{x},t_{p}) &=&
\left(\bea{cc} 0 & \hat{X}_{ss\ppr}(\vec{x},t_{p}) \\
\hat{X}_{ss\ppr}\pdag(\vec{x},t_{p})  &  0  \eea\right)^{ab}  \;; \\ \lb{s2_64}
\hat{X}_{ss\ppr}(\vec{x},t_{p}) &=& -\hat{X}_{ss\ppr}^{T}(\vec{x},t_{p}) =
\big(\hat{\tau}_{2}\big)_{ss\ppr}\;f(\vec{x},t_{p}) =
\im\;\left(\bea{cc} 0 & -f(\vec{x},t_{p})  \\ +f(\vec{x},t_{p}) & 0 \eea\right)\;; \\ \lb{s2_65}
\hat{X}_{ss\ppr}\pdag(\vec{x},t_{p}) &=&
\big(\hat{\tau}_{2}\big)_{ss\ppr}\;f^{*}(\vec{x},t_{p})\;;\hspace*{0.4cm}
f(\vec{x},t_{p}) = \big|f(\vec{x},t_{p})\,\big|\;\;\exp\big\{\im\;\phi(\vec{x},t_{p})\,\big\} \;; \\ \lb{s2_66}
\hat{T}_{0}(\vec{x},t_{p})  &=& \hat{T}(\vec{x},t_{p})\;\;
\hat{Q}^{-1}(\vec{x},t_{p}) \;.
\eeq
We briefly collect above relations (\ref{s2_33}-\ref{s2_65}) with incorporation of analogous results for the ortho-symplectic
case in Ref. \cite{pop1} and list the corresponding, anomalous doubled HST (\ref{s2_67}) for the quartic interaction
of fields, but under inclusion of the precise, essential, subsequent time shifts
in the complex parts \(\psi_{\vec{x},s}^{*}(t_{p}+\Delta t_{p})\).
Furthermore, the total HST has to encompass
imaginary increments with extension of the approximating potential parameter (\ref{s2_36})
\(V_{0}\rightarrow\hat{V}_{0}^{ab}\) to a matrix \(\hat{V}_{0}^{ab}\) in order to achieve converging
Gaussian integrations for the self-energy matrices \(\delta\wt{\Sigma}_{ss\ppr}^{ab}(\vec{x},t_{p})\) (\ref{s2_40}) and for the hermitian
self-energy density field \(\sigma_{D;ss\ppr}^{(0)a=b}(\vec{x},t_{p})\). (We again emphasize that it is a
convenient standard to simplify Eqs. (\ref{s2_33}-\ref{s2_35},\ref{s2_67})  of this HST to solely 'equal time' fields and relations without auxiliary, imaginary increments as e.g. described in Refs. \cite{pop1,pop2} !).
\beq \lb{s2_67}
\lefteqn{\exp\bigg\{-\frac{\im}{\hbar}\int_{C}d t_{p}\sum_{\vec{x},\vec{x}\ppr}\sumss
\psi_{\vec{x},s}^{*,T}(t_{p}+\Delta t_{p})\;\psi_{\vec{x}\ppr,s\ppr}^{*,T}(t_{p}+\Delta t_{p})\;V_{|\vec{x}\ppr-\vec{x}|}\;
\psi_{\vec{x}\ppr,s\ppr}(t_{p})\;\psi_{\vec{x},s}(t_{p})\bigg\}
\stackrel{V_{|\vec{x}-\vec{x}\ppr|}\approx V_{0}}{\approx}  }
\\ \no &=&
\int\;\;d[\sigma_{D;ss\ppr}^{(0)}(\vec{x},t_{p})]\;\;
\exp\bigg\{-\frac{\im}{2\hbar}\int_{C}d t_{p}\sum_{\vec{x};s,s\ppr=\uparrow,\downarrow}
\frac{\sigma_{D;ss\ppr}^{(0)}(\vec{x},t_{p})\;\;\sigma_{D;s\ppr s}^{(0)}(\vec{x},t_{p})}{V_{0}+\im\;\ve_{p}}\bigg\}
\\ \no &\times &
 \int d[\delta\wt{\Sigma}(\vec{x},t_{p})]\;\;
 \exp\Bigg\{-\frac{\im}{4\hbar}\int_{C}d t_{p}\sum_{\vec{x}}
 \TRS\bigg[\frac{\delta\wt{\Sigma}_{ss\ppr}^{ab}(\vec{x},t_{p})\;
 \delta\wt{\Sigma}_{s\ppr s}^{ba}(\vec{x},t_{p})}{V_{0}+\im\:\ve_{p}\;\big(\delta_{a=b}-\delta_{a\neq b}\big)}\bigg]\Bigg\}
\\ \no &\times&
 \exp\Bigg\{\frac{\im}{2\hbar}\int_{C}d t_{p}\sum_{\vec{x},\vec{x}\ppr}
 \TRS\Bigg[\left(
 \bea{cc}
 \breve{R}_{\vec{x},s;\vec{x}\ppr,s\ppr}^{11}(t_{p}) &
\breve{R}_{\vec{x},s;\vec{x}\ppr,s\ppr}^{12}(t_{p}) \\
\breve{R}_{\vec{x},s;\vec{x}\ppr,s\ppr}^{21}(t_{p}) &
\breve{R}_{\vec{x},s;\vec{x}\ppr,s\ppr}^{22}(t_{p})
 \eea\right)^{ab}\;\hat{S}\;\times\;
\overbrace{\frac{V_{0}}{V_{0}+\im\:\ve_{p}\;\big(\delta_{a=b}-\delta_{a\neq b}\big)}}^{:=1}
  \\ \no &\times&  \delta_{\vec{x},\vec{x}\ppr}\;\mcal{N}_{x} \left(  \bea{cc}
\sigma_{D;s\ppr s}^{(0)}(\vec{x},t_{p})+
 \delta\hat{\Sigma}_{s\ppr s}^{11}(\vec{x},t_{p}) &
\delta\hat{\Sigma}_{s\ppr s}^{12}(\vec{x},t_{p}) \\
 \delta\hat{\Sigma}_{s\ppr s}^{21}(\vec{x},t_{p}) &
-\sigma_{D;s\ppr s}^{(0)T}(\vec{x},t_{p})-
 \delta\hat{\Sigma}_{s\ppr s}^{22}(\vec{x},t_{p})
 \eea\right)^{ba}\;\hat{S}\Bigg]\Bigg\} \;;   \\   \no
\hat{V}_{0}^{ab} &=&  V_{0}+\im\:\ve_{p}\;\big(\delta_{a=b}-\delta_{a\neq b}\big) \;\;\;;\;\;\;\;
\ve_{p}=\eta_{p}\;\ve_{+}\;\;;\;\;\;\ve_{+}>0\;\;\;.
\eeq
A similar anomalous doubling of the one-particle part
\(\psi_{\vec{x},s}^{*,T}(t_{p}+\Delta t_{p})\;\hat{H}_{p}(\vec{x},t_{p})\;
\psi_{\vec{x},s}(t_{p})\) as for
the interaction part (\ref{s2_33}-\ref{s2_35},\ref{s2_67}) leads to relation (\ref{s2_68})
with the anomalous doubled, one-particle operator
\(\breve{\mscr{H}}_{\vec{x}\ppr, s\ppr;\vec{x}, s}^{ba}(t_{q}\ppr,t_{p})\) (\ref{s2_69})
which includes the transpose \(\breve{\mscr{H}}_{\vec{x}\ppr, s\ppr;\vec{x}, s}^{22}(t_{q}\ppr,t_{p})\)
(\ref{s2_71}) apart from \(\breve{\mscr{H}}_{\vec{x}\ppr, s\ppr;\vec{x}, s}^{11}(t_{q}\ppr,t_{p})\) (\ref{s2_70}).
Note that one has to use the doubled fields \(\breve{\Psi}_{\vec{x},s}^{a}(t_{p})\) (\ref{s2_13}) with the
hermitian conjugate \(\breve{\Psi}_{\vec{x}\ppr,s\ppr}^{\sharp b}(t_{q}\ppr)\) (\ref{s2_14}) in order to
maintain the ubiquitous time shifts '\(\Delta t_{p}\)' in the complex part \(\psi_{\vec{x},s}^{*}(t_{p}+\Delta t_{p})\)
(relative to \(\psi_{\vec{x},s}(t_{p})\)). We also list a 'lax' kind for the anomalous doubled one-particle
operator with \(\hat{\mscr{H}}_{\vec{x}\ppr, s\ppr;\vec{x}, s}^{ba}(t_{q}\ppr,t_{p})\)
in relations (\ref{s2_72}-\ref{s2_74})
\beq  \lb{s2_68}
\lefteqn{\hspace*{-3.6cm}\sum_{\vec{x}}\sums
\psi_{\vec{x}, s}^{*,T}(t_{p}+\Delta t_{p})\;\;\hat{H}_{p}(\vec{x},t_{p})\;\;
\psi_{\vec{x}, s}(t_{p})=
\sum_{\vec{x}}\sums
\psi_{\vec{x}, s}^{T}(t_{p})\;(-)\;\hat{H}_{p}^{T}(\vec{x},t_{p})\;\;
\psi_{\vec{x}, s}^{*}(t_{p}+\Delta t_{p}) =}    \\  \no  &=&
\int_{C}dt_{q}\ppr\;\;\sum_{\vec{x},\vec{x}\ppr}\sumss\frac{1}{2}\;
\breve{\Psi}_{\vec{x}\ppr,s\ppr}^{\sharp b}(t_{q}\ppr)\;\hat{S}\;
\breve{\mscr{H}}_{\vec{x}\ppr, s\ppr;\vec{x}, s}^{ba}(t_{q}\ppr,t_{p})\;
\breve{\Psi}_{\vec{x},s}^{a}(t_{p})  \;\;\;; \\    \lb{s2_69}
\breve{\mscr{H}}_{\vec{x}\ppr, s\ppr;\vec{x}, s}^{ba}(t_{q}\ppr,t_{p})&=&
\mbox{diag}\left(\breve{\mscr{H}}_{\vec{x}\ppr, s\ppr;\vec{x}, s}^{11}(t_{q}\ppr,t_{p})\;;\;
\breve{\mscr{H}}_{\vec{x}\ppr, s\ppr;\vec{x}, s}^{22}(t_{q}\ppr,t_{p})\right)  \;; \\  \lb{s2_70}
\breve{\mscr{H}}_{\vec{x}\ppr, s\ppr;\vec{x}, s}^{11}(t_{q}\ppr,t_{p}) &=&
\delta_{pq}\;\eta_{p}\;
\bigg(-\im\,\hbar\;\frac{\delta_{t_{q}\ppr,t_{p}-\Delta t_{p}}-\delta_{t_{q}\ppr,t_{p}}}{\Delta t_{p}}+
\big(\hat{h}(\vec{x}\ppr)-\im\;\hat{\ve}_{p}\big)\;\delta_{t_{q}\ppr,t_{p}}\bigg)
\;\delta_{\vec{x}\ppr,\vec{x}}\; \;; \\ \lb{s2_71}
\breve{\mscr{H}}_{\vec{x}\ppr, s\ppr;\vec{x}, s}^{22}(t_{q}\ppr,t_{p}) &=&
\delta_{pq}\;\eta_{p}\;
\bigg(-\im\,\hbar\;\frac{\delta_{t_{q}\ppr-\Delta t_{q}\ppr,t_{p}}-\delta_{t_{q}\ppr,t_{p}}}{\Delta t_{p}}+
\big(\hat{h}^{T}(\vec{x}\ppr)-\im\;\hat{\ve}_{p}\big)\;\delta_{t_{q}\ppr,t_{p}}\bigg)
\;\delta_{\vec{x}\ppr,\vec{x}}\; \\ \no &=&
\Big(\breve{\mscr{H}}_{\vec{x}\ppr, s\ppr;\vec{x}, s}^{11}(t_{q}\ppr,t_{p}) \Big)^{T}  \;; \\ \lb{s2_72}
\breve{\mscr{H}}_{\vec{x}\ppr, s\ppr;\vec{x}, s}^{ba}(t_{q}\ppr,t_{p}) &\simeq&
\hat{\mscr{H}}_{\vec{x}\ppr, s\ppr;\vec{x}, s}^{ba}(t_{q}\ppr,t_{p}) \;; \\ \no
\hat{\mscr{H}}_{\vec{x}\ppr, s\ppr;\vec{x}, s}^{ba}(t_{q}\ppr,t_{p})&=&
\delta_{pq}\;\eta_{p}\;
\mbox{diag}\left(\hat{H}_{p}(\vec{x}\ppr,t_{q}\ppr)\;\hat{1}_{N\times N}\;;\;
\hat{H}_{p}^{T}(\vec{x}\ppr,t_{q}\ppr)\;\hat{1}_{N\times N}\right)
\;\delta(t_{q}\ppr-t_{p})\;\delta_{\vec{x}\ppr,\vec{x}}\; \;; \\  \lb{s2_73}
\hat{H}_{p}(\vec{x},t_{p})&=&-\im\hbar\frac{\pp}{\pp t_{p}}-\im\;\ve_{p}+
\frac{\vec{p}^{\;2}}{2m}+u(\vec{x})-\mu_{0} \;; \\   \lb{s2_74}
\hat{H}_{p}^{T}(\vec{x},t_{p})&=&+\im\hbar\frac{\pp}{\pp t_{p}}-\im\;\ve_{p}+
\frac{\vec{p}^{\;2}}{2m}+u(\vec{x})-\mu_{0}\;\;\;.
\eeq
In correspondence to the interaction (\ref{s2_33}-\ref{s2_35},\ref{s2_67}) and
one-particle part (\ref{s2_68}-\ref{s2_74}), one has to perform an anomalous doubling
of the source fields \(j_{\psi;s}(\vec{x},t_{p})\), \(\hat{j}_{\psi\psi;ss\ppr}(\vec{x},t_{p})\) to
\(J_{\psi;s}^{a}(\vec{x},t_{p})\) (\ref{s2_75}) and
\(\hat{J}_{\psi\psi;ss\ppr}^{a\neq b}(\vec{x},t_{p})\) (\ref{s2_76},\ref{s2_77}) in the
'equal time' form (\ref{s2_10}) without any time shifts of the complex parts so that the 'equal time'
hermitian conjugation operations with '\(\pdag\)' (\ref{s2_11}) have to be applied. This anomalous doubled,
'equal time' form of the source fields \(j_{\psi;s}(\vec{x},t_{p})\) and \(\hat{j}_{\psi\psi;ss\ppr}(\vec{x},t_{p})\)
has to incorporate the symmetry relations (\ref{s2_37}-\ref{s2_48}) of the 'equal time' restricted, hermitian self-energies
\beq \lb{s2_75}
J_{\psi;s}^{a(=1/2)}(\vec{x},t_{p})&=&\Big(\underbrace{j_{\psi;s}(\vec{x},t_{p})}_{a=1}\;;\;
\underbrace{j_{\psi;s}^{*}(\vec{x},t_{p})}_{a=2}\Big)^{T} \;; \\  \lb{s2_76}
\hat{J}_{\psi\psi;ss\ppr}^{a\neq b}(\vec{x},t_{p}) &=&
\left( \bea{cc}
0 & \hat{j}_{\psi\psi;ss\ppr}(\vec{x},t_{p}) \\
\hat{j}_{\psi\psi;ss\ppr}\pdag(\vec{x},t_{p}) & 0
\eea\right)^{ab} \;;    \\   \lb{s2_77}
\hat{j}_{\psi\psi;ss\ppr}(\vec{x},t_{p})&=&-\hat{j}_{\psi\psi;ss\ppr}^{T}(\vec{x},t_{p})=
\big(\hat{\tau}_{2}\big)_{ss\ppr}\;\;j_{\psi\psi}(\vec{x},t_{p})\;; \\ \no
j_{\psi\psi}(\vec{x},t_{p})&=&\big|j_{\psi\psi}(\vec{x},t_{p})\big|\;\;\exp\{\im\:\gamma(\vec{x},t_{p})\}\;.
\eeq
Further collection of the above HST and the anomalous doubled one-particle and source field parts results into the
path integral (\ref{s2_78}) with only bilinear, anomalous doubled fields
which fulfill the requirement of additional time shifts
\(\Delta t_{p}\) in the complex parts \(\psi_{\vec{x},s}^{*}(t_{p}+\Delta t_{p})\) (relative to \(\psi_{\vec{x},s}(t_{p})\)).
This bilinear part of fields \(\breve{\Psi}_{\vec{x}\ppr,s\ppr}^{\sharp b}(t_{q}\ppr)\), \(\breve{\Psi}_{\vec{x},s}^{a}(t_{p})\)
comprises the matrix \(\breve{\mscr{M}}_{\vec{x}\ppr,s\ppr;\vec{x},s}^{ba}(t_{q}\ppr,t_{p})\) (\ref{s2_79})
with the sum of anomalous doubled one particle operator, source matrix
\(\hat{\mscr{J}}_{\vec{x},s;\vec{x}\ppr,s\ppr}^{ab}(t_{p},t_{q}\ppr)\) and
self-energies \(\sigma_{D;ss\ppr}^{(0)}(\vec{x},t_{p})\),
\(\delta\wt{\Sigma}_{ss\ppr}^{ab}(\vec{x},t_{p})\).
In order to accomplish the coherent state path integral (\ref{s2_78},\ref{s2_79}),
one has to transform by the metric \(\hat{I}\) (\ref{s2_80},\ref{s2_81}) with the imaginary units
in the second part (\(a=2\)) satisfying \(\hat{I}\cdot\hat{I}=\hat{S}\)
\beq \lb{s2_78}
\lefteqn{Z[\hat{\mscr{J}},J_{\psi},\im\,\hat{J}_{\psi\psi}]=
\int\;\;d[\sigma_{D;ss\ppr}^{(0)}(\vec{x},t_{p})]\;\; \exp\bigg\{-\frac{\im}{2\hbar}
\int_{C}d t_{p}\sum_{\vec{x};s,s\ppr=\uparrow,\downarrow}
\frac{\sigma_{D;ss\ppr}^{(0)}(\vec{x},t_{p})\;\;
\sigma_{D;s\ppr s}^{(0)}(\vec{x},t_{p})}{V_{0}+\im\;\ve_{p}}\bigg\} } \\ \no &\times &
\int d[\delta\wt{\Sigma}(\vec{x},t_{p})]\;\;
\exp\bigg\{-\frac{\im}{4\hbar}
\int_{C}d t_{p}\sum_{\vec{x}}\TRS\bigg[\frac{\delta\wt{\Sigma}_{ss\ppr}^{ab}(\vec{x},t_{p})\;
\delta\wt{\Sigma}_{s\ppr s}^{ba}(\vec{x},t_{p})}{V_{0}+\im\;\ve_{p}(\delta_{a=b}-\delta_{a\neq b})}\bigg]\bigg\} \;
\times\; \int d[\psi_{\vec{x}\ppr,s\ppr}^{*}(t_{p}),\psi_{\vec{x},s}(t_{p})]\;\times \\ \no &\times&
\exp\Bigg\{-\frac{\im}{2\hbar}\int_{\breve{C}}d t_{p}\;d t_{q}\ppr
\sum_{\vec{x},\vec{x}\ppr}\mcal{N}_{x}\sumss\sum_{a,b=1,2}
\breve{\Psi}_{\vec{x}\ppr,s\ppr}^{\sharp b}(t_{q}\ppr)\;\hat{I}\;
\breve{\mscr{M}}_{\vec{x}\ppr,s\ppr;\vec{x},s}^{ba}(t_{q}\ppr,t_{p})\;\hat{I}\;
\breve{\Psi}_{\vec{x},s}^{a}(t_{p})\Bigg\} \\ \no &\times &
\exp\bigg\{-\frac{\im}{2\hbar}\int_{\breve{C}}d t_{p}\sum_{\vec{x}}\sums\Big(
J_{\psi;s}^{\dagger a}(\vec{x},t_{p})\;\hat{S}\;\breve{\Psi}_{\vec{x},s}^{a}(t_{p})+
\breve{\Psi}_{\vec{x},s}^{\sharp a}(t_{p})\;\hat{S}\;J_{\psi;s}^{a}(\vec{x},t_{p})\Big)\bigg\}_{\mbox{;}}
\eeq
\beq\lb{s2_79}
\lefteqn{|\Delta t_{q}\ppr|\;\breve{\mscr{M}}_{\vec{x}\ppr,s\ppr;\vec{x},s}^{ba}(t_{q}\ppr,t_{p})=
\underbrace{\breve{\mscr{H}}_{\vec{x}\ppr,s\ppr;
\vec{x},s}^{ba}(t_{q}\ppr,t_{p})}_{\mbox{\scz(\ref{s2_69}-\ref{s2_71})}} +
\eta_{q}\;\hat{I}\;\hat{S}\;|\Delta t_{q}\ppr|\;
\frac{\hat{\mscr{J}}_{\vec{x}\ppr,s\ppr;\vec{x},s}^{ba}(t_{q}\ppr,t_{p})}{\mcal{N}_{x}}\;
\hat{S}\;\hat{I}\;\eta_{p}   +      } \\ \no &+&
\Bigg(\sigma_{D;s\ppr s}^{(0)ba}(\vec{x},t_{p})\;\delta_{ba} -
\im\;\underbrace{\hat{J}_{\psi\psi;s\ppr s}^{ba}(\vec{x},t_{p})}_{\mbox{\scz(\ref{s2_76},\ref{s2_77})}} +
\underbrace{\left(\bea{cc}
\delta\hat{\Sigma}_{s\ppr s}^{11}(\vec{x},t_{p}) & \im\;
\delta\hat{\Sigma}_{s\ppr s}^{12}(\vec{x},t_{p}) \\
\im\;\delta\hat{\Sigma}_{s\ppr s}^{21}(\vec{x},t_{p}) &
\delta\hat{\Sigma}_{s\ppr s}^{22}(\vec{x},t_{p})
\eea\right)}_{\delta\wt{\Sigma}_{s\ppr s}(\vec{x},t_{p})=\mbox{\scz(\ref{s2_40})}}\;\Bigg)
\delta_{\vec{x}\ppr,\vec{x}}\;\eta_{q}\;\delta_{qp}\;\delta_{t_{q}\ppr,t_{p}} \;;
\eeq
\beq
\sigma_{D;s\ppr s}^{(0)11}(\vec{x},t_{p}) &=& \sigma_{D;s\ppr s}^{(0)}(\vec{x},t_{p})\;;\hspace*{0.6cm}
\sigma_{D;s\ppr s}^{(0)22}(\vec{x},t_{p}) = \sigma_{D;s\ppr s}^{(0)T}(\vec{x},t_{p})\;; \\ \no
\sigma_{D;s\ppr s}^{(0)T}(\vec{x},t_{p}) &=& \sigma_{D;ss\ppr}^{(0)}(\vec{x},t_{p})\;;  \\  \lb{s2_80}
\hat{I}_{4\times 4}&=&\Big\{\underbrace{\hat{1}_{2\times 2}}_{a=1}\;;\;
\underbrace{\hat{\im}_{2\times 2}}_{a=2}\Big\}\;;\;\;\;
\hat{\im}_{2\times 2}=\im\;\hat{1}_{ss\ppr}\;;\;\;\;s,s\ppr=\uparrow,\downarrow \;; \\ \lb{s2_81}
\hat{I}_{4\times 4}\cdot\hat{I}_{4\times 4} &=&\hat{S}_{4\times 4} \;.
\eeq
It is straightforward to transform the contour time integrals in (\ref{s2_78}) to the
appropriate, discrete time step version having the essential time shifts
\(\Delta t_{p}\) in the complex fields \(\psi_{\vec{x},s}^{*}(t_{p}+ \Delta t_{p})\)
(relative to \(\psi_{\vec{x},s}(t_{p})\)) so that the original normal ordering of the
second quantized Hamilton operator (\ref{s2_3}-\ref{s2_6}) is taken into account and
the action of field operators with its hermitian conjugate at the same time contour point
is avoided. However, we have to mention the following detail for a somewhat modified
time contour integration path '\(\breve{C}\)' in (\ref{s2_78}) with (\ref{s2_79}) instead of '\(C\)' (\ref{s2_8},\ref{s2_9}) :
\begin{itemize}
  \item In order to include the {\it precise} boundary and discrete 'time step' conditions, one has to
extend the time contour integrations (\ref{s2_8},\ref{s2_9}) at the end points
'\(t_{p=\boldsymbol{+}}=T_{\mathsf{ini}}\)' and
'\(t_{p=\boldsymbol{-}}=T_{\mathsf{ini}}\)' of the two '\(p=\pm\)' branches
by an additional single time step (where we omit the further straightforward specification of the discrete
time steps between \(T_{\mathsf{ini}}\) and \(T_{\mathsf{fin}}\) for brevity)
\be \lb{s2_82}
\int_{\breve{C}}d t_{p}\;\ldots =\int_{T_{\mathsf{ini}}-|\Delta t_{p}|}^{T_{\mathsf{fin}}}dt_{+}\ldots -
\int_{T_{\mathsf{ini}}-|\Delta t_{p}|}^{T_{\mathsf{fin}}}dt_{-}\ldots
\ee
and has to set the following fields and sources at the extended time boundary points
'\(t_{p=\boldsymbol{+}}=T_{\mathsf{ini}}-|\Delta t_{p}|\)' , '\(t_{p=\boldsymbol{-}}=
T_{\mathsf{ini}}-|\Delta t_{p}|\)' of the two  branches
identical to zero
\beq \lb{s2_83}
\psi_{\vec{x},s}(t_{p=\boldsymbol{\pm}}=T_{\mathsf{ini}}-|\Delta t_{p}|) &=&
\psi_{\vec{x},s}^{*}(t_{p=\boldsymbol{\pm}}=T_{\mathsf{ini}}-|\Delta t_{p}|) \equiv 0 \;;  \\  \lb{s2_84}
J_{\psi;s}^{a(=1/2)}(\vec{x},t_{p=\boldsymbol{\pm}}=T_{\mathsf{ini}}-|\Delta t_{p}|) &\equiv& 0 \;; \hspace*{0.6cm}
J_{\psi\psi;s\ppr s}^{b\neq a}(\vec{x},t_{p=\boldsymbol{\pm}}=T_{\mathsf{ini}}-|\Delta t_{p}|)
 \equiv 0 \;; \\  \lb{s2_85}
\delta\wt{\Sigma}_{s\ppr s}^{ba}(\vec{x},t_{p=\boldsymbol{\pm}}=T_{\mathsf{ini}}-|\Delta t_{p}|) &\equiv& 0 \;; \hspace*{0.6cm}
\sigma_{D;s\ppr s}^{(0)}(\vec{x},t_{p=\boldsymbol{\pm}}=T_{\mathsf{ini}}-|\Delta t_{p}|) \equiv 0 \;.
\eeq
\item This only amounts to an extension from the anomalous doubled field
\(\breve{\Psi}_{\vec{x},s}^{a}(t_{p})=(\psi_{\vec{x},s}(t_{p})\,,\,\psi_{\vec{x},s}^{*}(t_{p}+\Delta t_{p})\,)^{T}\)
with non-zero entries for times '\(T_{\mathsf{fin}}\geq t_{p}\geq T_{\mathsf{ini}}\)' to the interval
'\(T_{\mathsf{fin}}\geq t_{p}\geq T_{\mathsf{ini}}-|\Delta t_{p}|\)' so that one also regards the complex
conjugated field \(\psi_{\vec{x},s}^{*}(t_{p}=T_{\mathsf{ini}})\) in the second part (\(a=2\)) of
\(\breve{\Psi}_{\vec{x},s}^{a}(t_{p}=T_{\mathsf{ini}}-|\Delta t_{p}|)=
(\psi_{\vec{x},s}(t_{p}=T_{\mathsf{ini}}-|\Delta t_{p}|)\equiv0\,\boldsymbol{,}\,
\psi_{\vec{x},s}^{*}(t_{p}=T_{\mathsf{ini}})\,)^{T}\)
with a vanishing first component at '\(t_{p}=T_{\mathsf{ini}}-|\Delta t_{p}|\)' according to (\ref{s2_83}-\ref{s2_85}).
Similar amendments have to be performed for the anomalous doubled one-particle operator
\(\breve{\mscr{H}}_{\vec{x}\ppr,s\ppr;\vec{x},s}^{ba}(t_{q}\ppr,t_{p})\) (\ref{s2_69}-\ref{s2_71}) concerning
this time point '\(t_{p},\,t_{q}\ppr=T_{\mathsf{ini}}-|\Delta t_{p}|\)'.
\end{itemize}
Since one has the following relation between \(\breve{\Psi}_{\vec{x},s}^{\sharp a}(t_{p})\) and
\(\breve{\Psi}_{\vec{x},s}^{a}(t_{p})\) with Pauli matrix \((\hat{\tau}_{1})^{ba}\)
for the exchange of the two components '\(a=1,2\)'
\be \lb{s2_86}
\breve{\Psi}_{\vec{x},s}^{\sharp b}(t_{p}) = \Big((\hat{\tau}_{1})^{ba}\;\breve{\Psi}_{\vec{x},s}^{a}(t_{p})\Big)^{T,b}\;,
\ee
the bilinear, anomalous doubled anti-commuting fields
\(\breve{\Psi}_{\vec{x}\ppr,s\ppr}^{\sharp b}(t_{q}\ppr)\), \(\breve{\Psi}_{\vec{x},s}^{a}(t_{p})\) in (\ref{s2_78})
can be removed by integration
so that one obtains the square root of the anomalous doubled fermion determinant with matrix
\(\breve{\mscr{M}}_{\vec{x}\ppr,s\ppr;\vec{x},s}^{ba}(t_{q}\ppr,t_{p})\) (\ref{s2_79})
and the propagator \(\breve{\mscr{M}}_{\vec{x}\ppr,s\ppr;\vec{x},s}^{\mathbf{-1};ba}(t_{q}\ppr,t_{p})\)
weighted by the anomalous doubled source fields \(J_{\psi;s\ppr}^{\dag b}(\vec{x}\ppr,t_{q}\ppr)\),
\(J_{\psi;s}^{a}(\vec{x},t_{p})\). The exchange matrix \((\hat{\tau}_{1})^{ba}\) of (\ref{s2_86}) can be omitted
in the path integral (\ref{s2_87}) because its combined appearance on the two branches of the time contour
does not affect the final weighting for observables, neither within the determinant nor with the propagator
of \(\breve{\mscr{M}}_{\vec{x}\ppr,s\ppr;\vec{x},s}^{ba}(t_{q}\ppr,t_{p})\) (\ref{s2_79})
\beq  \lb{s2_87}
\lefteqn{Z[\hat{\mscr{J}},J_{\psi},\im\hat{J}_{\psi\psi}]  =
\int\;\;d[\sigma_{D;ss\ppr}^{(0)}(\vec{x},t_{p})]\;\; \exp\bigg\{-\frac{\im}{2\hbar}
\int_{C}d t_{p}\sum_{\vec{x};s,s\ppr=\uparrow,\downarrow}
\frac{\sigma_{D;ss\ppr}^{(0)}(\vec{x},t_{p})\;\;
\sigma_{D;s\ppr s}^{(0)}(\vec{x},t_{p})}{V_{0}+\im\;\ve_{p}}\bigg\} }   \\ \no &\times&
\int d[\delta\wt{\Sigma}(\vec{x},t_{p})]\;\;\exp\bigg\{-\frac{\im}{4\hbar}
\int_{C}d t_{p}\sum_{\vec{x}}\TRS\bigg[\frac{\delta\wt{\Sigma}_{ss\ppr}^{ab}(\vec{x},t_{p})\;
\delta\wt{\Sigma}_{s\ppr s}^{ba}(\vec{x},t_{p})}{V_{0}+\im\;\ve_{p}(\delta_{a=b}-\delta_{a\neq b})}\bigg]\bigg\}
\\ \no &\times &
\Bigg\{{\raisebox{-5pt}{$\mbox{\large DET}\atop {\scriptstyle \breve{C}}$}}
\bigg[(\hat{\tau}_{1})^{bb\ppr}\breve{\mscr{M}}_{\vec{x}\ppr,s\ppr;\vec{x},s}^{b\ppr a}(t_{q}\ppr,t_{p})\bigg]
\Bigg\}^{\mathbf{1/2}} \times \\ \no &\times&
\exp\bigg\{\frac{\im}{2\hbar}\Omega^{2}\int_{\breve{C}}d t_{p}\;d t\ppr_{q}\sum_{\vec{x},\vec{x}\ppr}
\mcal{N}_{x}\sumss J_{\psi;s\ppr}^{\dagger b}(\vec{x}\ppr,t_{q}\ppr)\;\hat{I}\;
\breve{\mscr{M}}_{\vec{x}\ppr,s\ppr;\vec{x},s}^{\mathbf{-1};ba\ppr}(t_{q}\ppr,t_{p})\;
(\hat{\tau}_{1})^{a\ppr b\ppr}\hat{I}\;(\hat{\tau}_{1})^{b\ppr a}
J_{\psi;s}^{a}(\vec{x},t_{p})\bigg\}_{.}
\eeq
According to the particular relation (\ref{s2_86}) for the exchange between
\(\breve{\Psi}_{\vec{x},s}^{\sharp b}(t_{p})\) and \(\breve{\Psi}_{\vec{x},s}^{a}(t_{p})\) with Pauli-matrix
\((\hat{\tau}_{1})^{ba}\), we have applied the standard relation (\ref{s2_88}) for anti-commuting variables
\(\xi_{j}\) (\(j=1,\ldots,2\,N_{0}\)) in order to remove the bilinear fields
\(\breve{\Psi}_{\vec{x}\ppr,s\ppr}^{\sharp b}(t_{q}\ppr)\) and \(\breve{\Psi}_{\vec{x},s}^{a}(t_{p})\)
in (\ref{s2_78},\ref{s2_79}) by Gaussian integration. It has to be pointed out that the symmetric part of the
matrix \(\breve{\mscr{M}}_{ji}\) in (\ref{s2_88}) cancels in the exponent with the anti-commuting
variables so that the determinant '\(\mbox{DET}\)' in (\ref{s2_88}) only contains the anti-symmetric part
of the considered matrix \(\breve{\mscr{M}}_{ji}\)
\be \lb{s2_88}
\int d[\xi_{i}]\;\;\exp\bigg\{-\sum_{i,j=1}^{2\,N_{0}}\xi_{j}^{T}\;\breve{\mscr{M}}_{ji}\;\xi_{i}\bigg\}=
\Big\{\mbox{DET}\big[\breve{\mscr{M}}_{ji}\big]\Big\}^{1/2}\;.
\ee
The coset decomposition into densities and anomalous parts with matrix \(\hat{T}(\vec{x},t_{p})\)
yields the path integral (\ref{s2_89}) where the change of the Jacobian is already incorporated from the
'flat' Euclidean self-energy \(\delta\wt{\Sigma}_{ss\ppr}^{ab}(\vec{x},t_{p})\) to densities
and to the independent parameters of pair condensate fields of cosets within
\(d[\hat{T}^{-1}(\vec{x},t_{p})\;d\hat{T}(\vec{x},t_{p})]\) (see appendix \ref{sa}). Apart from
the 'source action' \(\mscr{A}_{\hat{J}_{\psi\psi}}[\hat{T}]\) for pair condensates, there appears a
logarithmic action \(\mscr{A}_{DET}[\hat{T},\hat{\sigma}_{D}^{(0)};\hat{\mscr{J}}]\propto
\mbox{Tr}\ln[\breve{\!\mscr{O}}_{\vec{x}\ppr,s\ppr;\vec{x},s}^{ba}(t_{q}\ppr,t_{p})]\) (\ref{s2_90})
from the determinant \((\,\mbox{DET}[\breve{\mscr{M}}_{\vec{x}\ppr,s\ppr;\vec{x},s}^{ba}(t_{q}\ppr,t_{p})]
\,)^{\mathbf{1/2}}\) and a propagator
\(\breve{\!\mscr{O}}_{\vec{x}\ppr,s_{2};\vec{x},s_{1}}^{\mathbf{-1};b\ppr a\ppr}(t_{q}\ppr,t_{p})\) weighted by the
source fields \(J_{\psi;s\ppr}^{\dag b}(\vec{x}\ppr,t_{q}\ppr)\),
\(J_{\psi;s}^{a}(\vec{x},t_{p})\) and additionally by the coset matrices
\(\hat{T}_{s\ppr s_{2}}^{bb\ppr}(\vec{x}\ppr,t_{q}\ppr)\),
\(\hat{T}_{s_{1}s}^{\mathbf{-1};a\ppr a}(\vec{x},t_{p})\). The two actions
\(\mscr{A}_{DET}[\hat{T},\hat{\sigma}_{D}^{(0)};\hat{\mscr{J}}]\) (\ref{s2_90}),
\(\mscr{A}_{J_{\psi}}[\hat{T},\hat{\sigma}_{D}^{(0)};\hat{\mscr{J}}]\) (\ref{s2_91}) are determined by the operator
\(\breve{\mscr{O}}_{\vec{x}\ppr,s\ppr;\vec{x},s}^{ba}(t_{q}\ppr,t_{p})\) (\ref{s2_92}) which consists
of the density part \(\breve{\mscr{H}}+\hat{\sigma}_{D}^{(0)}\) with one-particle operator
\(\breve{\!\mscr{H}}\) (\ref{s2_69}-\ref{s2_71}) and gradient term
\(\delta\breve{\!\mscr{H}}(\hat{T}^{-1},\hat{T})=\hat{T}^{-1}\;(\breve{\!\mscr{H}}+\hat{\sigma}_{D}^{(0)})\;\hat{T}-(\breve{\!\mscr{H}}+\hat{\sigma}_{D}^{(0)})\)
of coset matrices aside from the source term for generating correlation functions
\beq \lb{s2_89}
Z[\hat{\mscr{J}},J_{\psi},\im\;\hat{J}_{\psi\psi}] &=&
\int d[\hat{\sigma}_{D;ss\ppr}^{(0)}(\vec{x},t_{p})]\;\;
\exp\bigg\{-\frac{\im}{2\hbar}\int_{C}d t_{p}\sum_{\vec{x};s,s\ppr=\uparrow,\downarrow}
\frac{\sigma_{D;ss\ppr}^{(0)}(\vec{x},t_{p})\;
\sigma_{D;s\ppr s}^{(0)}(\vec{x},t_{p})}{V_{0}+\im\;\ve_{p}}\bigg\}   \\ \no &\times &
\int d\big[\hat{T}^{-1}(\vec{x},t_{p})\;d\hat{T}(\vec{x},t_{p})\big]\;\;
\exp\Big\{\im\;\mscr{A}_{\hat{J}_{\psi\psi}}\big[\hat{T}\big]\Big\}\;\;\times
\\ \no &\times&
\exp\Big\{\mscr{A}_{DET}\big[\hat{T},\hat{\sigma}_{D}^{(0)};\hat{\mscr{J}}\big]\Big\}\;\;
\exp\Big\{\im\;\mscr{A}_{J_{\psi}}\big[\hat{T},\hat{\sigma}_{D}^{(0)};\hat{\mscr{J}}\big]
\Big\}  \;;  \\   \lb{s2_90}
\mscr{A}_{DET}\big[\hat{T},\hat{\sigma}_{D}^{(0)};\hat{\mscr{J}}\big]
&=&\frac{1}{2}\int_{\breve{C}}\frac{d t_{p}}{\hbar}\eta_{p}\sum_{\vec{x}}\mcal{N}\;
\TRS\Big[\ln\Big(\breve{\mscr{O}}_{\vec{x}\ppr,s\ppr;\vec{x},s}^{ba}(t_{q}\ppr,t_{p})
\Big)\Big]  \;;   \\   \lb{s2_91}
\mscr{A}_{J_{\psi}}\big[\hat{T},\hat{\sigma}_{D}^{(0)};\hat{\mscr{J}}\big] &=&
\frac{\Omega^{2}}{2\hbar}\int_{\breve{C}}d t_{p}\;d t_{q}\ppr\sum_{\vec{x},\vec{x}\ppr}\mcal{N}_{x}
\sumss\sum_{a,b=1,2} \times \\ \no &\times &
J_{\psi; s\ppr}^{+b}(\vec{x}\ppr,t_{q}\ppr)\;\hat{I}\;
\bigg(\hat{T}_{s\ppr s_{2}}^{bb\ppr}(\vec{x}\ppr,t_{q}\ppr)\;\;
\breve{\mscr{O}}_{\vec{x}\ppr,s_{2};\vec{x},s_{1}}^{-1;b\ppr a\ppr}(t_{q}\ppr,t_{p})\;\;
\hat{T}_{s_{1}s}^{-1;a\ppr a}(\vec{x},t_{p})\bigg)_{\vec{x}\ppr,s\ppr;\vec{x},s}^{ba}\;\hat{I}\;
J_{\psi;s}^{a}(\vec{x},t_{p})  \;; \\   \lb{s2_92}
\breve{\mscr{O}}_{\vec{x}\ppr,s\ppr;\vec{x},s}^{ba}(t_{q}\ppr,t_{p})&=&\Big(\frac{1}{|\Delta t_{q}\ppr|}\Big)
\bigg[\Big(\breve{\mscr{H}}+\hat{\sigma}_{D}^{(0)}\Big)+
\overbrace{\Big(\hat{T}^{-1}(\vec{x}\ppr,t_{q}\ppr)\;
\big(\breve{\mscr{H}}+\hat{\sigma}_{D}^{(0)}\big)\;\hat{T}(\vec{x},t_{p})-
\big(\breve{\mscr{H}}+\hat{\sigma}_{D}^{(0)}\big) \Big)}^{\delta\breve{\mscr{H}}(\hat{T}^{-1},\hat{T})} +
\\  \no     &+& \underbrace{
\hat{T}^{-1}(\vec{x}\ppr,t_{q}\ppr)\;\hat{I}\;\hat{S}\;\eta_{q}\;|\Delta t_{q}\ppr|\;
\frac{\hat{\mscr{J}}_{\vec{x}\ppr,s_{1}\ppr;\vec{x},s_{1}}^{b\ppr a\ppr}(t_{q}\ppr;t_{p})}{\mcal{N}_{x}}\;
\eta_{p}\;\hat{S}\;\hat{I}\;
\hat{T}(\vec{x},t_{p})}_{\wt{\mscr{J}}(\hat{T}^{-1},\hat{T})}
\bigg]_{\vec{x}\ppr,s\ppr;\vec{x},s}^{ba}\hspace*{-0.64cm}(t_{q}\ppr,t_{p})\;\;\;\;\;\;;  \\ \no
\delta\breve{\mscr{H}}(\hat{T}^{-1},\hat{T}) &=&
\Big(\hat{T}^{-1}\;\big(\breve{\mscr{H}}+\hat{\sigma}_{D}^{(0)}\big)\;\hat{T}-
\big(\breve{\mscr{H}}+\hat{\sigma}_{D}^{(0)}\big)\Big) \;.
\eeq

\subsection{Determination of the 'pair condensate seed'
action $\mscr{A}_{\hat{J}_{\psi\psi}}[\hat{T}]$}  \lb{s23}

The 'pair condensate seed' \(\mscr{A}_{\hat{J}_{\psi\psi}}[\hat{T}]\) follows from the quadratic
term of self-energies originating from the HST of the quartic interaction of anti-commuting fields.
A shift (\ref{s2_93}) of the self-energy matrix is additionally performed in
\(\breve{\mscr{M}}_{\vec{x}\ppr,s\ppr;\vec{x},s}^{ba}(t_{q}\ppr,t_{p})\) (\ref{s2_79}) so that
the source matrix \(\hat{J}_{\psi\psi;s\ppr s}^{b\neq a}(\vec{x},t_{p})\) of pair condensate fields
only appears in the quadratic term \(\mscr{A}_{2}[\hat{T},\delta\hat{\Sigma};\im\:\hat{J}_{\psi\psi}]\) (\ref{s2_97})
of the self-energy remaining from the HST
\be \lb{s2_93}
\delta\wt{\Sigma}_{s\ppr s}^{ba}(\vec{x},t_{p})\rightarrow
\delta\wt{\Sigma}_{s\ppr s}^{ba}(\vec{x},t_{p})+\im\;\hat{J}_{\psi\psi;s\ppr s}^{b\neq a}(\vec{x},t_{p})\;.
\ee
The coset decomposition combined with the change of the integration measure leads to relations (\ref{s2_94}-\ref{s2_100})
with block diagonal self-energy densities \(\delta \hat{\Sigma}_{D;s\ppr s}^{aa}(\vec{x},t_{p})\) (\ref{s2_98})
and parameters (\ref{s2_99},\ref{s2_100})
which are further decomposed into eigenvalues \(\delta\hat{\lambda}_{s}(\vec{x},t_{p})\) and
diagonalizing matrices \(\hat{Q}_{s\ppr s}^{aa}(\vec{x},t_{p})\)
\beq \lb{s2_94}
\exp\Big\{\im\;\mscr{A}_{\hat{J}_{\psi\psi}}\big[\hat{T}\big]\Big\}  &=&
\int d\big[\delta\hat{\Sigma}_{D}(\vec{x},t_{p})\big]\;\;
\mscr{P}\big(\delta\hat{\lambda}(\vec{x},t_{p})\big)\;\;
\exp\Big\{\im\;\mscr{A}_{2}\big[\hat{T},\delta\hat{\Sigma}_{D};\im\hat{J}_{\psi\psi}\big]\Big\}
\\ \no &\hspace*{-0.4cm}=&\hspace*{-0.5cm}
\int d\big[d\hat{Q}(\vec{x},t_{p})\;\hat{Q}^{-1}(\vec{x},t_{p});
\delta\hat{\lambda}(\vec{x},t_{p})\big]\;\;
\mscr{P}\big(\delta\hat{\lambda}(\vec{x},t_{p})\big)\;\;
\exp\Big\{\im\;\mscr{A}_{2}\big[\hat{T},\hat{Q}^{-1}\;\delta\hat{\Lambda}\;\hat{Q};
\im\hat{J}_{\psi\psi}\big]\Big\}  \;;  \\  \lb{s2_95}
\mbox{det}\Big\{\delta\hat{\Sigma}_{D; s s\ppr}^{11}-\delta\lambda\;\;\delta_{ s s\ppr}\Big\}
&=&0\hspace*{1.5cm}
\mbox{det}\Big\{\delta\hat{\Sigma}_{D; s s\ppr}^{22}-
\big(-\delta\lambda\big)\;\;\delta_{ s s\ppr}\Big\}=0  \\   \lb{s2_96}
\delta\hat{\Sigma}_{D;ss\ppr}^{22}(\vec{x},t_{p})&=&-\delta\hat{\Sigma}_{D;ss\ppr}^{11,T}(\vec{x},t_{p})\;;
\\    \lb{s2_97}
\mscr{A}_{2}\big[\hat{T},\delta\hat{\Sigma}_{D};\im\hat{J}_{\psi\psi}\big] &=&  -\frac{1}{4\hbar}
\int_{C}d t_{p}\sum_{\vec{x}}  \times  \\   \no &\times&  \TRS\bigg[\frac{
\big(\delta\wt{\Sigma}_{ss\ppr}^{ab}(\vec{x},t_{p})+\im\:\hat{J}_{\psi\psi;ss\ppr}^{a\neq b}(\vec{x},t_{p})\,\big)\;
\big(\delta\wt{\Sigma}_{s\ppr s}^{ba}(\vec{x},t_{p})+\im\:\hat{J}_{\psi\psi;s\ppr s}^{b\neq a}(\vec{x},t_{p})\,\big)
}{V_{0}+\im\;\ve_{p}(\delta_{a=b}-\delta_{a\neq b})}\bigg]   \;;
\eeq
\beq  \lb{s2_98}
\delta\hat{\Sigma}_{D;ss\ppr}^{11}(\vec{x},t_{p})  &=&
\Big(\hat{Q}^{-1;11}(\vec{x},t_{p})\;\delta\hat{\lambda}(\vec{x},t_{p})\;
\hat{Q}^{11}(\vec{x},t_{p})\Big)_{ss\ppr}^{11} \;;  \\  \lb{s2_99}
\delta\hat{\Sigma}_{D;ss\ppr}^{11}(\vec{x},t_{p})  &=&
\left(\bea{cc} \delta\Sigma_{D;\uparrow\uparrow}(\vec{x},t_{p})  & \delta\Sigma_{D;\uparrow\downarrow}(\vec{x},t_{p}) \\
\delta\Sigma_{D;\uparrow\downarrow}^{*}(\vec{x},t_{p})  & \delta\Sigma_{D;\downarrow\downarrow}(\vec{x},t_{p})
\eea\right)_{ss\ppr}^{11} \;;    \\   \no
\delta\hat{\Sigma}_{D;ss\ppr}^{11}(\vec{x},t_{p}) &=&\frac{1}{2}\bigg[
\big(\hat{\tau}_{0}\big)_{ss\ppr}\big(\delta\lambda_{\uparrow}(\vec{x},t_{p})+
\delta\lambda_{\downarrow}(\vec{x},t_{p})\big) +
\big(\hat{\tau}_{3}\big)_{ss\ppr}\;\cos\big(2|\mscr{F}_{\uparrow\downarrow}(\vec{x},t_{p})|\big)\;
\big(\delta\lambda_{\uparrow}(\vec{x},t_{p})-
\delta\lambda_{\downarrow}(\vec{x},t_{p})\big) +  \\  \lb{s2_100}   &-&
\big(\hat{\tau}_{1}\big)_{ss\ppr}\;\sin\big(2|\mscr{F}_{\uparrow\downarrow}(\vec{x},t_{p})|\big)\;
\sin\big(\varphi(\vec{x},t_{p})\big)\;\big(\delta\lambda_{\uparrow}(\vec{x},t_{p})-
\delta\lambda_{\downarrow}(\vec{x},t_{p})\big) +  \\ \no &-&
\big(\hat{\tau}_{2}\big)_{ss\ppr}\;\sin\big(2|\mscr{F}_{\uparrow\downarrow}(\vec{x},t_{p})|\big)\;
\cos\big(\varphi(\vec{x},t_{p})\big)\;\big(\delta\lambda_{\uparrow}(\vec{x},t_{p})-
\delta\lambda_{\downarrow}(\vec{x},t_{p})\big)  \bigg]_{ss\ppr}^{11} \;.
\eeq
As one collects the various parameters of the total self-energy \(\delta\wt{\Sigma}_{ss\ppr}^{ab}(\vec{x},t_{p})\)
with pair condensate field \(f(\vec{x},t_{p})\) and density fields
\(\delta\lambda_{\uparrow}(\vec{x},t_{p})\), \(\delta\lambda_{\downarrow}(\vec{x},t_{p})\),
\(\mscr{F}_{\uparrow\downarrow}(\vec{x},t_{p})\) (\ref{s2_101}-\ref{s2_104}), one finally achieves the quadratic
term (\ref{s2_105}) of the self-energy which is caused by the HST in terms of six real field variables of the
\(\mbox{so}(4)\) generators with four real density variables for the \(\mbox{u}(2)\) generators and two real
field degrees of freedom within the coset generators \(\mbox{so}(4)\,/\,\mbox{u}(2)\) or pair condensate terms
\beq  \lb{s2_101}
\delta\wt{\Sigma}_{ss\ppr}^{ab}(\vec{x},t_{p})  &=&
\Big(\hat{T}(\vec{x},t_{p})\Big)_{ss_{1}}^{aa_{1}}\;\delta\hat{\Sigma}_{D;s_{1}s_{2}}^{a_{1}=a_{2}}(\vec{x},t_{p})\;
\Big(\hat{T}^{-1}(\vec{x},t_{p})\Big)_{s_{2}s\ppr}^{a_{2}b} =
\left(\bea{cc} \delta\hat{\Sigma}_{ss\ppr}^{11}(\vec{x},t_{p})  &
\delta\wt{\Sigma}_{ss\ppr}^{12}(\vec{x},t_{p})  \\  \delta\wt{\Sigma}_{ss\ppr}^{21}(\vec{x},t_{p}) &
\delta\hat{\Sigma}_{ss\ppr}^{22}(\vec{x},t_{p}) \eea\right)_{ss\ppr}^{ab}\hspace*{-0.5cm}; \\  \lb{s2_102}
\delta\hat{\Sigma}_{ss\ppr}^{22}(\vec{x},t_{p}) &=& -\big(\delta\hat{\Sigma}_{ss\ppr}^{11}(\vec{x},t_{p})\big)^{T}
\;;\hspace*{0.3cm}\delta\wt{\Sigma}_{ss\ppr}^{21}(\vec{x},t_{p})
= -\big(\delta\wt{\Sigma}_{ss\ppr}^{12}(\vec{x},t_{p})\big)\pdag\;;  \\  \lb{s2_103}
\delta\hat{\Sigma}_{ss\ppr}^{11}(\vec{x},t_{p}) &=&\frac{1}{2}\bigg[
\big(\hat{\tau}_{0}\big)_{ss\ppr}\;\cosh(2|f(\vec{x},t_{p})|)\;
\big(\delta\lambda_{\uparrow}(\vec{x},t_{p})+
\delta\lambda_{\downarrow}(\vec{x},t_{p})\big) + \\ \no &+&
\big(\hat{\tau}_{3}\big)_{ss\ppr}\;\cos\big(2|\mscr{F}_{\uparrow\downarrow}(\vec{x},t_{p})|\big)\;
\big(\delta\lambda_{\uparrow}(\vec{x},t_{p})-
\delta\lambda_{\downarrow}(\vec{x},t_{p})\big) +  \\ \no &-&2\;
\big(\hat{\tau}_{1}\big)_{ss\ppr}\;\sin\big(2|\mscr{F}_{\uparrow\downarrow}(\vec{x},t_{p})|\big)\;
\sin\big(\varphi(\vec{x},t_{p})\big)\;\big(\delta\lambda_{\uparrow}(\vec{x},t_{p})-
\delta\lambda_{\downarrow}(\vec{x},t_{p})\big) +  \\ \no &-&2\;
\big(\hat{\tau}_{2}\big)_{ss\ppr}\;\sin\big(2|\mscr{F}_{\uparrow\downarrow}(\vec{x},t_{p})|\big)\;
\cos\big(\varphi(\vec{x},t_{p})\big)\;\big(\delta\lambda_{\uparrow}(\vec{x},t_{p})-
\delta\lambda_{\downarrow}(\vec{x},t_{p})\big)  \bigg]_{ss\ppr}^{11} \;;  \\  \lb{s2_104}
\delta\wt{\Sigma}_{ss\ppr}^{12}(\vec{x},t_{p}) &=&\frac{1}{2}
\big(\hat{\tau}_{2}\big)_{ss\ppr}\;\sinh(2|f(\vec{x},t_{p})|)\;\exp\{\im\;\phi(\vec{x},t_{p})\}\;
\big(\delta\lambda_{\uparrow}(\vec{x},t_{p})+\delta\lambda_{\downarrow}(\vec{x},t_{p})\big) \;;
\eeq
\beq \lb{s2_105}
\lefteqn{\im\;\mscr{A}_{2}\big[\hat{T},\hat{Q}^{-1}\;\im\;\delta\hat{\Lambda}\;\hat{Q};
\im\hat{J}_{\psi\psi}\big]  =  -\frac{\im}{4\hbar}
\int_{C}d t_{p}\sum_{\vec{x}}  \times } \\ \no &\times& \TRS\bigg[\frac{
\big(\delta\wt{\Sigma}_{ss\ppr}^{ab}(\vec{x},t_{p})+\im\:\hat{J}_{\psi\psi;ss\ppr}^{a\neq b}(\vec{x},t_{p})\,\big)\;
\big(\delta\wt{\Sigma}_{s\ppr s}^{ba}(\vec{x},t_{p})+\im\:\hat{J}_{\psi\psi;s\ppr s}^{b\neq a}(\vec{x},t_{p})\,\big)
}{V_{0}+\im\;\ve_{p}(\delta_{a=b}-\delta_{a\neq b})}\bigg]  = -\frac{\im}{4\hbar} \int_{C}d t_{p}\sum_{\vec{x}}
\times \\ \no &\times&  \Bigg(
\frac{-\im\,\ve_{+}\,\eta_{p}+V_{0}}{\ve_{+}^{2}+V_{0}^{2}}
\Big[\cosh^{\boldsymbol{2}}(2|f(\vec{x},t_{p})|)\;
\big(\delta\lambda_{\uparrow}(\vec{x},t_{p})+\delta\lambda_{\downarrow}(\vec{x},t_{p})\big)^{2} +
\big(\delta\lambda_{\uparrow}(\vec{x},t_{p})-\delta\lambda_{\downarrow}(\vec{x},t_{p})\big)^{2} \Big] + \\ \no &-&4\;
\frac{\im\,\ve_{+}\,\eta_{p}+V_{0}}{\ve_{+}^{2}+V_{0}^{2}}\sinh(2|f(\vec{x},t_{p})|)\;
\sin\big(\phi(\vec{x},t_{p})-\gamma(\vec{x},t_{p})\big)\;\big|j_{\psi\psi}(\vec{x},t_{p})\big|\;
\big(\delta\lambda_{\uparrow}(\vec{x},t_{p})+\delta\lambda_{\downarrow}(\vec{x},t_{p})\big) + \\ \no
&-&4\;\frac{\im\,\ve_{+}\,\eta_{p}+V_{0}}{\ve_{+}^{2}+V_{0}^{2}}\;\big|j_{\psi\psi}(\vec{x},t_{p})\big|^{2}
\Bigg)\;.
\eeq
According to the imaginary increments \(\im\:\ve_{p}\,(\delta_{a=b}-\delta_{a\neq b})\) of the parameter \(V_{0}\)
for an effective, short-ranged interaction, the integrations of \(\delta\hat{\lambda}_{s}(\vec{x},t_{p})\),
\(\hat{Q}_{s\ppr s}^{aa}(\vec{x},t_{p})\) or \(\delta \hat{\Sigma}_{D;s\ppr s}^{aa}(\vec{x},t_{p})\)
can be performed in (\ref{s2_94}) so that the 'pair condensate seed' action
\(\exp\{\im\,\mscr{A}_{\hat{J}_{\psi\psi}}[\hat{T}]\}\)
simplifies to the relation (\ref{s2_106}) with remaining pair condensate field degrees of freedom
\(f(\vec{x},t_{p})=|f(\vec{x},t_{p})|\;\exp\{\im\:\phi(\vec{x},t_{p})\}\). One has to note from the integration of the density variables the appearance of the nontrivial factors \(\cosh^{\boldsymbol{-3}}(2|f(\vec{x},t_{p})|)\)
which have additionally to be considered in the coset integration measure of pair condensate field variables
\beq \lb{s2_106}
\exp\big\{\im\;\mscr{A}_{\hat{J}_{\psi\psi}}[\hat{T}]\big\}&=&\hspace*{-0.3cm}
\prod_{\{\vec{x},t_{p}\}}\bigg(\pi^{2}\bigg(\frac{4\,\hbar\,\mcal{N}_{x}\,|V_{0}|^{3}}{|\Delta t_{p}|}\bigg)^{3}\;
\frac{1}{\cosh^{\boldsymbol{3}}(2\,|f(\vec{x},t_{p})|)}\bigg)\;
\exp\Big\{\im\int_{C}d t_{p}\sum_{\vec{x}}\big|j_{\psi\psi}(\vec{x},t_{p})\big|^{2}\Big\} \\ \no &\times&
\exp\Big\{\im\;\mfrak{a}_{\hat{j}_{\psi\psi}}[\hat{T}]\Big\}\;
\Big(1+2\im\;\mfrak{a}_{\hat{j}_{\psi\psi}}[\hat{T}]\Big) \;;  \\  \lb{s2_107}
\mfrak{a}_{\hat{j}_{\psi\psi}}[\hat{T}] &=&\int_{C}d t_{p}\sum_{\vec{x}}
\big|j_{\psi\psi}(\vec{x},t_{p})\big|^{\boldsymbol{2}}\;\tanh^{\boldsymbol{2}}\!\big(2\,|f(\vec{x},t_{p})|\big)\;
\sin^{\boldsymbol{2}}\!\big(\phi(\vec{x},t_{p})-\gamma(\vec{x},t_{p})\big) \;.
\eeq
We briefly list the involved integration measure for density terms and the pair condensate with corresponding
parameters which are specified in Eqs. (\ref{s2_108}-\ref{s2_110}), respectively (compare appendix \ref{sa})
\beq \lb{s2_108}
\mscr{P}\big(\delta\hat{\lambda}(\vec{x},t_{p})\big)  &=&
\prod_{\vec{x},t_{p}} \big|\delta\lambda_{\uparrow}(\vec{x},t_{p})+\delta\lambda_{\downarrow}(\vec{x},t_{p})\big|^{2} \;; \\  \lb{s2_109}
d[\delta\hat{\Sigma}_{D}(\vec{x},t_{p})]\;\;\mscr{P}\big(\delta\hat{\lambda}(\vec{x},t_{p})\big) &=&
d\big[d\hat{Q}(\vec{x},t_{p})\;\hat{Q}^{-1}(\vec{x},t_{p});\delta\hat{\lambda}(\vec{x},t_{p})\big]\;\;
\mscr{P}\big(\delta\hat{\lambda}(\vec{x},t_{p})\big) =  \\ \no &=&
\prod_{\vec{x},t_{p}} \Bigg\{8\;d\big(\delta\lambda_{\uparrow}(\vec{x},t_{p})\big)\;
d\big(\delta\lambda_{\downarrow}(\vec{x},t_{p})\big)\;\;
\big|\delta\lambda_{\uparrow}^{2}(\vec{x},t_{p})-
\delta\lambda_{\downarrow}^{2}(\vec{x},t_{p})\big|^{\boldsymbol{2}}\;\times \\ \no &\times&
d\big(|\mscr{F}(\vec{x},t_{p})|\big)\;\sin\big(|\mscr{F}(\vec{x},t_{p})|\big)\;
\cos\big(|\mscr{F}(\vec{x},t_{p})|\big)\;\;d\varphi(\vec{x},t_{p}) \Bigg\} \;;  \\ \lb{s2_110}
d\big[\hat{T}^{-1}(\vec{x},t_{p})\;d\hat{T}(\vec{x},t_{p})\big] &=&
\prod_{\{\vec{x},t_{p}\}} \bigg\{8\;d\big(|f(\vec{x},t_{p})|\big)\;\sinh\big(|f(\vec{x},t_{p})|\big)\;
\cosh\big(|f(\vec{x},t_{p})|\big)\;\;d\phi(\vec{x},t_{p})\bigg\}_{\mbox{.}}
\eeq
The combination of (\ref{s2_110}) with the factors \(\cosh^{\boldsymbol{-3}}(2|f(\vec{x},t_{p})|)\)
of (\ref{s2_106},\ref{s2_107}) finally results into the total integration measure (\ref{s2_111})
for the pair condensate field variables
\beq \lb{s2_111}
\lefteqn{d\big[\hat{T}^{-1}(\vec{x},t_{p})\;d\hat{T}(\vec{x},t_{p})\big]\;\times\; \prod_{\{\vec{x},t_{p}\}}\bigg(
\frac{1}{\cosh^{\boldsymbol{3}}(2\,|f(\vec{x},t_{p})|)}\bigg)= } \\ \no &=&
\prod_{\{\vec{x},t_{p}\}} \bigg\{8\;d\big(|f(\vec{x},t_{p})|\big)\;\sinh\big(|f(\vec{x},t_{p})|\big)\;
\cosh\big(|f(\vec{x},t_{p})|\big)\;\;d\phi(\vec{x},t_{p})\bigg\}\;\times\; \prod_{\{\vec{x},t_{p}\}}\bigg(
\frac{1}{\cosh^{\boldsymbol{3}}(2\,|f(\vec{x},t_{p})|)}\bigg)  \\ \no &=&\prod_{\{\vec{x},t_{p}\}} \bigg\{4\;
d\big(|f(\vec{x},t_{p})|\big)\;\frac{\sinh\big(2\,|f(\vec{x},t_{p})|\big)}{\cosh^{\boldsymbol{3}}(2\,|f(\vec{x},t_{p})|)}
\;\;d\phi(\vec{x},t_{p})\bigg\} = \prod_{\vec{x},t_{p}} \bigg\{
-d\Big(\cosh^{\boldsymbol{-2}}\big(2\,|f(\vec{x},t_{p})|\big)\Big)\;\;d\phi(\vec{x},t_{p})\bigg\}  \\ \no &=&
d\big[\tanh^{\boldsymbol{2}}(2|f(\vec{x},t_{p})|)-1\big]\;\;
d\big[\phi(\vec{x},t_{p})\big] = d\big[\tanh^{\boldsymbol{2}}(2|f(\vec{x},t_{p})|)\big]\;\;
d\big[\phi(\vec{x},t_{p})\big]\;.
\eeq
It is instructive to point out the combination
\(\exp\{\im\,\mfrak{a}_{\hat{j}_{\psi\psi}}[\hat{T}]\}\;(1+2\im\,\mfrak{a}_{\hat{j}_{\psi\psi}}[\hat{T}]\,)\)
in (\ref{s2_106},\ref{s2_107}) which can also be obtained from transforming the term
\(|\delta\lambda_{\uparrow}^{2}(\vec{x},t_{p})-\delta\lambda_{\downarrow}^{2}(\vec{x},t_{p})\,|^{\boldsymbol{2}}\)
to a Vandermonde determinant with Hermite polynomials and Gaussian weights
for orthogonal basis functions \cite{mehta}.
The above mentioned term with action \(\mfrak{a}_{\hat{j}_{\psi\psi}}[\hat{T}]\) (\ref{s2_107}) can therefore be
regarded as a Hermite polynomial \(H_{2}(x)\) with argument \(x^{2}=-\im\,\mfrak{a}_{\hat{j}_{\psi\psi}}[\hat{T}]\)
and corresponding Gaussian weight
\be \lb{s2_112}
\exp\Big\{\im\;\mfrak{a}_{\hat{j}_{\psi\psi}}[\hat{T}]\Big\}\;
\Big(1+2\im\;\mfrak{a}_{\hat{j}_{\psi\psi}}[\hat{T}]\Big) \propto
\exp\{-x^{2}\}\;(-2)\;(1-2\,x^{2})\propto \exp\{-x^{2}\}\;H_{2}(x)\;; \;\;
x^{2}=-\im\;\mfrak{a}_{\hat{j}_{\psi\psi}}[\hat{T}]\;.
\ee

\subsection{Gradient expansion of the fermion determinant} \lb{s24}

The separation of the integration measure from the 'flat' Euclidean self-energy
\(\delta\wt{\Sigma}_{ss\ppr}^{ab}(\vec{x},t_{p})\) to the product of a density and
pair condensate part can be merged with a division of the actions
\(\mscr{A}_{DET}[\hat{T},\hat{\sigma}_{D}^{(0)};\hat{\mscr{J}}]\) (\ref{s2_90}),
\(\mscr{A}_{J_{\psi}}[\hat{T},\hat{\sigma}_{D}^{(0)};\hat{\mscr{J}}]\) (\ref{s2_91})
into a pure 'density path integral' \(Z[\hat{\sigma}_{D}^{(0)};j_{\psi}]\) (\ref{s2_113})
with the hermitian self-energy field variables \(\sigma_{D;ss\ppr}^{(0)}(\vec{x},t_{p})\)
and remaining coset parts
\(\mscr{A}_{DET}\ppr[\hat{T};\hat{\mscr{J}}]\), \(\mscr{A}_{J_{\psi}}\ppr[\hat{T};\hat{\mscr{J}}]\)
(see following Eqs. (\ref{s2_115}-\ref{s2_117}))
\beq  \lb{s2_113}
Z[\hat{\sigma}_{D}^{(0)};j_{\psi}] &=& \int d[\sigma_{D;ss\ppr}^{(0)}(\vec{x},t_{p})]\;\;
\exp\bigg\{-\frac{\im}{2\hbar}\int_{C}d t_{p}\sum_{\vec{x};s,s\ppr=\uparrow,\downarrow}
\frac{\sigma_{D;ss\ppr}^{(0)}(\vec{x},t_{p})\;
\sigma_{D;s\ppr s}^{(0)}(\vec{x},t_{p})}{V_{0}+\im\;\hat{\ve}_{p}}\bigg\}
\\ \no &\times& \exp\bigg\{\int_{\breve{C}}\frac{d t_{p}}{\hbar}\eta_{p}\sum_{\vec{x}}\mcal{N}
\trs\ln\Big[\breve{\mscr{H}}^{11}+\hat{\sigma}_{D}^{(0)11}\Big]\bigg\} \\ \no &\times&
\exp\bigg\{\im\frac{\Omega}{\hbar}\int_{\breve{C}}d t_{p}\;d t_{q}\ppr\sum_{\vec{x},\vec{x}\ppr}
\mcal{N}_{x}\sum_{s,s\ppr=\uparrow,\downarrow}j_{\psi;s\ppr}\pdag(\vec{x}\ppr,t_{q}\ppr)
\Big(\breve{\mscr{H}}^{11}+\hat{\sigma}_{D}^{(0)11}\Big)_{\vec{x}\ppr,s\ppr;\vec{x},s}^{-1}
\hspace*{-0.3cm}(t_{q}\ppr,t_{p})\;\;j_{\psi;s}(\vec{x},t_{p})\bigg\}\;.
\eeq
One can use a saddle point approximation in (\ref{s2_114}) of the pure 'density path integral'
\(Z[\hat{\sigma}_{D}^{(0)};j_{\psi}]\) (\ref{s2_113}) for extracting
classical, complex fields \(\langle\sigma_{D;ss\ppr}^{(0)}(\vec{x},t_{p})\rangle\) whose imaginary parts of the
corresponding eigenvalues
comply with the imaginary increment \(\im\:\ve_{p}\) of the anomalous doubled one-particle operator
\(\breve{\!\mscr{H}}\) (\ref{s2_69}-\ref{s2_71}). This classical approximation
\(\langle\sigma_{D;ss\ppr}^{(0)}(\vec{x},t_{p})\rangle\) can be inserted into
\(\mscr{A}_{DET}\ppr[\hat{T};\hat{\mscr{J}}]\), \(\mscr{A}_{J_{\psi}}\ppr[\hat{T};\hat{\mscr{J}}]\) so that
the independent parameters of the coset matrix \(\hat{T}(\vec{x},t_{p})\) are the only remaining
'path integration fields' (\ref{s2_115}-\ref{s2_117})
\beq \lb{s2_114}
Z[\hat{\mscr{J}},J_{\psi},\im\:\hat{J}_{\psi\psi}]&=&
\boldsymbol{\bigg\langle}Z[\hat{\sigma}_{D}^{(0)};j_{\psi}] \;\times
\int\prod_{\{\vec{x},t_{p}\}} \Big(d\big[\tanh^{\boldsymbol{2}}(2|f(\vec{x},t_{p})|)\big]\;
d\big[\phi(\vec{x},t_{p})\big]\Big)\;\;\\ \no &\times&
\exp\Big\{\im\;\mfrak{a}_{\hat{j}_{\psi\psi}}[\hat{T}]\Big\}\;
\Big(1+2\im\;\mfrak{a}_{\hat{j}_{\psi\psi}}[\hat{T}]\Big)\;
\exp\Big\{\im\int_{C}d t_{p}\sum_{\vec{x}}\big|j_{\psi\psi}(\vec{x},t_{p})\big|^{2}\Big\}    \\ \no &\times&
\exp\bigg\{\mscr{A}_{DET}[\hat{T},\hat{\sigma}_{D}^{(0)};\hat{\mscr{J}}]-
\int_{\breve{C}}\frac{d t_{p}}{\hbar}\eta_{p}\sum_{\vec{x}}\mcal{N}
\trs\ln\Big[\breve{\mscr{H}}^{11}+\hat{\sigma}_{D}^{(0)11}\Big]\bigg\} \\ \no &\times&
\exp\bigg\{\im\;\mscr{A}_{J_{\psi}}[\hat{T},\hat{\sigma}_{D}^{(0)};\hat{\mscr{J}}]-
\im\frac{\Omega}{\hbar}\int_{\breve{C}}d t_{p}\;d t_{q}\ppr\sum_{\vec{x},\vec{x}\ppr}
\mcal{N}_{x}\sum_{s,s\ppr=\uparrow,\downarrow}\times \\ \no &\times& j_{\psi;s\ppr}\pdag(\vec{x}\ppr,t_{q}\ppr)\;\;
\Big(\breve{\mscr{H}}^{11}+\hat{\sigma}_{D}^{(0)11}\Big)_{\vec{x}\ppr,s\ppr;\vec{x},s}^{-1}
\hspace*{-0.3cm}(t_{q}\ppr,t_{p})\;\;j_{\psi;s}(\vec{x},t_{p})\bigg\}
\boldsymbol{\bigg\rangle_{\hat{\sigma}_{D}^{(0)}}}\;; \\ \no &\Longrightarrow&
\mbox{saddle point approximation for }\; \langle\sigma_{D;ss\ppr}^{(0)}(\vec{x},t_{p})\rangle\;\Longrightarrow
 \\   \lb{s2_115}   Z[\hat{\mscr{J}},J_{\psi},\im\:\hat{J}_{\psi\psi}]&\approx&
\int\prod_{\{\vec{x},t_{p}\}} \Big(d\big[\tanh^{\boldsymbol{2}}(2|f(\vec{x},t_{p})|)\big]\;
d\big[\phi(\vec{x},t_{p})\big]\Big)\;\;\\ \no &\times&
\exp\Big\{\im\;\mfrak{a}_{\hat{j}_{\psi\psi}}[\hat{T}]\Big\}\;
\Big(1+2\im\;\mfrak{a}_{\hat{j}_{\psi\psi}}[\hat{T}]\Big)\;
\exp\Big\{\im\int_{C}d t_{p}\sum_{\vec{x}}\big|j_{\psi\psi}(\vec{x},t_{p})\big|^{2}\Big\}    \\ \no &\times&
\exp\bigg\{\mscr{A}_{DET}\ppr[\hat{T};\hat{\mscr{J}}]+\im\;\mscr{A}_{J_{\psi}}\ppr[\hat{T};\hat{\mscr{J}}]\bigg\} \;;   \\    \lb{s2_116}
\mscr{A}_{DET}\ppr[\hat{T};\hat{\mscr{J}}]&=&\frac{1}{2}\int_{\breve{C}}\frac{d t_{p}}{\hbar}\eta_{p}
\sum_{\vec{x}}\mcal{N}\TRS\ln\bigg[\hat{1}+\Big(\breve{\mscr{H}}+
\langle\hat{\sigma}_{D}^{(0)}\rangle\Big)^{-1}\;\delta \breve{\mscr{H}}(\hat{T}^{-1},\hat{T})
+  \\ \no &+&  \Big(\breve{\mscr{H}}+\langle\hat{\sigma}_{D}^{(0)}\rangle\Big)^{-1}\;
\wt{\mscr{J}}(\hat{T}^{-1},\hat{T})\bigg]_{ss\ppr}^{ab} \;; \\ \lb{s2_117}
\mscr{A}_{J_{\psi}}\ppr[\hat{T};\hat{\mscr{J}}]&=&\frac{\Omega}{2\hbar}
\int_{\breve{C}}d t_{p}\;d t_{q}\ppr\sum_{\vec{x},\vec{x}\ppr}\mcal{N}_{x}\sum_{a,b=1,2}\;\sum_{s,s\ppr=\uparrow,\downarrow}
J_{\psi;s\ppr}^{\dagger,b}(\vec{x}\ppr,t_{q}\ppr)\;\hat{I}\times
\\ \no &\times&\hspace*{-0.2cm} \bigg[\hat{T}(\vec{x}\ppr,t_{q}\ppr)
\Big[\hat{1}+\Big(\breve{\mscr{H}}+\langle\hat{\sigma}_{D}^{(0)}\rangle\Big)^{-1}
\delta \breve{\mscr{H}}(\hat{T}^{-1},\hat{T})+
\Big(\breve{\mscr{H}}+\langle\hat{\sigma}_{D}^{(0)}\rangle\Big)^{-1}\;
\wt{\mscr{J}}(\hat{T}^{-1},\hat{T})\Big]^{-1}\times \\ \no &\times&
\Big(\breve{\mscr{H}}+\langle\hat{\sigma}_{D}^{(0)}\rangle\Big)^{-1}\;\hat{T}^{-1}(\vec{x},t_{p}) -
\Big(\breve{\mscr{H}}+\langle\hat{\sigma}_{D}^{(0)}\rangle\Big)^{-1}\;
\bigg]_{\vec{x}\ppr,s\ppr;\vec{x},s}^{ba}\hspace*{-0.3cm}(t_{q}\ppr,t_{p})\;\;\hat{I}\;
J_{\psi;s}^{a}(\vec{x},t_{p}) \;.
\eeq
This allows to extract an effective Lagrangian of coset fields restricted to
finite order gradients.
If the combination \((\breve{\!\mscr{H}}+\langle\hat{\sigma}_{D}^{(0)}\rangle)\)
of one-particle operator
\(\breve{\!\mscr{H}}\) (\ref{s2_69}-\ref{s2_71}) with the classical density field
\(\langle\sigma_{D;ss\ppr}^{(0)}(\vec{x},t_{p})\rangle\) contains comparable momentum values as the
gradient term \(\delta\breve{\!\mscr{H}}(\hat{T}^{-1},\hat{T})\), one can apply following identities
for the logarithm and the inverse of an operator \(\breve{\mfrak{O}}\)
\beq \lb{s2_118}
\Big(\ln\breve{\mfrak{O}}\Big)&=&\bigg(\int_{0}^{+\infty}d v\;\;
\frac{\exp\{-v\:\hat{1}\}-\exp\{-v\:\breve{\mfrak{O}}\}}{v}\bigg)\;;
\\ \lb{s2_119} \Big(\breve{\mfrak{O}}^{-1}\Big) &=& \bigg(\int_{0}^{+\infty}d v\;\;
\exp\{-v\:\breve{\mfrak{O}}\}\bigg)\;,
\eeq
so that one also achieves valid expansions of the term
\beq \lb{s2_120}
\breve{\mfrak{O}}&=&\boldsymbol{\hat{1}}+\delta\breve{\!\mscr{H}}(\hat{T}^{-1},\hat{T})\;
\Big(\breve{\!\mscr{H}}+\langle\hat{\sigma}_{D}^{(0)}\rangle\Big)^{-1}  = \boldsymbol{\hat{1}}+
\Big(\hat{T}^{-1}\:\big(\breve{\mscr{H}}+\langle\hat{\sigma}_{D}^{(0)}\rangle\big)\:\hat{T}-
\big(\breve{\mscr{H}}+\langle\hat{\sigma}_{D}^{(0)}\rangle\big)\Big)\;
\Big(\breve{\!\mscr{H}}+\langle\hat{\sigma}_{D}^{(0)}\rangle\Big)^{-1}  \\ \no &=&
\boldsymbol{\hat{1}}+
\Big(\hat{T}^{-1}\:\big(\breve{\mscr{H}}+\langle\hat{\sigma}_{D}^{(0)}\rangle\big)\:\hat{T}\:
\big(\breve{\!\mscr{H}}+\langle\hat{\sigma}_{D}^{(0)}\rangle\big)^{-1} - \boldsymbol{\hat{1}}\Big)  \\ \no &=&
\hat{T}^{-1}\:\big(\breve{\mscr{H}}+\langle\hat{\sigma}_{D}^{(0)}\rangle\big)\:\hat{T}\;
\Big(\breve{\!\mscr{H}}+\langle\hat{\sigma}_{D}^{(0)}\rangle\Big)^{-1} \;,
\eeq
beyond the order unity in comparison to the standard expansion of the logarithm and the inverse
of an operator \(\breve{\mfrak{O}}\) (as e.g. in
\(\ln\breve{\mfrak{O}}=\ln[\hat{1}+(\breve{\mfrak{O}}-\hat{1})]\) or
\(\breve{\mfrak{O}}^{\mathbf{-1}}=[\hat{1}+(\breve{\mfrak{O}}-\hat{1})]^{\mathbf{-1}}\)).

One might infer that the hermitian part of the rather general operator \(\breve{\mfrak{O}}\), defined
in terms of the coset matrices and saddle point fields (\ref{s2_120}), could have negative eigenvalues
so that the given relations (\ref{s2_118},\ref{s2_119}) do not converge for the logarithm and the
inverse of the operator \(\breve{\mfrak{O}}\) (\ref{s2_120}); although there are no restrictions
onto the values of the hermitian part of \(\breve{\mfrak{O}}\) (\ref{s2_120}), the anti-hermitian
part of \(\breve{\mfrak{O}}\) (\ref{s2_120}) has to comply with
the infinitesimal imaginary increment \(-\im\;\ve_{p}\) in the one-particle operator \(\breve{\!\mscr{H}}\).
It determines the sign of the eigenvalues of the anti-hermitian part in \(\breve{\mfrak{O}}\) (\ref{s2_120})
to be negative valued (\(\times\;\im\)) because this infinitesimal 
imaginary increment \(-\im\;\ve_{p}\) chooses a time direction 
in an otherwise time-reversal invariant system and therefore also fixes the sign of the imaginary parts
of eigenvalues within the saddle point solution \(\langle\sigma_{D;ss\ppr}^{(0)}(\vec{x},t_{p})\rangle\).
Since the determinant and the corresponding logarithm of the total operator \(\breve{\mfrak{O}}\) (\ref{s2_120})
follow from bilinear integration with the anomalous doubled, anti-commuting fields over
the one-particle operator with imaginary increment \(-\im\;\ve_{p}\), one has to require the identical
sign of eigenvalues of the anti-hermitian part of \(\breve{\mfrak{O}}\); if this requirement fails to hold
at any step of derivations and involved approximations, one has obtained an unstable system due to increasing
field values within the time development. This sign convention, following from \(-\im\;\ve_{p}\)
within \(\breve{\!\mscr{H}}\), has also to be fulfilled for fermionic systems apart from bosonic systems
with negative powers of the boson-determinant because the chosen imaginary increment \(-\im\;\ve_{p}\)
assigns a time direction and time development from small, real to larger, real time values on
physical grounds. Therefore, the imaginary, infinitesimal increment can be regarded as an artificial,
dissipative part in an otherwise hermitian system where subsequent approximations (as e.g. the saddle
point solution \(\langle\sigma_{D;ss\ppr}^{(0)}(\vec{x},t_{p})\rangle\) or even the total operator
\(\breve{\mfrak{O}}\) (\ref{s2_120})) have to stay in accordance with '\(-\im\;\ve_{p}\)'.

Since the eigenvalues of the anti-hermitian part of \(\breve{\mfrak{O}}\) (\ref{s2_120})
are determined to be negative valued (\(\times\;\im\)), we can simply multiply the operator
\(\breve{\mfrak{O}}\) by the imaginary unit operator modified with the contour time metric,
which has unit determinant and appropriate inverse, so that the change of the operator
\(\breve{\mfrak{O}}\) (\ref{s2_120}) to the operator
\be \lb{s2_121}
\breve{\mfrak{O}}\;\rightarrow\;\big(\im\;\hat{1}\;\eta_{p}\big)\;\breve{\mfrak{O}}\;,
\ee
always results into positive eigenvalues in the hermitian part of the total operator
'\((\im\;\hat{1}\;\eta_{p}\,)\;\breve{\mfrak{O}}\)'. Therefore, this modified operator can be replaced in
(\ref{s2_118},\ref{s2_119}) for convergent integral relations
\beq \lb{s2_122}
\mbox{DET}\big\{(\im\;\hat{1}\;\eta_{p}\,)\big\} &\equiv& 1\;;\hspace*{0.6cm}
(\im\;\hat{1}\;\eta_{p}\,)^{-1}\times(\im\;\hat{1}\;\eta_{p}\,)=
(-\im\;\hat{1}\;\eta_{p}\,)\times(\im\;\hat{1}\;\eta_{p}\,)\equiv 1\;; \\ \lb{s2_123}
\Big(\ln(\im\;\hat{1}\;\eta_{p}\,)\breve{\mfrak{O}}\Big)&=&\bigg(\int_{0}^{+\infty}d v\;\;
\frac{\exp\{-v\:\hat{1}\}-\exp\{-v\:(\im\;\hat{1}\;\eta_{p}\,)\breve{\mfrak{O}}\}}{v}\bigg)\;;
\\ \lb{s2_124}\Big(\breve{\mfrak{O}}\Big)^{-1} &=&
 \Big((\im\;\hat{1}\;\eta_{p}\,)\breve{\mfrak{O}}\Big)^{-1}\;(\im\;\hat{1}\;\eta_{p}\,) = \bigg(\int_{0}^{+\infty}d v\;\;\exp\{-v\:(\im\;\hat{1}\;\eta_{p}\,)\breve{\mfrak{O}}\}\bigg)\;(\im\;\hat{1}\;\eta_{p}\,)\;.
\eeq
Further simplifications for a finite or infinite order gradient expansion are attained if the total
number \(\mcal{N}_{x}\) (\ref{s1_23}) of involved grid points is properly taken
into account for an expansion.

\section{Extension to nonlocal self-energies for long-ranged interactions} \lb{s3}

\subsection{Field variables and dimensions of the spatially nonlocal self-energy} \lb{s31}

In contrast to sections \ref{s22} to \ref{s24} with approximation
\(V_{|\vec{x}-\vec{x}\ppr|}\approx \delta_{\vec{x},\vec{x}\ppr}\;V_{0}\) (\ref{s2_36}), we extend
to the case of an arbitrary long-ranged interaction potential \(V_{|\vec{x}-\vec{x}\ppr|}\)
and specify the independent field variables of the self-energy with adaption of the coset
decomposition and corresponding parameters. Since we start out from a Hamiltonian (\ref{s2_3},\ref{s2_4})
without an ensemble average for disorder, the extended self-energy depends on a single
time contour variable '\(t_{p}\)' so that the nonlocal self-energy (\ref{s3_1}) has only to be specified
by adding vector 'indices' \(\vec{x}\), \(\vec{x}\ppr\) apart from the
spin indices \(s,\,s\ppr=\uparrow,\downarrow\). We also split the total self-energy
\(\wt{\Sigma}_{\vec{x},s;\vec{x}\ppr,s\ppr}^{ab}(t_{p})\) (\ref{s3_1}) into a hermitian,
density related self-energy \(\sigma_{D;ss\ppr}^{(0)}(\vec{x},\vec{x}\ppr,t_{p})\) as in sections \ref{s22}-\ref{s24}
and deviations \(\delta\wt{\Sigma}_{\vec{x},s;\vec{x}\ppr,s\ppr}^{ab}(t_{p})\)
with {\it anti-hermitian} pair condensate terms according to the additional, imaginary unit factor
in the off-diagonal blocks \(a\neq b\)
\beq \lb{s3_1}
\wt{\Sigma}_{\vec{x},s;\vec{x}\ppr,s\ppr}^{ab}(t_{p}) &=&
\sigma_{D;ss\ppr}^{(0)ab}(\vec{x},\vec{x}\ppr,t_{p})\;\delta_{ab} +
\delta\wt{\Sigma}_{\vec{x},s;\vec{x}\ppr,s\ppr}^{ab}(t_{p}) =
\left(\bea{cc} \hat{\Sigma}_{\vec{x},s;\vec{x}\ppr,s\ppr}^{11}(t_{p})  &
\im\;\delta\hat{\Sigma}_{\vec{x},s;\vec{x}\ppr,s\ppr}^{12}(t_{p})  \\
\im\;\delta\hat{\Sigma}_{\vec{x},s;\vec{x}\ppr,s\ppr}^{21}(t_{p}) &
\hat{\Sigma}_{\vec{x},s;\vec{x}\ppr,s\ppr}^{22}(t_{p}) \eea\right)^{ab}\;; \\  \lb{s3_2}
\hat{\Sigma}_{\vec{x},s;\vec{x}\ppr,s\ppr}^{11}(t_{p}) &=&
\sigma_{D;ss\ppr}^{(0)11}(\vec{x},\vec{x}\ppr,t_{p}) +
\delta\hat{\Sigma}_{\vec{x},s;\vec{x}\ppr,s\ppr}^{11}(t_{p}) \;;  \\ \lb{s3_3}
\hat{\Sigma}_{\vec{x},s;\vec{x}\ppr,s\ppr}^{22}(t_{p}) &=&
\sigma_{D;ss\ppr}^{(0)22}(\vec{x},\vec{x}\ppr,t_{p}) +
\delta\hat{\Sigma}_{\vec{x},s;\vec{x}\ppr,s\ppr}^{22}(t_{p}) \;; \\
\sigma_{D;ss\ppr}^{(0)11}(\vec{x},\vec{x}\ppr,t_{p}) &=& \sigma_{D;ss\ppr}^{(0)}(\vec{x},\vec{x}\ppr,t_{p})\;;
\hspace*{0.6cm}\sigma_{D;ss\ppr}^{(0)22}(\vec{x},\vec{x}\ppr,t_{p}) =
\sigma_{D;ss\ppr}^{(0)T}(\vec{x},\vec{x}\ppr,t_{p})\;;   \\ \no
\sigma_{D;ss\ppr}^{(0)T}(\vec{x},\vec{x}\ppr,t_{p}) &=&
\sigma_{D;s\ppr s}^{(0)}(\vec{x}\ppr,\vec{x},t_{p})\;;   \\ \lb{s3_4}
\delta\wt{\Sigma}_{\vec{x},s;\vec{x}\ppr,s\ppr}^{ab}(t_{p}) &=&
\left(\bea{cc} \delta\hat{\Sigma}_{\vec{x},s;\vec{x}\ppr,s\ppr}^{11}(t_{p})  &
\im\;\delta\hat{\Sigma}_{\vec{x},s;\vec{x}\ppr,s\ppr}^{12}(t_{p})  \\
\im\;\delta\hat{\Sigma}_{\vec{x},s;\vec{x}\ppr,s\ppr}^{21}(t_{p}) &
\delta\hat{\Sigma}_{\vec{x},s;\vec{x}\ppr,s\ppr}^{22}(t_{p}) \eea\right)^{ab}\;.
\eeq
In the remainder the summation convention over multiply
occurring spatial vectors is always implied for the presented case of nonlocal self-energies,
unless the spatial vectors as indices
are set into parenthesis (as e.\ g.\ \(\delta\wt{\Sigma}_{(\vec{x}),s;(\vec{x}\ppr),s\ppr}^{ab}(t_{p})\)).
In analogy to (\ref{s2_41}-\ref{s2_46}) for the short-ranged interaction
\(\delta_{\vec{x},\vec{x}\ppr}\;V_{0}\) (\ref{s2_36}), we take the factorization of
\(\delta\wt{\Sigma}_{\vec{x},s;\vec{x}\ppr,s\ppr}^{ab}(t_{p})\) into block diagonal
densities \(\delta\hat{\Sigma}_{D;\vec{x}_{1},s_{1};\vec{x}_{2},s_{2}}^{a_{1}=a_{2}}(t_{p})\),
with further diagonalizing matrices \(\hat{Q}_{\vec{x}\ppr,s\ppr;\vec{x},s}^{aa}(t_{p})\) and local
eigenvalue densities \(\delta\hat{\Lambda}_{s}^{a=a\ppr}(\vec{x},t_{p})\) in its anomalous doubled
kind, and into off-diagonal (\(a\neq b\)), anti-hermitian pair condensate field degrees of freedom
which are regarded by the coset matrices
\((\hat{T}(t_{p})\,)_{\vec{x},s;\vec{x}_{1},s_{1}}^{aa_{1}}\),
\((\hat{T}^{-1}(t_{p})\,)_{\vec{x}_{2},s_{2};\vec{x}\ppr,s\ppr}^{a_{2}b}\) with extension to
the spatially nonlocal case. Note that the symmetries require extension of the hermitian
conjugation and transpose between the various block parts of
\(\delta\wt{\Sigma}_{\vec{x},s;\vec{x}\ppr,s\ppr}^{ab}(t_{p})\) so that the spatial vectors
\(\vec{x}\), \(\vec{x}\ppr\) have to be incorporated into these matrix operations aside from
the spin space
\beq \lb{s3_5}
\wt{\Sigma}_{\vec{x},s;\vec{x}\ppr,s\ppr}^{ab}(t_{p}) &=&
\left(\bea{cc}
\sigma_{D;ss\ppr}^{(0)}(\vec{x},\vec{x}\ppr,t_{p})+
\delta\hat{\Sigma}_{\vec{x},s;\vec{x}\ppr,s\ppr}^{11}(t_{p})  &
\im\;\delta\hat{\Sigma}_{\vec{x},s;\vec{x}\ppr,s\ppr}^{12}(t_{p})  \\
\im\;\delta\hat{\Sigma}_{\vec{x},s;\vec{x}\ppr,s\ppr}^{21}(t_{p}) &
\sigma_{D;ss\ppr}^{(0)T}(\vec{x},\vec{x}\ppr,t_{p})+
\delta\hat{\Sigma}_{\vec{x},s;\vec{x}\ppr,s\ppr}^{22}(t_{p}) \eea\right)^{ab}  \\ \no &=&
\sigma_{D;ss\ppr}^{(0)ab}(\vec{x},\vec{x}\ppr,t_{p})\;\delta_{ab}+
\delta\wt{\Sigma}_{\vec{x},s;\vec{x}\ppr,s\ppr}^{ab}(t_{p}) \\  \no &=&
\sigma_{D;ss\ppr}^{(0)ab}(\vec{x},\vec{x}\ppr,t_{p})\;\delta_{ab}+
\Big(\hat{T}(t_{p})\Big)_{\vec{x},s;\vec{x}_{1},s_{1}}^{aa_{1}}\;
\delta\hat{\Sigma}_{D;\vec{x}_{1},s_{1};\vec{x}_{2},s_{2}}^{a_{1}=a_{2}}(t_{p})\;
\Big(\hat{T}^{-1}(t_{p})\Big)_{\vec{x}_{2},s_{2};\vec{x}\ppr,s\ppr}^{a_{2}b}  \\  \no &=&
\sigma_{D;ss\ppr}^{(0)ab}(\vec{x},\vec{x}\ppr,t_{p})\;\delta_{ab}+  \\ \no &+&
\Big(\hat{T}(t_{p})\Big)_{\vec{x},s;\vec{x}_{1},s_{1}}^{aa_{1}}\;
\hat{Q}_{\vec{x}_{1},s_{1};\vec{x}_{2}\ppr,s_{2}\ppr}^{-1;a_{1}a_{1}}(t_{p})\;\;
\delta\hat{\Lambda}_{s_{2}\ppr}^{a_{1}=a_{2}}(\vec{x}_{2}\ppr,t_{p})\;\;
\hat{Q}_{\vec{x}_{2}\ppr,s_{2}\ppr;\vec{x}_{2},s_{2}}^{a_{2}a_{2}}(t_{p})\;\;
\Big(\hat{T}^{-1}(t_{p})\Big)_{\vec{x}_{2},s_{2};\vec{x}\ppr,s\ppr}^{a_{2}b}  \\ \no &=&
\sigma_{D;ss\ppr}^{(0)ab}(\vec{x},\vec{x}\ppr,t_{p})\;\delta_{ab}+
\Big(\hat{T}_{0}(t_{p})\Big)_{\vec{x},s;\vec{x}_{1},s_{1}}^{aa_{1}}\;
\delta\hat{\Lambda}_{s_{1}=s_{2}}^{a_{1}=a_{2}}(\vec{x}_{1},t_{p})\;\delta_{\vec{x}_{1},\vec{x}_{2}}\;
\Big(\hat{T}_{0}^{-1}(t_{p})\Big)_{\vec{x}_{2},s_{2};\vec{x}\ppr,s\ppr}^{a_{2}b} \;;  \\ \lb{s3_6}
\delta\hat{\Sigma}_{\vec{x},s;\vec{x}\ppr,s\ppr}^{11}(t_{p}) &=&
\delta\hat{\Sigma}_{\vec{x},s;\vec{x}\ppr,s\ppr}^{11,\dagger}(t_{p}) \;;\hspace*{0.6cm}
\delta\hat{\Sigma}_{\vec{x},s;\vec{x}\ppr,s\ppr}^{22}(t_{p}) =
\delta\hat{\Sigma}_{\vec{x},s;\vec{x}\ppr,s\ppr}^{22,\dagger}(t_{p}) \;; \\ \lb{s3_7}
\delta\hat{\Sigma}_{\vec{x},s;\vec{x}\ppr,s\ppr}^{22}(t_{p}) &=&-
\delta\hat{\Sigma}_{\vec{x},s;\vec{x}\ppr,s\ppr}^{11,T}(t_{p}) \;;   \\  \lb{s3_8}
\delta\hat{\Sigma}_{\vec{x},s;\vec{x}\ppr,s\ppr}^{12}(t_{p}) &=&
-\delta\hat{\Sigma}_{\vec{x},s;\vec{x}\ppr,s\ppr}^{12,T}(t_{p})\;;\hspace*{0.6cm}
\delta\hat{\Sigma}_{\vec{x},s;\vec{x},s}^{12}(t_{p})\equiv 0 \;;  \\ \lb{s3_9}
\delta\hat{\Sigma}_{\vec{x},s;\vec{x}\ppr,s\ppr}^{21}(t_{p}) &=&
\delta\hat{\Sigma}_{\vec{x},s;\vec{x}\ppr,s\ppr}^{12,\dagger}(t_{p}) \;;  \\ \lb{s3_10}
\delta\hat{\Sigma}_{\vec{x},s;\vec{x}\ppr,s\ppr}^{21}(t_{p}) &=&
-\delta\hat{\Sigma}_{\vec{x},s;\vec{x}\ppr,s\ppr}^{21,T}(t_{p}) \;; \hspace*{0.6cm}
\delta\hat{\Sigma}_{\vec{x},s;\vec{x},s}^{21}(t_{p})\equiv 0 \;.
\eeq
According to the nonlocal form of \(\delta\wt{\Sigma}_{\vec{x},s;\vec{x}\ppr,s\ppr}^{ab}(t_{p})\),
the dimensions of the coset decomposition increase from the \(\mbox{so}(4)\) generators
and \(\mbox{u}(2)\) densities (\ref{s2_47},\ref{s2_48}) with factoring into
\(\mbox{SO}(4)\simeq\mbox{SO}(4)\,/\,\mbox{U}(2)\otimes\mbox{U}(2)\) (\ref{s2_60},\ref{s2_61})
for the local case to the extended dimensions of \(\mbox{so}(4\mcal{N}_{x})\) generators and
\(\mbox{u}(2\mcal{N}_{x})\) densities (\ref{s3_11},\ref{s3_12}) with corresponding factoring into
\(\mbox{SO}(4\mcal{N}_{x})\,/\,\mbox{U}(2\mcal{N}_{x})\otimes\mbox{U}(2\mcal{N}_{x})\)
(\ref{s3_13},\ref{s3_14}). We emphasize the appearance of the total number of spatial grid points
\(\mcal{N}_{x}\) (\ref{s1_23}) which determine the number of summands involved over the
summations of the spatial vector 'indices'
\beq \lb{s3_11}
\delta\wt{\Sigma}_{\vec{x},s;\vec{x}\ppr,s\ppr}^{ab}(t_{p}) &\simeq&
\left(\bea{cc}  \im\;\delta\hat{\Sigma}_{\vec{x},s;\vec{x}\ppr,s\ppr}^{12}(t_{p}) &
\delta\hat{\Sigma}_{\vec{x},s;\vec{x}\ppr,s\ppr}^{11}(t_{p})  \\
\underbrace{-\delta\hat{\Sigma}_{\vec{x},s;\vec{x}\ppr,s\ppr}^{11,T}(t_{p})}_{
\delta\hat{\Sigma}_{\vec{x},s;\vec{x}\ppr,s\ppr}^{22}(t_{p})} &
\im\;\delta\hat{\Sigma}_{\vec{x},s;\vec{x}\ppr,s\ppr}^{21}(t_{p}) \eea\right)\in \mbox{so}(4\;\mcal{N}_{x}) \;; \\ \lb{s3_12}
\delta\hat{\Sigma}_{\vec{x},s;\vec{x}\ppr,s\ppr}^{11}(t_{p}) &\simeq&
\Big(\delta\hat{\Sigma}_{\vec{x},s;\vec{x}\ppr,s\ppr}^{22}(t_{p})=
-\delta\hat{\Sigma}_{\vec{x},s;\vec{x}\ppr,s\ppr}^{11,T}(t_{p}) \Big)\in\mbox{u}(2\;\mcal{N}_{x}) \;;
\eeq
\beq \lb{s3_13}
\underbrace{
\overbrace{\mbox{SO}(4\;\mcal{N}_{x})}^{\delta\wt{\Sigma}_{\vec{x},s;\vec{x}\ppr,s\ppr}^{ab}}\,/\,
\overbrace{\mbox{U}(2\;\mcal{N}_{x})}^{\delta\hat{\Sigma}_{\vec{x},s;\vec{x}\ppr,s\ppr}^{aa}}}_{
\delta\hat{\Sigma}_{\vec{x},s;\vec{x}\ppr,s\ppr}^{12}\,,\,
\delta\hat{\Sigma}_{\vec{x},s;\vec{x}\ppr,s\ppr}^{12,*}}  &\otimes&
\overbrace{\mbox{U}(2\;\mcal{N}_{x})}^{\delta\hat{\Sigma}_{\vec{x},s;\vec{x}\ppr,s\ppr}^{aa}} \;; \\
\lb{s3_14}
\bigg(\underbrace{\underbrace{\mbox{SO}(4\;\mcal{N}_{x})}_{\mbox{\scz $4\mcal{N}_{x}(4\mcal{N}_{x}-1)/2$ parameters}}\,/\,
\underbrace{\mbox{U}(2\;\mcal{N}_{x})}_{\mbox{\scz $(2\mcal{N}_{x})^{2}$ parameters}}}_{
\mbox{\scz $(2\mcal{N}_{x})^{2}-2\mcal{N}_{x}$ remaining, real parameters}}\bigg)
&\simeq&  \delta\hat{\Sigma}_{\vec{x},s;\vec{x}\ppr,s\ppr}^{12}(t_{p}),\;
\big(\delta\hat{\Sigma}_{\vec{x},s;\vec{x}\ppr,s\ppr}^{12,*}(t_{p})\big) \;.
\eeq
Similar to the case (\ref{s2_49}-\ref{s2_59}) for \(\delta\hat{\Sigma}_{D;ss\ppr}^{aa}(\vec{x},t_{p})\),
we introduce anomalous doubled parameters and fields for the nonlocal case
\(\delta\hat{\Sigma}_{D;\vec{x},s;\vec{x}\ppr,s\ppr}(t_{p})\) where the symmetry operations
of hermitian conjugation and the transpose of a matrix have to comprise the spatial vector
'indices' \(\vec{x}\), \(\vec{x}\ppr\) apart from the spin space \(s,s\ppr=\uparrow,\downarrow\)
\beq \lb{s3_15}
\delta\hat{\Sigma}_{D;\vec{x},s;\vec{x}\ppr,s\ppr}^{ab}(t_{p}) &=&\delta_{ab}\;\;
\delta\hat{\Sigma}_{D;\vec{x},s;\vec{x}\ppr,s\ppr}^{aa}(t_{p})  \;;  \\ \lb{s3_16}
\delta\hat{\Sigma}_{D;\vec{x},s;\vec{x}\ppr,s\ppr}^{aa}(t_{p}) &=&
\delta\hat{\Sigma}_{D;\vec{x},s;\vec{x}\ppr,s\ppr}^{aa,\dagger}(t_{p})  \;;  \\ \lb{s3_17}
\delta\hat{\Sigma}_{D;\vec{x},s;\vec{x}\ppr,s\ppr}^{22}(t_{p}) &=&
-\delta\hat{\Sigma}_{D;\vec{x},s;\vec{x}\ppr,s\ppr}^{11,T}(t_{p})  \;;  \\ \lb{s3_18}
\delta\hat{\Sigma}_{D;\vec{x}_{1},s_{1};\vec{x}_{2},s_{2}}^{aa}(t_{p}) &=&
\hat{Q}_{\vec{x}_{1},s_{1};\vec{x}\ppr,s\ppr}^{-1;aa}(t_{p})\;\;
\delta\hat{\Lambda}_{s\ppr=s}^{aa}(\vec{x},t_{p})\;\delta_{\vec{x}\ppr,\vec{x}}\;
\hat{Q}_{\vec{x},s;\vec{x}_{2},s_{2}}^{aa}(t_{p})  \;; \\ \lb{s3_19}
\delta\hat{\Lambda}_{s\ppr s}^{aa}(\vec{x},t_{p}) &=& \delta_{s\ppr s}\;\;
\mbox{diag}\Big(\underbrace{\delta\hat{\lambda}_{s}(\vec{x},t_{p})}_{a=1}\;;\;
\underbrace{-\delta\hat{\lambda}_{s}(\vec{x},t_{p})}_{a=2} \Big) \;; \\ \lb{s3_20}
\delta\hat{\lambda}_{s}(\vec{x},t_{p}) &=& \mbox{diag}\Big(\delta\hat{\lambda}_{s=\uparrow}(\vec{x},t_{p})\,,\,
\delta\hat{\lambda}_{s=\downarrow}(\vec{x},t_{p})\Big)\;;   \\  \lb{s3_21}
\hat{Q}_{\vec{x},s;\vec{x}\ppr,s\ppr}^{ab}(t_{p}) &=&\delta_{ab}\;\;
\left(\bea{cc}  \hat{Q}_{\vec{x},s;\vec{x}\ppr,s\ppr}^{11}(t_{p})  &  \\ &
\hat{Q}_{\vec{x},s;\vec{x}\ppr,s\ppr}^{22}(t_{p})  \eea \right)^{ab} \;; \\  \lb{s3_22}
\hat{Q}_{\vec{x},s;\vec{x}\ppr,s\ppr}^{11}(t_{p})  &=&
\bigg(\exp\Big\{\im\;\hat{\mscr{F}}_{D;\vec{x}_{1},s_{1};\vec{x}_{2},s_{2}}(t_{p})\Big\}\bigg)_{
\vec{x},s;\vec{x}\ppr,s\ppr}\;;  \\   \lb{s3_23}
\hat{Q}_{\vec{x},s;\vec{x}\ppr,s\ppr}^{22}(\vec{x},t_{p})  &=&
\bigg(\exp\Big\{-\im\;\hat{\mscr{F}}_{D;\vec{x}_{1},s_{1};\vec{x}_{2},s_{2}}^{T}(t_{p})\Big\}\bigg)_{
\vec{x},s;\vec{x}\ppr,s\ppr}\;; \\  \lb{s3_24}
\hat{\mscr{F}}_{D;\vec{x},s;\vec{x},s}(t_{p}) &\equiv& 0\;;\hspace*{0.4cm}
\hat{\mscr{F}}_{D;\vec{x},s;\vec{x}\ppr,s\ppr}(t_{p})=\hat{\mscr{F}}_{D;\vec{x},s;\vec{x}\ppr,s\ppr}\pdag(t_{p})
\;;  \\  \lb{s3_25}
\hat{\mscr{F}}_{D;\vec{x},s;\vec{x}\ppr,s\ppr}(t_{p}) &\stackrel{\wedge}{=}&
\mbox{u}(2\;\mcal{N}_{x})\,\big/\,
\Big(\mbox{{\small$\prod_{\{\mcal{N}_{x}\}}$}}\big(\mbox{u}(1)\otimes\mbox{u}(1)\big)\Big)
 \\ \no &\stackrel{\wedge}{=}&
\hat{\mscr{F}}_{D;\vec{x},s;\vec{x}\ppr,s\ppr}(t_{p}) \;,\;\;
\big(\hat{\mscr{F}}_{D;\vec{x},s;\vec{x}\ppr,s\ppr}^{*}(t_{p})\big)\;,\;\;
(\mbox{$(2\,\mcal{N}_{x})^{2}-2\,\mcal{N}_{x}$ real parameters})\;.
\eeq
The coset matrix \(\hat{T}_{\vec{x},s;\vec{x}\ppr,s\ppr}^{ab}(t_{p})\) (\ref{s3_26}) of the
pair condensates is specified by the spatially nonlocal generators
\(\hat{Y}_{\vec{x},s;\vec{x}\ppr,s\ppr}^{a\neq b}(t_{p})\) (\ref{s3_27}) with complex valued,
anti-symmetric sub-generators \(\hat{X}_{\vec{x},s;\vec{x}\ppr,s\ppr}(t_{p})\),
\(\hat{X}_{\vec{x},s;\vec{x}\ppr,s\ppr}\pdag(t_{p})\) (\ref{s3_28}) in the off-diagonal blocks
(\(a\neq b\)), each having \((2\mcal{N}_{x})^{2}-2\mcal{N}_{x}\) remaining, real parameters.
We have also included the modified coset matrix
\((\hat{T}_{0}(t_{p})\,)_{\vec{x},s;\vec{x}\ppr,s\ppr}^{ab}\) (\ref{s3_29}) for the calculation of
the invariant coset integration measure whose computation follows in an analogous manner from the
local case of appendix \ref{sa} or can be taken from the more general case of an ortho-symplectic
integration measure with super-symmetry described in detail in Ref. \cite{pop1}
\beq  \lb{s3_26}
\hat{T}_{\vec{x},s;\vec{x}\ppr,s\ppr}^{ab}(t_{p}) &=&
\bigg(\exp\Big\{-\hat{Y}_{\vec{x}_{1},s_{1};\vec{x}_{2},s_{2}}^{a_{1}\neq b_{1}}(t_{p})\Big\}\bigg)_{
\vec{x},s;\vec{x}\ppr,s\ppr}^{ab}
\;;  \\ \lb{s3_27}  \hat{Y}_{\vec{x},s;\vec{x}\ppr,s\ppr}^{a\neq b}(t_{p}) &=&
\left(\bea{cc} 0 & \hat{X}_{\vec{x},s;\vec{x}\ppr,s\ppr}(t_{p}) \\
\hat{X}_{\vec{x},s;\vec{x}\ppr,s\ppr}\pdag(t_{p})  &  0  \eea\right)^{ab}  \;; \\  \lb{s3_28}
\hat{X}_{\vec{x},s;\vec{x}\ppr,s\ppr}(t_{p}) &=& -\hat{X}_{\vec{x},s;\vec{x}\ppr,s\ppr}^{T}(t_{p}) \;;\hspace*{0.6cm}
\hat{X}_{\vec{x},s;\vec{x},s}(t_{p})\equiv 0 \;; \\  \lb{s3_29}
\Big(\hat{T}_{0}(t_{p})\Big)_{\vec{x},s;\vec{x}\ppr,s\ppr}^{ab}  &=&
\hat{T}_{\vec{x},s;\vec{x}_{1},s_{1}}^{ab}(t_{p})\;\;
\hat{Q}_{\vec{x}_{1},s_{1};\vec{x}\ppr,s\ppr}^{-1;bb}(t_{p}) \;.
\eeq

\subsection{HST and separation into actions of density and pair condensate terms} \lb{s32}

In correspondence to the precise, discrete time steps within (\ref{s2_67}), one can accomplish
the HST with nonlocal self-energies where we exactly distinguish between 'equal time',
anomalous doubled fields and 'equal time' hermitian conjugation '\(\pdag\)' (\ref{s2_10},\ref{s2_11})
and between the hermitian conjugation '\(^{\sharp}\)' with 'time shift' correction '\(\Delta t_{p}\)' in
the resulting complex part (\ref{s2_13},\ref{s2_14}). The latter kind has to be applied for the
transformation to density and pair condensate matrices following from dyadic products of anomalous doubled fields
whereas the 'equal time' form has to be used for self-energies and its coset decomposition
\beq \lb{s3_30}
\lefteqn{\exp\bigg\{-\frac{\im}{\hbar}\int_{C}d t_{p}\sum_{\vec{x},\vec{x}\ppr}\sumss
\psi_{\vec{x},s}^{*}(t_{p}+\Delta t_{p})\;\psi_{\vec{x}\ppr,s\ppr}^{*}(t_{p}+\Delta t_{p})\;
V_{|\vec{x}\ppr-\vec{x}|}\;
\psi_{\vec{x}\ppr,s\ppr}(t_{p})\;\psi_{\vec{x},s}(t_{p})\bigg\} =  }
\\ \no &=&
\int\;\;d[\sigma_{D;ss\ppr}^{(0)}(\vec{x},\vec{x}\ppr,t_{p})]\;\;
\exp\bigg\{-\frac{\im}{2\hbar}\int_{C}d t_{p}\sum_{\vec{x},s;\vec{x}\ppr,s\ppr}
\frac{\sigma_{D;ss\ppr}^{(0)}(\vec{x},\vec{x}\ppr,t_{p})\;\;
\sigma_{D;s\ppr s}^{(0)}(\vec{x}\ppr,\vec{x},t_{p})}{V_{|\vec{x}\ppr-\vec{x}|}+
\im\:\ve_{p}}\bigg\}
\\ \no &\times &
 \int d[\delta\wt{\Sigma}_{\vec{x},s;\vec{x}\ppr,s\ppr}^{ab}(t_{p})]\;\;
 \exp\Bigg\{-\frac{\im}{4\hbar}\int_{C}d t_{p}\sum_{\vec{x},\vec{x}\ppr}
 \TRS\bigg[\frac{\delta\wt{\Sigma}_{\vec{x},s;\vec{x}\ppr,s\ppr}^{ab}(t_{p})\;
 \delta\wt{\Sigma}_{\vec{x}\ppr,s\ppr;\vec{x},s}^{ba}(t_{p})}{V_{|\vec{x}\ppr-\vec{x}|}+
\im\:\ve_{p}\:\big(\delta_{a=b}-\delta_{a\neq b}\big)}
\bigg]\Bigg\}
\\ \no &\times&
 \exp\Bigg\{\frac{\im}{2\hbar}\int_{\breve{C}}d t_{p}\sum_{\vec{x},\vec{x}\ppr}
 \TRS\Bigg[\left(
 \bea{cc}
 \breve{R}_{\vec{x},s;\vec{x}\ppr,s\ppr}^{11}(t_{p}) &
\breve{R}_{\vec{x},s;\vec{x}\ppr,s\ppr}^{12}(t_{p}) \\
\breve{R}_{\vec{x},s;\vec{x}\ppr,s\ppr}^{21}(t_{p}) &
\breve{R}_{\vec{x},s;\vec{x}\ppr,s\ppr}^{22}(t_{p})
 \eea\right)\;\hat{S}\;\times \overbrace{\frac{V_{|\vec{x}\ppr-\vec{x}|}}{V_{|\vec{x}\ppr-\vec{x}|}
+\im\:\ve_{p}\:\big(\delta_{a=b}-\delta_{a\neq b}\big)}}^{:=1}
\\ \no &\times&   \left(  \bea{cc}
\sigma_{D;s\ppr s}^{(0)}(\vec{x}\ppr,\vec{x},t_{p})+
 \delta\hat{\Sigma}_{\vec{x}\ppr,s\ppr;\vec{x},s}^{11}(t_{p}) &
\delta\hat{\Sigma}_{\vec{x}\ppr,s\ppr;\vec{x},s}^{12}(t_{p}) \\
 \delta\hat{\Sigma}_{\vec{x}\ppr,s\ppr;\vec{x},s}^{21}(t_{p}) &   \hspace*{-0.6cm}
-\sigma_{D;s\ppr s}^{(0)T}(\vec{x}\ppr,\vec{x},t_{p})-\delta\hat{\Sigma}_{\vec{x}\ppr,s\ppr;\vec{x},s}^{22}(t_{p})
 \eea\right)\;\hat{S}\Bigg]\Bigg\} \;;  \\  \no
\hat{V}_{|\vec{x}\ppr-\vec{x}|}^{ab}&=&V_{|\vec{x}\ppr-\vec{x}|}+
\im\:\ve_{p}\:\big(\delta_{a=b}-\delta_{a\neq b}\big)
\;\;\;;\;\;\;\ve_{p}=\eta_{p}\;\ve_{+}\;\;;\;\;
\ve_{+}>0\;\;\;.
\eeq
Since the precise, anomalous doubling of one-particle parts (\ref{s2_68}-\ref{s2_74}) is not affected
by the extension to nonlocal self-energies, we have only to generalize the source matrix of pair
condensates to its nonlocal kind \(\hat{J}_{\psi\psi;\vec{x},s;\vec{x}\ppr,s\ppr}^{a\neq b}(t_{p})\)
with anti-symmetric sub-matrices \(\hat{j}_{\psi\psi;\vec{x},s;\vec{x}\ppr,s\ppr}(t_{p})\),
\(\hat{j}_{\psi\psi;\vec{x},s;\vec{x}\ppr,s\ppr}\pdag(t_{p})\) whereas the anti-commuting source
field \(\hat{J}_{\psi;s}^{a}(\vec{x},t_{p})\) exactly remains for a coherent,
macroscopic wavefunction in its manner
\beq \lb{s3_31}
\hat{J}_{\psi\psi;\vec{x},s;\vec{x}\ppr,s\ppr}^{a\neq b}(t_{p}) &=&
\left( \bea{cc}
0 & \hat{j}_{\psi\psi;\vec{x},s;\vec{x}\ppr,s\ppr}(t_{p}) \\
\hat{j}_{\psi\psi;\vec{x},s;\vec{x}\ppr,s\ppr}\pdag(t_{p}) & 0
\eea\right)^{ab} \;; \\  \lb{s3_32} \hat{j}_{\psi\psi;\vec{x},s;\vec{x}\ppr,s\ppr}(t_{p})
&=&-\hat{j}_{\psi\psi;\vec{x},s;\vec{x}\ppr,s\ppr}^{T}(t_{p})  \;;  \\ \lb{s3_33}
J_{\psi;s}^{a(=1/2)}(\vec{x},t_{p})&=&\Big(\underbrace{j_{\psi;s}(\vec{x},t_{p})}_{a=1}\;;\;
\underbrace{j_{\psi;s}^{*}(\vec{x},t_{p})}_{a=2}\Big)^{T} \;.
\eeq
After collection of terms with nonlocal self-energies, one finally achieves relation (\ref{s3_34})
with bilinear, anomalous doubled fields \(\breve{\Psi}_{\vec{x}\ppr,s\ppr}^{\sharp b}(t_{q}\ppr)\),
\(\breve{\Psi}_{\vec{x},s}^{a}(t_{p})\) with exact discrete time steps and overall matrix
\(\breve{\mscr{M}}_{\vec{x}\ppr,s\ppr;\vec{x},s}^{ba}(t_{q}\ppr,t_{p})\) (\ref{s3_35})
which is only modified by the nonlocal self-energies
\(\delta\wt{\Sigma}_{\vec{x}\ppr,s\ppr;\vec{x},s}^{ba}(t_{p})\) and source matrix
\(\im\:\hat{J}_{\psi\psi;\vec{x}\ppr,s\ppr;\vec{x},s}^{b\neq a}(t_{p})\)
in comparison to the local case (\ref{s2_78}-\ref{s2_81})
\beq \lb{s3_34}
\lefteqn{Z[\hat{\mscr{J}},J_{\psi},\im\,\hat{J}_{\psi\psi}]=
\int\;\;d[\sigma_{D;ss\ppr}^{(0)}(\vec{x},\vec{x}\ppr,t_{p})]\;\;
\exp\bigg\{-\frac{\im}{2\hbar}\int_{C}d t_{p}\sum_{\vec{x},s;\vec{x}\ppr,s\ppr}
\frac{\sigma_{D;ss\ppr}^{(0)}(\vec{x},\vec{x}\ppr,t_{p})\;\;
\sigma_{D;s\ppr s}^{(0)}(\vec{x}\ppr,\vec{x},t_{p})}{V_{|\vec{x}\ppr-\vec{x}|}+
\im\:\ve_{p} }  \bigg\} }
\\ \no &\times &
 \int d[\delta\wt{\Sigma}_{\vec{x},s;\vec{x}\ppr,s\ppr}^{ab}(t_{p})]\;\;
 \exp\Bigg\{-\frac{\im}{4\hbar}\int_{C}d t_{p}\sum_{\vec{x},\vec{x}\ppr}
 \TRS\bigg[\frac{\delta\wt{\Sigma}_{\vec{x},s;\vec{x}\ppr,s\ppr}^{ab}(t_{p})\;
 \delta\wt{\Sigma}_{\vec{x}\ppr,s\ppr;\vec{x},s}^{ba}(t_{p})}{V_{|\vec{x}\ppr-\vec{x}|}+
\im\:\ve_{p}\:\big(\delta_{a=b}-\delta_{a\neq b}\big)}
\bigg]\Bigg\}       \\  \no &\times&\hspace*{-0.4cm}
\int d[\psi_{\vec{x}\ppr,s\ppr}^{*}(t_{p}),\psi_{\vec{x},s}(t_{p})]\:
\exp\Bigg\{-\frac{\im}{2\hbar}\int_{\breve{C}}d t_{p}\;d t_{q}\ppr
\sum_{\vec{x},\vec{x}\ppr}\mcal{N}_{x}\!\!\sumss\sum_{a,b=1,2}
\breve{\Psi}_{\vec{x}\ppr,s\ppr}^{\sharp b}(t_{q}\ppr)\;\hat{I}\;
\breve{\mscr{M}}_{\vec{x}\ppr,s\ppr;\vec{x},s}^{ba}(t_{q}\ppr,t_{p})\;\hat{I}\;
\breve{\Psi}_{\vec{x},s}^{a}(t_{p})\Bigg\} \\ \no &\times &
\exp\bigg\{-\frac{\im}{2\hbar}\int_{\breve{C}}d t_{p}\sum_{\vec{x}}\sums\Big(
J_{\psi;s}^{\dagger a}(\vec{x},t_{p})\;\hat{S}\;\breve{\Psi}_{\vec{x},s}^{a}(t_{p})+
\breve{\Psi}_{\vec{x},s}^{\sharp a}(t_{p})\;\hat{S}\;J_{\psi;s}^{a}(\vec{x},t_{p})\Big)\bigg\}_{\mbox{;}}
\eeq
\beq\lb{s3_35}
\lefteqn{|\Delta t_{q}\ppr|\;\breve{\mscr{M}}_{\vec{x}\ppr,s\ppr;\vec{x},s}^{ba}(t_{q}\ppr,t_{p})=
\underbrace{\breve{\mscr{H}}_{\vec{x}\ppr,s\ppr;
\vec{x},s}^{ba}(t_{q}\ppr,t_{p})}_{\mbox{\scz(\ref{s2_69}-\ref{s2_71})}} +
\eta_{q}\;\hat{I}\;\hat{S}\;|\Delta t_{q}\ppr|\;
\frac{\hat{\mscr{J}}_{\vec{x}\ppr,s\ppr;\vec{x},s}^{ba}(t_{q}\ppr,t_{p})}{\mcal{N}_{x}}\;
\hat{S}\;\hat{I}\;\eta_{p}   +      } \\ \no &+&
\Bigg(\sigma_{D;s\ppr s}^{(0)ba}(\vec{x}\ppr,\vec{x},t_{p})\;\delta_{ba} -
\frac{\im}{\mcal{N}_{x}}\;\underbrace{\hat{J}_{\psi\psi;\vec{x}\ppr,s\ppr;\vec{x},s}^{b\neq a}(t_{p})}_{
\mbox{\scz(\ref{s3_31},\ref{s3_32})}} +
\underbrace{\left(\bea{cc}
\delta\hat{\Sigma}_{\vec{x}\ppr,s\ppr;\vec{x},s}^{11}(t_{p}) & \im\;
\delta\hat{\Sigma}_{\vec{x}\ppr,s\ppr;\vec{x},s}^{12}(t_{p}) \\
\im\;\delta\hat{\Sigma}_{\vec{x}\ppr,s\ppr;\vec{x},s}^{21}(t_{p}) &
\delta\hat{\Sigma}_{\vec{x}\ppr,s\ppr;\vec{x},s}(t_{p})
\eea\right)}_{\delta\wt{\Sigma}_{\vec{x}\ppr,s\ppr;\vec{x},s}(t_{p})=\mbox{\scz(\ref{s3_4})}}\;\Bigg)
\eta_{q}\;\delta_{qp}\;\frac{\delta_{t_{q}\ppr,t_{p}}}{\mcal{N}_{x}} \;;
\eeq
\beq
\sigma_{D;s\ppr s}^{(0)11}(\vec{x}\ppr,\vec{x},t_{p}) &=& \sigma_{D;s\ppr s}^{(0)}(\vec{x}\ppr,\vec{x},t_{p})\;;
\hspace*{0.6cm}\sigma_{D;s\ppr s}^{(0)22}(\vec{x}\ppr,\vec{x},t_{p}) =
\sigma_{D;s\ppr s}^{(0)T}(\vec{x}\ppr,\vec{x},t_{p})\;;   \\ \no
\sigma_{D;s\ppr s}^{(0)T}(\vec{x}\ppr,\vec{x},t_{p}) &=&
\sigma_{D;ss\ppr}^{(0)}(\vec{x},\vec{x}\ppr,t_{p})\;;   \\   \lb{s3_36}
\hat{I}_{4\times 4}&=&\Big\{\underbrace{\hat{1}_{2\times 2}}_{a=1}\;;\;
\underbrace{\hat{\im}_{2\times 2}}_{a=2}\Big\}\;;\;\;\;
\hat{\im}_{2\times 2}=\im\;\hat{1}_{ss\ppr}\;;\;\;\;s,s\ppr=\uparrow,\downarrow \;; \\ \lb{s3_37}
\hat{I}_{4\times 4}\cdot\hat{I}_{4\times 4} &=&\hat{S}_{4\times 4} \;.
\eeq
After integration over bilinear, anti-commuting fields
\(\breve{\Psi}_{\vec{x}\ppr,s\ppr}^{\sharp b}(t_{q}\ppr)\), \(\breve{\Psi}_{\vec{x},s}^{a}(t_{p})\)
as in (\ref{s2_78}-\ref{s2_88}), we obtain the path integral
\(Z[\hat{\mscr{J}},J_{\psi},\hat{J}_{\psi\psi}]\) (\ref{s3_38}) which can be reordered into the form
(\ref{s3_39}) with the actions \(\mscr{A}_{DET}[\hat{T},\hat{\sigma}_{D}^{(0)};\hat{\mscr{J}}]\)
(\ref{s3_40}), \(\mscr{A}_{J_{\psi}}[\hat{T},\hat{\sigma}_{D}^{(0)};\hat{\mscr{J}}]\) (\ref{s3_41})
of the common operator \(\breve{\mscr{O}}_{\vec{x}\ppr,s\ppr;\vec{x},s}^{ba}(t_{q}\ppr,t_{p})\) (\ref{s3_42})
containing the gradient term of nonlocal coset matrices
\beq \lb{s3_38}
\lefteqn{\hspace*{-1.6cm}Z[\hat{\mscr{J}},J_{\psi},\im\,\hat{J}_{\psi\psi}]=
\int\;\;d[\sigma_{D;ss\ppr}^{(0)}(\vec{x},\vec{x}\ppr,t_{p})]\;\;
\exp\bigg\{-\frac{\im}{2\hbar}\int_{C}d t_{p}\sum_{\vec{x},s;\vec{x}\ppr,s\ppr}
\frac{\sigma_{D;ss\ppr}^{(0)}(\vec{x},\vec{x}\ppr,t_{p})\;\;
\sigma_{D;s\ppr s}^{(0)}(\vec{x}\ppr,\vec{x},t_{p})}{V_{|\vec{x}\ppr-\vec{x}|}+
\im\:\ve_{p} }\bigg\} }
\\ \no &\times &
 \int d[\delta\wt{\Sigma}_{\vec{x},s;\vec{x}\ppr,s\ppr}^{ab}(t_{p})]\;\;
 \exp\Bigg\{-\frac{\im}{4\hbar}\int_{C}d t_{p}\sum_{\vec{x},\vec{x}\ppr}
 \TRS\bigg[\frac{\delta\wt{\Sigma}_{\vec{x},s;\vec{x}\ppr,s\ppr}^{ab}(t_{p})\;
 \delta\wt{\Sigma}_{\vec{x}\ppr,s\ppr;\vec{x},s}^{ba}(t_{p})}{V_{|\vec{x}\ppr-\vec{x}|}+
\im\:\ve_{p}\:\big(\delta_{a=b}-\delta_{a\neq b}\big)}
\bigg]\Bigg\}  \\ \no &\times &
\Bigg\{{\raisebox{-5pt}{$\mbox{\large DET}\atop {\scriptstyle \breve{C}}$}}
\bigg[\breve{\mscr{M}}_{\vec{x}\ppr,s\ppr;\vec{x},s}^{ba}(t_{q}\ppr,t_{p})\bigg]
\Bigg\}^{\mathbf{1/2}}     \\ \no &\times&
\exp\bigg\{\frac{\im}{2\hbar}\Omega^{2}\int_{\breve{C}}d t_{p}\;d t\ppr_{q}\sum_{\vec{x},\vec{x}\ppr}
\mcal{N}_{x}\sumss J_{\psi;s\ppr}^{\dagger b}(\vec{x}\ppr,t_{q}\ppr)\;\hat{I}\;
\breve{\mscr{M}}_{\vec{x}\ppr,s\ppr;\vec{x},s}^{\mathbf{-1};ba}(t_{q}\ppr,t_{p})\;\hat{I}\;
J_{\psi;s}^{a}(\vec{x},t_{p})\bigg\}_{;}
\eeq
\beq \lb{s3_39}
Z[\hat{\mscr{J}},J_{\psi},\im\,\hat{J}_{\psi\psi}] &=&
\int\;\;d[\sigma_{D;ss\ppr}^{(0)}(\vec{x},\vec{x}\ppr,t_{p})]\;\;
\exp\bigg\{-\frac{\im}{2\hbar}\int_{C}d t_{p}\sum_{\vec{x},s;\vec{x}\ppr,s\ppr}
\frac{\sigma_{D;ss\ppr}^{(0)}(\vec{x},\vec{x}\ppr,t_{p})\;\;
\sigma_{D;s\ppr s}^{(0)}(\vec{x}\ppr,\vec{x},t_{p})}{V_{|\vec{x}\ppr-\vec{x}|}+
\im\:\ve_{p} }  \bigg\}  \\ \no &\times &
\int d\big[\hat{T}_{\vec{x},s;\vec{x}_{1},s_{1}}^{-1;aa\ppr}(t_{p})\;
d\hat{T}_{\vec{x}_{1},s_{1};\vec{x}\ppr,s\ppr}^{a\ppr b}(t_{p})\big]\;\;
\exp\Big\{\im\;\mscr{A}_{\hat{J}_{\psi\psi}}\big[\hat{T}\big]\Big\}\;\;\times
\\ \no &\times&
\exp\Big\{\mscr{A}_{DET}\big[\hat{T},\hat{\sigma}_{D}^{(0)};\hat{\mscr{J}}\big]\Big\}\;\;
\exp\Big\{\im\;\mscr{A}_{J_{\psi}}\big[\hat{T},\hat{\sigma}_{D}^{(0)};\hat{\mscr{J}}\big]
\Big\} \;;
\eeq
\beq\lb{s3_40}
\mscr{A}_{DET}\big[\hat{T},\hat{\sigma}_{D}^{(0)};\hat{\mscr{J}}\big]
&=&\frac{1}{2}\int_{\breve{C}}\frac{d t_{p}}{\hbar}\eta_{p}\sum_{\vec{x}}\mcal{N}\;
\TRS\Big[\ln\Big(\hat{\mscr{O}}_{\vec{x}, s;\vec{x}\ppr, s\ppr}^{ab}(t_{p},t_{q}\ppr)
\Big)\Big]  \;;
\eeq
\beq \lb{s3_41}
\lefteqn{\mscr{A}_{J_{\psi}}\big[\hat{T},\hat{\sigma}_{D}^{(0)};\hat{\mscr{J}}\big]=
\frac{\Omega^{2}}{2\hbar}\int_{\breve{C}}d t_{p}\;d t_{q}\ppr\sum_{\vec{x},\vec{x}\ppr}\mcal{N}_{x}
\sumss\sum_{a,b=1,2} \times}
\\ \no &\times &
J_{\psi; s\ppr}^{+b}(\vec{x}\ppr,t_{q}\ppr)\;\hat{I}\;
\bigg(\hat{T}_{\vec{x}\ppr,s\ppr;\vec{x}_{2},s_{2}}^{bb\ppr}(t_{q}\ppr)\;\;
\hat{\mscr{O}}_{\vec{x}_{2},s_{2};\vec{x}_{1},s_{1}}^{-1;b\ppr a\ppr}(t_{q}\ppr,t_{p})\;\;
\hat{T}_{\vec{x}_{1},s_{1};\vec{x},s}^{-1;a\ppr a}(t_{p})\bigg)_{\vec{x}\ppr,s\ppr;\vec{x},s}^{ba}\;\hat{I}\;
J_{\psi;s}^{a}(\vec{x},t_{p})\;\;;
\eeq
\beq \lb{s3_42}
\hat{\mscr{O}}_{\vec{x},s;\vec{x}\ppr,s\ppr}^{ab}(t_{p},t_{q}\ppr)&=&\Big(\frac{1}{|\Delta t_{q}\ppr|}\Big)
\bigg[\Big(\hat{\mscr{H}}+\hat{\sigma}_{D}^{(0)}\Big)+
\Big(\hat{T}^{-1}(t_{p})\;\big(\hat{\mscr{H}}+\hat{\sigma}_{D}^{(0)}\big)\;
\hat{T}(t_{q}\ppr)-\big(\hat{\mscr{H}}+\hat{\sigma}_{D}^{(0)}\big)\Big)_{\vec{x},s;\vec{x}\ppr,s\ppr}^{ab} +
\\  \no     &+& \underbrace{
\hat{T}_{\vec{x},s;\vec{x}_{1},s_{1}}^{-1;aa\ppr}(t_{p})\;\hat{I}\;\hat{S}\;\eta_{p}\;|\Delta t_{q}\ppr|\;
\frac{\hat{\mscr{J}}_{\vec{x}_{1},s_{1};\vec{x}_{2},s_{2}}^{a\ppr b\ppr}(t_{p};
t_{q}\ppr)}{\mcal{N}_{x}}\;
\eta_{q}\;\hat{S}\;\hat{I}\;
\hat{T}_{\vec{x}_{2},s_{2};\vec{x}\ppr,s\ppr}^{b\ppr b}(t_{q}\ppr)}_{\wt{\mscr{J}}(\hat{T}^{-1},\hat{T})}
\bigg]_{\vec{x},s;\vec{x}\ppr,s\ppr}^{ab}\hspace*{-0.64cm}(t_{p},t_{q}\ppr)\;\;\;\;\;\;.
\eeq
In analogy to relations in section \ref{s23}, the pair condensate 'seed' action term
\(\exp\{\im\;\mscr{A}_{\hat{J}_{\psi\psi}}[\hat{T}]\,\}\) is given by Eqs. (\ref{s3_43}) to (\ref{s3_47})
and can be computed by application of Vandermonde determinants for the integration measure
of the eigenvalues \(\delta\lambda_{s}(\vec{x},t_{p})\) with the orthogonal properties of
Hermite polynomials and corresponding Gaussian weights
\beq \lb{s3_43}
\lefteqn{\exp\Big\{\im\;\mscr{A}_{\hat{J}_{\psi\psi}}\big[\hat{T}\big]\Big\}  =
\int d\big[\delta\hat{\Sigma}_{D;\vec{x},s;\vec{x}\ppr,s\ppr}(t_{p})\big]\;\;
\mscr{P}\big(\delta\hat{\lambda}(\vec{x},t_{p})\big)\;\;
\exp\Big\{\im\;\mscr{A}_{2}\big[\hat{T},\delta\hat{\Sigma}_{D};\im\hat{J}_{\psi\psi}\big]\Big\} = }
\\ \no &=&
\int d\big[d\hat{Q}_{\vec{x},s;\vec{x}_{1},s_{1}}(t_{p})\;\hat{Q}_{\vec{x}_{1},s_{1};\vec{x}\ppr,s\ppr}^{-1}(t_{p});
\delta\hat{\lambda}(\vec{x},t_{p})\big]\;\;
\mscr{P}\big(\delta\hat{\lambda}(\vec{x},t_{p})\big)\;\;
\exp\Big\{\im\;\mscr{A}_{2}\big[\hat{T},\hat{Q}^{-1}\;\delta\hat{\Lambda}\;\hat{Q};
\im\hat{J}_{\psi\psi}\big]\Big\}  \;\;;
\eeq
\beq \lb{s3_44}
\mbox{det}\Big\{\delta\hat{\Sigma}_{D;\vec{x},s;\vec{x}\ppr,s\ppr}^{11}-
\delta\lambda\;\delta_{\vec{x},\vec{x}\ppr}\;\delta_{ s s\ppr}\Big\}
&=&0\;;\hspace*{1.5cm}
\mbox{det}\Big\{\delta\hat{\Sigma}_{D;\vec{x},s;\vec{x}\ppr,s\ppr}^{22}-
\big(-\delta\lambda\big)\;\delta_{\vec{x},\vec{x}\ppr}\;\delta_{ s s\ppr}\Big\}=0 \;\;; \\ \lb{s3_45}
\delta\hat{\Sigma}_{D;\vec{x},s;\vec{x}\ppr,s\ppr}^{22}(\vec{x},t_{p})&=&-
\delta\hat{\Sigma}_{D;\vec{x},s;\vec{x}\ppr,s\ppr}^{11,T}(\vec{x},t_{p})\;;
\eeq
\beq \lb{s3_46}
\lefteqn{\im\,\mscr{A}_{2}\big[\hat{T},\delta\hat{\Sigma}_{D};\im\hat{J}_{\psi\psi}\big] = }  \\ \no &=&
-\frac{\im}{4\hbar}\int_{C}d t_{p}\sum_{\vec{x},\vec{x}\ppr}
 \TRS\bigg[\frac{\big(\delta\wt{\Sigma}_{\vec{x},s;\vec{x}\ppr,s\ppr}^{ab}(t_{p})+
\im\,\hat{J}_{\psi\psi;\vec{x},s;\vec{x}\ppr,s\ppr}^{a\neq b}(t_{p})\,\big)\;
\big(\delta\wt{\Sigma}_{\vec{x}\ppr,s\ppr;\vec{x},s}^{ba}(t_{p})+
\im\,\hat{J}_{\psi\psi;\vec{x}\ppr,s\ppr;\vec{x},s}^{b\neq a}(t_{p})\,\big)}{V_{|\vec{x}\ppr-\vec{x}|}+
\im\:\ve_{p}\:\big(\delta_{a=b}-\delta_{a\neq b}\big)}
\bigg]
\\ \no &=&
-\frac{\im}{4\hbar}\int_{C}d t_{p}\sum_{\vec{x},\vec{x}\ppr}  \times \\ \no &\times&\hspace*{-0.2cm}
\Bigg\{\hspace*{-0.1cm} \TRS\bigg[\frac{
\hat{T}_{\vec{x},s;\vec{x}_{3},s_{3}}^{ab_{3}} (t_{p})\;
\delta\hat{\Sigma}_{D;\vec{x}_{3},s_{3};\vec{x}_{4},s_{4}}^{b_{3}=b_{4}}(t_{p})\;
\hat{T}_{\vec{x}_{4},s_{4};\vec{x}\ppr,s\ppr}^{-1;b_{4}b}(t_{p})\;
\hat{T}_{\vec{x}\ppr,s\ppr;\vec{x}_{1},s_{1}}^{ba_{1}}(t_{p})\;
\delta\hat{\Sigma}_{D;\vec{x}_{1},s_{1};\vec{x}_{2},s_{2}}^{a_{1}=a_{2}}(t_{p})\;
\hat{T}_{\vec{x}_{2},s_{2};\vec{x},s}^{-1;a_{2}a}(t_{p})}{V_{|\vec{x}\ppr-\vec{x}|}+
\im\:\ve_{p}\:\big(\delta_{a=b}-\delta_{a\neq b}\big)}
\bigg]  +       \\ \no &+& 2\,\im
 \TRS\bigg[\frac{\hat{J}_{\psi\psi;\vec{x},s;\vec{x}\ppr,s\ppr}^{a\neq b}(t_{p})\;
\hat{T}_{\vec{x}\ppr,s\ppr;\vec{x}_{1},s_{1}}^{ba_{1}}(t_{p})\;
\delta\hat{\Sigma}_{D;\vec{x}_{1},s_{1};\vec{x}_{2},s_{2}}^{a_{1}=a_{2}}(t_{p})\;
\hat{T}_{\vec{x}_{2},s_{2};\vec{x},s}^{-1;a_{2}a}(t_{p})}{V_{|\vec{x}\ppr-\vec{x}|}-
\im\:\ve_{p}\:\delta_{a\neq b}}
\bigg]   +      \\ \no &-&
 \TRS\bigg[\frac{\hat{J}_{\psi\psi;\vec{x},s;\vec{x}\ppr,s\ppr}^{a\neq b}(t_{p})\;
\hat{J}_{\psi\psi;\vec{x}\ppr,s\ppr;\vec{x},s}^{b\neq a}(t_{p})}{V_{|\vec{x}\ppr-\vec{x}|}-
\im\:\ve_{p}\:\delta_{a\neq b}}
\bigg]    \Bigg\}\;;
\eeq
\beq \lb{s3_47}
\lefteqn{\im\,\mscr{A}_{2}\big[\hat{T},\hat{Q}^{-1}\;\delta\hat{\Lambda}\;\hat{Q};
\im\hat{J}_{\psi\psi}\big]  =
-\frac{\im}{4\hbar}\int_{C}d t_{p}\sum_{\vec{x},\vec{x}\ppr}  \times } \\ \no &\times&\hspace*{-0.2cm}
\Bigg\{\hspace*{-0.1cm} \TRS\bigg[\frac{
\hat{T}_{0;\vec{x},s;\vec{x}_{3},s_{3}}^{ab_{3}} (t_{p})\;
\delta\hat{\Lambda}_{s_{3}}^{b_{3}}(\vec{x}_{3},t_{p})\;
\hat{T}_{0;\vec{x}_{3},s_{3};\vec{x}\ppr,s\ppr}^{-1;b_{3}b}(t_{p})\;
\hat{T}_{0;\vec{x}\ppr,s\ppr;\vec{x}_{1},s_{1}}^{ba_{1}}(t_{p})\;
\delta\hat{\Lambda}_{s_{1}}^{a_{1}}(\vec{x}_{1},t_{p})\;
\hat{T}_{0;\vec{x}_{1},s_{1};\vec{x},s}^{-1;a_{1}a}(t_{p})}{V_{|\vec{x}\ppr-\vec{x}|}+
\im\:\ve_{p}\:\big(\delta_{a=b}-\delta_{a\neq b}\big)}
\bigg]  +       \\ \no &+& 2\,\im
 \TRS\bigg[\frac{\hat{J}_{\psi\psi;\vec{x},s;\vec{x}\ppr,s\ppr}^{a\neq b}(t_{p})\;
\hat{T}_{0;\vec{x}\ppr,s\ppr;\vec{x}_{1},s_{1}}^{ba_{1}}(t_{p})\;
\delta\hat{\Lambda}_{s_{1}}^{a_{1}}(\vec{x}_{1},t_{p})\;
\hat{T}_{0;\vec{x}_{1},s_{1};\vec{x},s}^{-1;a_{1}a}(t_{p})\big)}{V_{|\vec{x}\ppr-\vec{x}|}-
\im\:\ve_{p}\:\delta_{a\neq b}}
\bigg]   +      \\ \no &-&
 \TRS\bigg[\frac{\hat{J}_{\psi\psi;\vec{x},s;\vec{x}\ppr,s\ppr}^{a\neq b}(t_{p})\;
\hat{J}_{\psi\psi;\vec{x}\ppr,s\ppr;\vec{x},s}^{b\neq a}(t_{p})}{V_{|\vec{x}\ppr-\vec{x}|}-
\im\:\ve_{p}\:\delta_{a\neq b}}
\bigg]    \Bigg\}\;;    \\   \lb{s3_48} &&
\hat{T}_{0;\vec{x},s;\vec{x}\ppr,s\ppr}^{ab} (t_{p})=
\hat{T}_{\vec{x},s;\vec{x}_{1},s_{1}}^{ab\ppr}(t_{p})\;
\hat{Q}_{\vec{x}_{1},s_{1};\vec{x}\ppr,s\ppr}^{-1;b\ppr=b}(t_{p})\;\;.
\eeq
The analogous gradient expansion can be performed as described in section \ref{s24} for the short-ranged
interaction case \(V_{0}\;\delta_{\vec{x},\vec{x}\ppr}\). The involved saddle point approximation
for the self-energy density field \(\langle\sigma_{D;ss\ppr}^{(0)}(\vec{x},\vec{x}\ppr,t_{p})\rangle\) is similar
and can be taken from the pure 'density path integral'
\(Z[\hat{\sigma}_{D}^{(0)};j_{\psi}]\) (\ref{s2_113}), but with nonlocal interaction \(V_{|\vec{x}\ppr-\vec{x}|}\)
in the Gaussian part; the expansion of the logarithm and the
inverse of an operator \(\breve{\mfrak{O}}\) as in (\ref{s2_118}-\ref{s2_124}) straightforwardly generalizes for the case
with the actions \(\mscr{A}_{DET}[\hat{T},\hat{\sigma}_{D}^{(0)};\hat{\mscr{J}}]\)
(\ref{s3_40}), \(\mscr{A}_{J_{\psi}}[\hat{T},\hat{\sigma}_{D}^{(0)};\hat{\mscr{J}}]\) (\ref{s3_41})
with common operator \(\breve{\mscr{O}}_{\vec{x}\ppr,s\ppr;\vec{x},s}^{ba}(t_{q}\ppr,t_{p})\) (\ref{s3_42})
consisting of the nonlocal coset matrices \((\hat{T}(t_{p})\,)_{\vec{x},s;\vec{x}\ppr,s\ppr}^{ab}=
(\,\exp\{-\hat{Y}_{\vec{x}_{1},s_{1};\vec{x}_{2},s_{2}}^{a_{1}\neq b_{1}}(t_{p})\,\}\,)_{
\vec{x},s;\vec{x}\ppr,s\ppr}^{ab}\) (\ref{s3_26}-\ref{s3_29}).

\section{Summary and conclusion}  \lb{s4}

\subsection{Transformation from the spatially Euclidean fields in the path integration
to spherical variables} \lb{s41}

As one transforms to other coordinates, one has to define invariant principles which should
hold in all coordinate systems. In general we have to distinguish between Euclidean spaces,
which even remain 'flat' under transformation to curvilinear coordinates, and spaces
with nontrivial curvature which keep this property under arbitrary transformations to other
coordinates. The latter spaces with curvature cause the known problems of operator ordering
in the quantization procedure so that it is a priori not obvious how to transform the path
integration fields in a curved manifold of spacetime. This turns out in particular because
the path integral follows from the discrete time step development of the quantized Hamilton
operator which already has the problem of the mentioned ambiguous transitions from the
classical to a quantum system. Since we only use Euclidean, spatial base manifolds,
we extend the invariant length \((ds_{\mbox{\scz SO}(4)})^{2}\)
(\ref{sa_1},\ref{sa_4},\ref{sa_6}) of internal degrees
of freedom to the invariant length \((dS)^{2}\) of coordinates which introduces the
additional summation over the space '\(\sum_{\vec{x}}\)' with \(|\vec{x}|<L\).
Therefore, we have to include two independent, spatial summations in the case
of the nonlocal self-energy
\beq \lb{s4_1}
\Big(dS\big(\psi_{\vec{x},s}^{*}(t_{p})\,,\,\psi_{\vec{x},s}(t_{p})\,\big)\Big)^{2} &=&
\sum_{\vec{x}}^{|\vec{x}|<L}\sum_{s=\uparrow,\downarrow}
d\psi_{\vec{x},s}^{*}(t_{p})\;\;d\psi_{\vec{x},s}(t_{p}) \;; \\  \lb{s4_2}
\Big(dS\big(\delta\wt{\Sigma}_{ss\ppr}^{ab}(\vec{x},t_{p})\,\big)\Big)^{2} &=&
\sum_{\vec{x}}^{|\vec{x}|<L} \TRS\Big[d \big(\delta \wt{\Sigma}_{ss\ppr}^{ab}(\vec{x},t_{p})\big)\;
d \big(\delta \wt{\Sigma}_{s\ppr s}^{ba}(\vec{x},t_{p})\big) \Big] \;;  \\ \lb{s4_3}
\Big(dS\big(\delta\wt{\Sigma}_{\vec{x},s;\vec{x}\ppr,s\ppr}^{ab}(t_{p})\,\big)\Big)^{2} &=&
\sum_{\vec{x},\vec{x}\ppr}^{(|\vec{x}|,|\vec{x}\ppr|)<L}
\TRS\Big[d \big(\delta \wt{\Sigma}_{\vec{x},s;\vec{x}\ppr,s\ppr}^{ab}(t_{p})\big)\;
d \big(\delta \wt{\Sigma}_{\vec{x}\ppr,s\ppr;\vec{x},s}^{ba}(t_{p})\big)\Big]  \;.
\eeq
Since spaces with curvature also involve the problem of determining a unique operator ordering,
we suggest to incorporate a confining potential embedded in a higher dimensional, Euclidean
base space where the quantization procedure is obvious. The confinement potential, inserted into the
classical Lagrangian and action, should then cause the restriction to the physical sub-space
with the required curvature even after a quantization. The wavefunctions of the quantized system
may extend above the entire, higher dimensional, Euclidean space, however, the confining potential
has to be chosen in such a manner that the wavefunctions are squeezed to the lower dimensional,
physical sub-space with the nontrivial curvature. This seems to be very feasible for a constant,
static curvature by static confinement potentials; but, one has to find the appropriate
dynamics of the confining potential in the nonstatic case, e.\ g.\ if one requires the
confinement potential to restrict the higher dimensional, Euclidean system to the dynamics
of the Einstein field equations in lower sub-dimensions in a classical limit. We can simply introduce
a confinement potential for the \(\mbox{S}^{1}\)-(\(\mbox{S}^{2}\)-)sphere embedded in a two-(three-)
dimensional Euclidean space in order to obtain quantization into unique coherent state path
integrals with the appropriate definition of the invariant lengths given in Eqs. (\ref{s4_1}-\ref{s4_3}).
In this case the definition (\ref{s4_1}-\ref{s4_3}) is not affected for deriving the integration measure
because the nontrivial curvature of the \(\mbox{S}^{1}\)-(\(\mbox{S}^{2}\)-) sphere only originates
from the confinement potential which restricts the two-(three-) dimensional
Euclidean space to the physically relevant, lower dimensional spheres.

Since we have only considered Euclidean coordinate systems in this paper,
one can easily transform to curvilinear coordinates from the invariant lengths (\ref{s4_1}-\ref{s4_3}).
In order to preserve identical scaling and renormalization procedures, one has to require that the
identical number of independent integration variables at the \(\mcal{N}_{x}\) (\ref{s1_23}) spatial
grid points of spheres with radial length $L$ stays invariant under transformation to the
\(D=2\) or \(D=3\) spherical coordinates. We distinguish the discrete coordinates of vectors
\(\vec{x}_{ij}\) and \(\vec{x}_{ijk}\) by the indices '\(i,j\)' and '\(i,j,k\)', respectively
\beq \lb{s4_4}
\mcal{N}_{x}&=& \frac{\Omega_{D}}{D}\:\bigg(\frac{L}{\Delta x}\bigg)^{D}\;;\;\;\;
\Omega_{D=2}=2\pi\;;\;\;\;\Omega_{D=3}=4\pi\;; \\  \lb{s4_5}
\mbox{D=2 case} &:& \vec{x}_{ij}=\big\{x_{i}^{(1)}\,,\,x_{j}^{(2)}\big\}=
\big\{i\cdot \Delta x\,,\,j\cdot\Delta x\big\} \;; \\   \lb{s4_6}  &&
r_{ij}=\sqrt{(x_{i}^{(1)})^{2}+(x_{j}^{(2)})^{2}} <L\;;\;\;\;
\tan\varphi_{ij}=\frac{x_{j}^{(2)}}{x_{i}^{(1)}}\;; \\   \lb{s4_7}
\mbox{D=3 case} &:& \vec{x}_{ijk}=\big\{x_{i}^{(1)}\,,\,x_{j}^{(2)}\,,\,\,x_{k}^{(3)}\big\}=
\big\{i\cdot \Delta x\,,\,j\cdot\Delta x\,,\,k\cdot\Delta x\big\} \;; \\  \lb{s4_8}   &&
r_{ijk}=\sqrt{(x_{i}^{(1)})^{2}+(x_{j}^{(2)})^{2}+(x_{k}^{(3)})^{2}} <L\;;\;\;\;
\tan\varphi_{ijk}=\frac{x_{j}^{(2)}}{x_{i}^{(1)}}\;;\;\;\;
\cos\vartheta_{ijk}=\frac{x_{k}^{(3)}}{r_{ijk}}\;;  \\  \lb{s4_9}
\mcal{N}_{x}&=&\mcal{N}_{r}\cdot\mcal{N}_{\Omega_{D}}\;;\;\;\;
\mcal{N}_{r}=\frac{L}{\Delta x}\;;\;\;\;\mcal{N}_{\Omega_{D}}=\frac{\Omega_{D}}{D}\;
\bigg(\frac{L}{\Delta x}\bigg)^{D-1}\;.
\eeq
If one chooses in the radial directions the same discrete interval as in the \(D=2\) and \(D=3\)
Cartesian systems, one remains with \(\mcal{N}_{\Omega_{D}}\) grid points for the spherical degrees
of freedom, although other decompositions of the total number \(\mcal{N}_{x}\) of grid points
may also be meaningful. Therefore, we introduce following transformations to
spherical coordinates in \(D=2\) (\ref{s4_10}-\ref{s4_13})
and \(D=3\) (\ref{s4_14}-\ref{s4_17}) for anti-commuting fields and for local, nonlocal
self-energies, respectively, where the number of independent spherical integration variables is adapted to
\(\mcal{N}_{\Omega_{D}}\) with \(\mcal{N}_{r}=L/\Delta x\) radial points of identical,
discrete intervals \(\Delta x\) as in the Cartesian systems
(\(\theta(x):=\) Heaviside step function with \(\theta(x)=1\) for \(x\geq 0\) and \(\theta(x)=0\)
for \(x<0\))
\beq \lb{s4_10}
d\psi_{\vec{x}_{ij},s}(t_{p}) &\stackrel{D=2}{=}&\sum_{n_{r}=0}^{\mcal{N}_{r}}
\sum_{m=-m_{0}}^{+m_{0}}d\psi_{n_{r},m,s}(t_{p}) \;\;
\frac{\exp\big\{\im\:\varphi_{ij}\cdot m\big\}}{\sqrt{2\pi}}\;\times  \\   \no  &\times&
\Big[\theta\big({\ts\sqrt{i^{2}+j^{2}}-(n_{r}-{\ts\frac{1}{2}})}\,\big)-
\theta\big({\ts\sqrt{i^{2}+j^{2}}-(n_{r}+{\ts\frac{1}{2}})}\,\big)\Big] \;;  \\  \lb{s4_11}
d\big(\delta\wt{\Sigma}_{ss\ppr}^{ab}(\vec{x}_{ij},t_{p})\,\big)
&\stackrel{D=2}{=}&\sum_{n_{r}=0}^{\mcal{N}_{r}}
\sum_{m=-m_{0}}^{+m_{0}} d\big(\delta\wt{\Sigma}_{ss\ppr}^{ab}(n_{r},m,t_{p})\,\big)\;
\frac{\exp\big\{\im\:\varphi_{ij}\cdot m\big\}}{\sqrt{2\pi}}\;\times  \\  \no  &\times&
\Big[\theta\big({\ts\sqrt{i^{2}+j^{2}}-(n_{r}-{\ts\frac{1}{2}})}\,\big)-
\theta\big({\ts\sqrt{i^{2}+j^{2}}-(n_{r}+{\ts\frac{1}{2}})}\,\big)\Big]  \;;   \\   \lb{s4_12}
d\big(\delta\wt{\Sigma}_{\vec{x}_{ij},s;\vec{x}_{i\ppr j\ppr}\ppr,s\ppr}^{ab}(t_{p})\,\big)
&\stackrel{D=2}{=}& \hspace*{-0.6cm} \sum_{n_{r},n_{r}\ppr=0}^{\mcal{N}_{r}}\;
\sum_{m,m\ppr=-m_{0}}^{+m_{0}} \hspace*{-0.55cm}
d \big(\delta\wt{\Sigma}_{n_{r},m,s;n_{r}\ppr,m\ppr,s\ppr}^{ab}(t_{p})\,\big) \,
\frac{\exp\big\{\im\:\varphi_{ij}\cdot m\big\}}{\sqrt{2\pi}}\,
\frac{\exp\big\{-\im\:\varphi_{i\ppr j\ppr}\cdot m\ppr\big\}}{\sqrt{2\pi}} \\ \no &\times&
\Big[\theta\big({\ts\sqrt{i^{2}+j^{2}}-(n_{r}-{\ts\frac{1}{2}})}\,\big)-
\theta\big({\ts\sqrt{i^{2}+j^{2}}-(n_{r}+{\ts\frac{1}{2}})}\,\big)\Big] \;\times \\ \no &\times&
\Big[\theta\big({\ts\sqrt{i^{\prime 2}+j^{\prime 2}}-(n_{r}\ppr-{\ts\frac{1}{2}})}\,\big)-
\theta\big({\ts\sqrt{i^{\prime 2}+j^{\prime 2}}-(n_{r}\ppr+{\ts\frac{1}{2}})}\,\big)\Big] \;;
  \\ \lb{s4_13}   &&
\mcal{N}_{r}=\frac{L}{\Delta x}\;;\;\;\;\mcal{N}_{\Omega_{D=2}}=
\frac{\Omega_{D}}{D}\;\bigg(\frac{L}{\Delta x}\bigg)^{D-1}=\pi\;\frac{L}{\Delta x}:=2\:m_{0}+1 \;; \\   \lb{s4_14}
d\psi_{\vec{x}_{ijk},s}(t_{p}) &\stackrel{D=3}{=}&\sum_{n_{r}=0}^{\mcal{N}_{r}}
\sum_{l=0}^{l_{0}}\sum_{m=-l}^{+l}d\psi_{n_{r},l,m,s}(t_{p}) \;\;
Y_{lm}(\vartheta_{ijk},\varphi_{ijk}) \;\times  \\  \no &\times&
\Big[\theta\big({\ts\sqrt{i^{2}+j^{2}+k^{2}}-(n_{r}-{\ts\frac{1}{2}})}\,\big)-
\theta\big({\ts\sqrt{i^{2}+j^{2}+k^{2}}-(n_{r}+{\ts\frac{1}{2}})}\,\big)\Big] \;;  \\  \lb{s4_15}
d\big(\delta\wt{\Sigma}_{ss\ppr}^{ab}(\vec{x}_{ijk},t_{p})\,\big)
&\stackrel{D=3}{=}&\sum_{n_{r}=0}^{\mcal{N}_{r}}
\sum_{l=0}^{l_{0}}\sum_{m=-l}^{+l}  d\big(\delta\wt{\Sigma}_{ss\ppr}^{ab}(n_{r},l,m,t_{p})\,\big)\;
Y_{lm}(\vartheta_{ijk},\varphi_{ijk}) \;\times  \\  \no &\times&
\Big[\theta\big({\ts\sqrt{i^{2}+j^{2}+k^{2}}-(n_{r}-{\ts\frac{1}{2}})}\,\big)-
\theta\big({\ts\sqrt{i^{2}+j^{2}+k^{2}}-(n_{r}+{\ts\frac{1}{2}})}\,\big)\Big]  \;;   \\   \lb{s4_16}
d\big(\delta\wt{\Sigma}_{\vec{x}_{ijk},s;\vec{x}_{i\ppr j\ppr k\ppr}\ppr,s\ppr}^{ab}(t_{p})\,\big)
&\stackrel{D=3}{=}&\sum_{n_{r},n_{r}\ppr=0}^{\mcal{N}_{r}}
\sum_{l,l\ppr=0}^{l_{0}}\sum_{m=-l}^{+l} \sum_{m\ppr=-l\ppr}^{+l\ppr}
d\big(\delta\wt{\Sigma}_{n_{r},l,m,s;n_{r}\ppr,l\ppr,m\ppr,s\ppr}^{ab}(t_{p})\,\big) \\  \no &\times&
Y_{lm}(\vartheta_{ijk},\varphi_{ijk})\;
Y_{l\ppr m\ppr}^{*}(\vartheta_{i\ppr j\ppr k\ppr},\varphi_{i\ppr j\ppr k\ppr})  \; \\ \no &\times&
\Big[\theta\big({\ts\sqrt{i^{2}+j^{2}+k^{2}}-(n_{r}-{\ts\frac{1}{2}})}\,\big)-
\theta\big({\ts\sqrt{i^{2}+j^{2}+k^{2}}-(n_{r}+{\ts\frac{1}{2}})}\,\big)\Big] \;\times \\ \no &\times&
\Big[\theta\big({\ts\sqrt{i^{\prime 2}+j^{\prime 2}+k^{\prime 2}}-(n_{r}\ppr-{\ts\frac{1}{2}})}\,\big)-
\theta\big({\ts\sqrt{i^{\prime 2}+j^{\prime 2}+k^{\prime 2}}-(n_{r}\ppr+{\ts\frac{1}{2}})}\,\big)\Big]
\;;   \\    \lb{s4_17}   && \mcal{N}_{r}=\frac{L}{\Delta x}\;;\;\;\;\mcal{N}_{\Omega_{D=3}}=
\frac{\Omega_{D}}{D}\;\bigg(\frac{L}{\Delta x}\bigg)^{D-1}=
\frac{4\pi}{3}\,\bigg(\frac{L}{\Delta x}\bigg)^{2}:=\big(l_{0}+1)^{2}   .
\eeq
After we have substituted the transformed variables into the defining invariant lengths
(\ref{s4_1}-\ref{s4_3}) and have reordered summations, in particular above \(i,j\) in \(D=2\)
and \(i,j,k\) in \(D=3\), one eventually achieves following metric tensors (\ref{s4_21},\ref{s4_25})
in the transformed, spherical integration variables
\beq \lb{s4_18}
\Big(dS\big(\psi_{\vec{x},s}^{*}(t_{p})\,,\,\psi_{\vec{x},s}(t_{p})\,\big)\Big)^{2}
&\stackrel{D=2}{=}&
\sum_{n_{r},n_{r}\ppr=0}^{\mcal{N}_{r}} \sum_{m,m\ppr=-m_{0}}^{+m_{0}}
\sum_{s=\uparrow,\downarrow}d\psi_{n_{r}\ppr,m\ppr,s}^{*}(t_{p})\;\;
\hat{g}_{n_{r}\ppr,m\ppr;n_{r},m}^{(D=2)}\;\;d\psi_{n_{r},m,s}(t_{p}) \;;  \\  \lb{s4_19}
\Big(dS\big(\delta\wt{\Sigma}_{ss\ppr}^{ab}(\vec{x},t_{p})\,\big)\Big)^{2} &\stackrel{D=2}{=}&\hspace*{-0.6cm}
\sum_{n_{r},n_{r}\ppr=0}^{\mcal{N}_{r}} \sum_{m,m\ppr=-m_{0}}^{+m_{0}}\hspace*{-0.3cm}
 \TRS\hspace*{-0.1cm}\Big[d \big(\delta \wt{\Sigma}_{ss\ppr}^{ab}(n_{r}\ppr,m\ppr,t_{p})\big)\,
d \big(\delta \wt{\Sigma}_{s\ppr s}^{ba}(n_{r},m,t_{p})\big)\Big]\,
\hat{g}_{n_{r}\ppr,m\ppr;n_{r},m}^{(D=2)};   \\    \lb{s4_20}
\Big(dS\big(\delta\wt{\Sigma}_{\vec{x},s;\vec{x}\ppr,s\ppr}^{ab}(t_{p})\,\big)\Big)^{2}
&\stackrel{D=2}{=}&
\sum_{n_{r}^{(1)}\!,n_{r}^{(2)}\!,n_{r}^{(3)}\!,n_{r}^{(4)}\!=0}^{\mcal{N}_{r}}\;
\sum_{m_{1},m_{2},m_{3},m_{4}=-m_{0}}^{+m_{0}}
\hat{g}_{n_{r}^{(2)}\!,m_{2};n_{r}^{(3)}\!,m_{3}}^{(D=2)}\;
\hat{g}_{n_{r}^{(4)}\!,m_{4};n_{r}^{(1)}\!,m_{1}}^{(D=2)}
\\ \no  &\times&
 \TRS\Big[d \big(\delta \wt{\Sigma}_{n_{r}^{(1)}\!,m_{1},s;n_{r}^{(2)}\!,m_{2},s\ppr}^{ab}(t_{p})\,\big)\;\;
d \big(\delta \wt{\Sigma}_{n_{r}^{(3)}\!,m_{3},s\ppr;n_{r}^{(4)}\!,m_{4},s}^{ba}(t_{p})\,\big)\;\Big] \;;  \\  \lb{s4_21}
\hat{g}_{n_{r}\ppr,m\ppr;n_{r},m}^{(D=2)} &=& \sum_{i,j}^{|\vec{x}_{ij}|<L}
\frac{\exp\big\{\im\;\varphi_{ij}\:(m-m\ppr)\big\}}{2\pi}\;\times  \\ \no &\times&
\Big[\theta\big({\ts\sqrt{i^{2}+j^{2}}-(n_{r}-{\ts\frac{1}{2}})}\,\big)-
\theta\big({\ts\sqrt{i^{2}+j^{2}}-(n_{r}+{\ts\frac{1}{2}})}\,\big)\Big] \;\times \\ \no &\times&
\Big[\theta\big({\ts\sqrt{i^{\prime 2}+j^{\prime 2}}-(n_{r}\ppr-{\ts\frac{1}{2}})}\,\big)-
\theta\big({\ts\sqrt{i^{\prime 2}+j^{\prime 2}}-(n_{r}\ppr+{\ts\frac{1}{2}})}\,\big)\Big] \;;  \\  \lb{s4_22}
\Big(dS\big(\psi_{\vec{x},s}^{*}(t_{p})\,,\,\psi_{\vec{x},s}(t_{p})\,\big)\Big)^{2}
&\stackrel{D=3}{=}&
\sum_{n_{r},n_{r}\ppr=0}^{\mcal{N}_{r}}
\sum_{l,l\ppr=0}^{l_{0}}\sum_{m=-l}^{+l}\sum_{m\ppr=-l\ppr}^{+l\ppr}
\sum_{s=\uparrow,\downarrow} \;\times  \\   \no &\times& d\psi_{n_{r}\ppr,l\ppr,m\ppr,s}^{*}(t_{p})\;\;
\hat{g}_{n_{r}\ppr,l\ppr,m\ppr;n_{r},l,m}^{(D=3)}\;\;d\psi_{n_{r},l,m,s}(t_{p}) \;;  \\  \lb{s4_23}
\Big(dS\big(\delta\wt{\Sigma}_{ss\ppr}^{ab}(\vec{x},t_{p})\,\big)\Big)^{2} &\stackrel{D=3}{=}&
\sum_{n_{r},n_{r}\ppr=0}^{\mcal{N}_{r}}
\sum_{l,l\ppr=0}^{l_{0}}\sum_{m=-l}^{+l}\sum_{m\ppr=-l\ppr}^{+l\ppr}\;\times \\  \no &\times&
 \TRS\Big[d \big(\delta \wt{\Sigma}_{ss\ppr}^{ab}(n_{r}\ppr,l\ppr,m\ppr,t_{p})\,\big)\;\;
d \big(\delta \wt{\Sigma}_{s\ppr s}^{ba}(n_{r},l,m,t_{p})\,\big)\Big]\;\;
\hat{g}_{n_{r}\ppr,l\ppr,m\ppr;n_{r},l,m}^{(D=3)}\;;   \\    \lb{s4_24}
\Big(dS\big(\delta\wt{\Sigma}_{\vec{x},s;\vec{x}\ppr,s\ppr}^{ab}(t_{p})\,\big)\Big)^{2}
&\stackrel{D=3}{=}&
\sum_{n_{r}^{(1)}\!,n_{r}^{(2)}\!,n_{r}^{(3)}\!,n_{r}^{(4)}\!=0}^{\mcal{N}_{r}} \sum_{l_{1},l_{2},l_{3},l_{4}=0}^{l_{0}}
\sum_{m_{1}=-l_{1}}^{+l_{1}}\sum_{m_{2}=-l_{2}}^{+l_{2}}\sum_{m_{3}=-l_{3}}^{+l_{3}}
\sum_{m_{4}=-l_{4}}^{+l_{4}}\times \\ \no  &\times&
 \TRS\Big[d \big(\delta \wt{\Sigma}_{n_{r}^{(1)}\!,l_{1},m_{1},s;n_{r}^{(2)}\!,l_{2},m_{2},s\ppr}^{ab}(t_{p})\,\big)\;\;
d \big(\delta \wt{\Sigma}_{n_{r}^{(3)}\!,l_{3},m_{3},s\ppr;n_{r}^{(4)}\!,l_{4},m_{4},s}^{ba}(t_{p})\,\big)\;\Big]
\;\times \\   \no &\times&
\hat{g}_{n_{r}^{(2)}\!,l_{2},m_{2};n_{r}^{(3)}\!,l_{3},m_{3}}^{(D=2)}\;\;
\hat{g}_{n_{r}^{(4)}\!,l_{4},m_{4};n_{r}^{(1)}\!,l_{1},m_{1}}^{(D=2)}\;;      \\    \lb{s4_25}
\hat{g}_{n_{r}\ppr,l\ppr,m\ppr;n_{r},l,m}^{(D=3)} &=& \sum_{i,j,k}^{|\vec{x}_{ijk}|<L}
Y_{lm}(\vartheta_{ijk},\varphi_{ijk})\;Y_{l\ppr m\ppr}^{*}(\vartheta_{ijk},\varphi_{ijk})\;\times  \\ \no &\times&
\Big[\theta\big({\ts\sqrt{i^{2}+j^{2}+k^{2}}-(n_{r}-{\ts\frac{1}{2}})}\,\big)-
\theta\big({\ts\sqrt{i^{2}+j^{2}+k^{2}}-(n_{r}+{\ts\frac{1}{2}})}\,\big)\Big] \;\times \\ \no &\times&
\Big[\theta\big({\ts\sqrt{i^{2}+j^{2}+k^{2}}-(n_{r}\ppr-{\ts\frac{1}{2}})}\,\big)-
\theta\big({\ts\sqrt{i^{2}+j^{2}+k^{2}}-(n_{r}\ppr+{\ts\frac{1}{2}})}\,\big)\Big] \;.
\eeq
The Jacobian for the change to spherical coordinates is given by the (inverse) square roots of the
determinant of the metric tensor. The nonlocal self-energies are weighted by the square of the integration measure compared to the local self-energies because of the square of independent
spatial integration variables.

\subsection{Generic use of the described coherent state path integral
with precise, discrete time steps} \lb{s42}

In this paper we have mainly described technical details of coherent state path integrals;
however, it has to be pointed out that the necessary limits and time step correction in
the resulting complex fields are usually omitted for brevity. Some references \cite{neg,klein}
even state insurmountable problems with coherent state path integrals, due to ambiguous
choices of discrete time steps; however, as we have already mentioned in sections \ref{s1}
and \ref{s2}, \ref{s3} with nontrivial SSB, one should not be guided by the appealing
property for hermitian actions in coherent state path integrals because this implies
simultaneous action of field operators with their hermitian conjugated operators at the
same spacetime point which involves infinities from the (anti-)commutators. Therefore,
the chosen normal ordering in this paper is particularly suited for the coherent state
path integrals and causes the hermitian conjugated field operators to transform to
complex conjugated fields \(\psi_{\vec{x},s}^{*}(t_{p}+\Delta t_{p})\) acting a discrete
time step '\(\Delta t_{p}\)' later than the corresponding fields \(\psi_{\vec{x},s}(t_{p})\)
which are located on the right-hand side of the hermitian conjugated operators.
However, since we aim to achieve a coset decomposition of self-energies into
block diagonal density and off-diagonal pair condensate parts, one has to take the 'equal time'
hermitian and transposition operations for the self-energy which follow from
the symmetries of the 'equal time' dyadic products of anomalous doubled, anti-commuting fields.
In comparison to the 'lax', {\it sloppy}, hermitian form of coherent state path integrals,
the essential changes only result from the originally defined (\ref{s2_13}),
'time shifted' \(\Delta t_{p}\), anomalous doubled field \(\breve{\Psi}_{\vec{x},s}^{a}(t_{p})\)
with the hermitian conjugation '\(^{\sharp}\)', \(\breve{\Psi}_{\vec{x},s}^{\sharp a}(t_{p})\) (\ref{s2_14})
and time shift correction in the resulting complex part.
Straightforward use of these time shift corrections \(\Delta t_{p}\)
only causes a few amendments so that the abridged 'lax' hermitian forms can be conveyed
to the precise, discrete steps of the time development. Therefore, one can directly read from
previous articles \cite{neg,physica6,pop1,pop2,physica1,repbm} the exact, precise non-hermitian forms
with the correct, discrete time steps.

The precise, discrete time steps of this paper can also be transferred to the case of disordered
systems with an ensemble average of a random potential \cite{pop2}. Further applications of
coherent state path integrals are possible for quantum systems with fixed particle- and
symmetry quantum-numbers where we take the trace with coherent states over delta functions
of second quantized operators which determine the maximal commuting set of symmetry transformations
\beq \lb{s4_26}
\lefteqn{\varrho(E,l,s;n_{0},l_{z},s_{z}) = \mbox{Coherent state path integral of the trace of}} \\ \no
&:=&\mbox{Tr}\bigg[\delta\Big(\hbar\,s_{z}-\boldsymbol{\hat{S}_{z}}(\hat{\psi}\pdag,\hat{\psi})\,\Big)\;
\delta\Big(\hbar\,l_{z}-\boldsymbol{\hat{L}_{z}}(\hat{\psi}\pdag,\hat{\psi})\,\Big)\;
\delta\Big(n_{0}-\boldsymbol{\hat{N}}(\hat{\psi}\pdag,\hat{\psi})\,\Big)\;\times \\ \no &\times&
\delta\Big(\hbar^{2}\,s(s+1)-\boldsymbol{\vec{S}}(\hat{\psi}\pdag,\hat{\psi})\cdot
\boldsymbol{\vec{S}}(\hat{\psi}\pdag,\hat{\psi})\,\Big)\;
\delta\Big(\hbar^{2}\,l(l+1)-\boldsymbol{\vec{L}}(\hat{\psi}\pdag,\hat{\psi})\cdot
\boldsymbol{\vec{L}}(\hat{\psi}\pdag,\hat{\psi})\,\Big)\;
\delta\Big(E-\boldsymbol{\hat{H}}(\hat{\psi}\pdag,\hat{\psi};B_{z})\,\Big)\bigg] \;;  \\  \lb{s4_27}
\lefteqn{\ovv{\varrho(E_{1},E_{2},l,s;n_{0},l_{z},s_{z})} =
\mbox{disordered system with ensemble average for $B_{z}$}   } \\ \no
&:=&\ovv{\mbox{Tr}\bigg[\delta\Big(\hbar\,s_{z}-\boldsymbol{\hat{S}_{z}}(\hat{\psi}\pdag,\hat{\psi})\,\Big)\;
\delta\Big(\hbar\,l_{z}-\boldsymbol{\hat{L}_{z}}(\hat{\psi}\pdag,\hat{\psi})\,\Big)\;
\delta\Big(n_{0}-\boldsymbol{\hat{N}}(\hat{\psi}\pdag,\hat{\psi})\,\Big)\;\times} \\ \no &&\ovv{\times\;
\delta\Big(\hbar^{2}\,s(s+1)-\boldsymbol{\vec{S}}(\hat{\psi}\pdag,\hat{\psi})\cdot
\boldsymbol{\vec{S}}(\hat{\psi}\pdag,\hat{\psi})\,\Big)\;
\delta\Big(\hbar^{2}\,l(l+1)-\boldsymbol{\vec{L}}(\hat{\psi}\pdag,\hat{\psi})\cdot
\boldsymbol{\vec{L}}(\hat{\psi}\pdag,\hat{\psi})\,\Big)\;\times} \\ \no &&\ovv{\times\;
\delta\Big(E_{2}-\boldsymbol{\hat{H}}(\hat{\psi}\pdag,\hat{\psi};B_{z})\,\Big)\;
\delta\Big(E_{1}-\boldsymbol{\hat{H}}(\hat{\psi}\pdag,\hat{\psi};B_{z})\,\Big)\bigg]} \;.
\eeq
The delta functions have standard representations with the Dirac identity and integrations over exponentials
where one has to take into account the described, precise steps of the time development
in this article
\beq   \lb{s4_28}
\lim_{\ve_{+}\rightarrow0_{+}}\frac{1}{E-\boldsymbol{\hat{H}}(\hat{\psi}\pdag,\hat{\psi};B_{z})\mp\im\,\ve_{+}} &=&
\frac{\mbox{Principal value}}{E-\boldsymbol{\hat{H}}(\hat{\psi}\pdag,\hat{\psi};B_{z})}\pm\im\,\pi\;
\delta\big(E-\boldsymbol{\hat{H}}(\hat{\psi}\pdag,\hat{\psi};B_{z})\,\big) \;;      \\    \lb{s4_29}
\delta\big(E-\boldsymbol{\hat{H}}(\hat{\psi}\pdag,\hat{\psi};B_{z})\,\big)  &=&
\lim_{|\ve_{p_{E}}|\rightarrow 0} \;\;\lim_{T^{(E)}\rightarrow+\infty}\sum_{p_{E}=\pm}\int_{0}^{T^{(E)}}
\frac{dt_{p_{E}}^{(E)}}{2\pi\,\hbar}\;\times \\  \no &\times&
\overleftarrow{\exp}\bigg\{-\im\,\eta_{p_{E}}\:\frac{t_{p_{E}}^{(E)}}{\hbar}\;
\Big(E-\boldsymbol{\hat{H}}(\hat{\psi}\pdag,\hat{\psi};B_{z})\,\big)-\im\:\ve_{p_{E}}\Big)\bigg\} \;.
\eeq
However, as we have verified in previous sections, one has only to amend a few changes
in order to accomplish the exact time development for the case of fixed particle- and
symmetry quantum numbers. This allows to derive nonlinear sigma models from coherent
state path integrals (even for the relativistic case) with the given description of
the HST and coset decomposition for constrained cases of second quantized operators.

\begin{appendix}

\section{Jacobian and integration measure $\mbox{SO}(4)\,/\,\mbox{U}(2)\otimes\mbox{U}(2)$} \lb{sa}

In this appendix we outline a few details for deriving the integration measure of a coset
decomposition of \(\mbox{so}(4)\) generators and corresponding fields. The reader is referred
for more details to Ref. \cite{pop1,hobson,hua} where the more general case of an ortho-symplectic
super-group is investigated and the super-symmetric integration measure is attained from
the square root of the super-determinant of the
\(\mbox{Osp}(S,S|2L)\,/\,\mbox{U}(L|S)\otimes \mbox{U}(L|S)\) metric tensor (\(S\)=even integer,
\(L\)=odd integer, both related to angular momentum degrees of freedom of fermions and bosons,
respectively.).

We begin with the 'flat', Euclidean form of the self-energy of \(\mbox{so}(4)\) generators and
their fields and define an invariant length \((ds_{\mbox{\scz SO}(4)}\,)^{2}\) (\ref{sa_1})
of internal degrees of freedom which is accomplished by taking the trace over spin space and
the anomalous doubling of fields
\beq  \lb{sa_1}
\big(ds_{\mbox{\scz SO}(4)}\big)^{2} &=&-\TRS\big[d(\delta\wt{\Sigma})\;d(\delta\wt{\Sigma})\big]  =
-\TRS\big[d(\hat{T}\;\delta\hat{\Sigma}_{D}^{aa}\;\hat{T}^{-1})\;d(\hat{T}\;\delta\hat{\Sigma}_{D}^{bb}\;\hat{T}^{-1})\big]
\\  \no &=& -\TRS\big[d(\underbrace{\hat{T}\;\hat{Q}^{-1}}_{\hat{T}_{0}}\;\delta\hat{\Lambda}_{4\times 4}\;
\underbrace{\hat{Q}\;\hat{T}^{-1}}_{\hat{T}_{0}^{-1}})\;\;
d(\underbrace{\hat{T}\;\hat{Q}^{-1}}_{\hat{T}_{0}}\;\delta\hat{\Lambda}_{4\times 4}\;
\underbrace{\hat{Q}\;\hat{T}^{-1}}_{\hat{T}_{0}^{-1}})\big]  \;; \\  \lb{sa_2}
\hat{T}_{0;4\times 4} &=&\hat{T}_{4\times 4}\;\hat{Q}_{4\times 4}^{-1} \;; \\ \lb{sa_3}
d(\delta\wt{\Sigma})&=&
d(\hat{T}_{0}\;\delta\hat{\Lambda}\;\hat{T}_{0}^{-1}\big) =
\hat{T}_{0}\bigg(\Big[\hat{T}_{0}^{-1}\;d\hat{T}_{0}\;\boldsymbol{,}\;\delta\hat{\Lambda}\Big]_{\boldsymbol{-}}+
d(\delta\hat{\Lambda})\bigg)\hat{T}_{0}^{-1} \;.
\eeq
Using the definition (\ref{sa_2}) and the relation (\ref{sa_3}) for the self-energy, we
transform the coset decomposition \(\mbox{SO}(4)\,/\,\mbox{U}(2)\otimes\mbox{U}(2)\),
already inserted into (\ref{sa_1}), to the traces of the eigenvalues, to the traces of block diagonal
'11', '22' density parts as \(\mbox{u}(2)\) sub-algebra elements and to the traces
'12$\cdot$21', '21$\cdot$12' of coset generators and pair condensate terms. Note that the coset part
of the complete \(\mbox{SO}(4)\) integrations also contains products of the eigenvalues which are
finally to be shifted to the \(\mbox{u}(2)\) density part of the integration measure
\beq \lb{sa_4}
\big(ds_{\mbox{\scz SO}(4)}\big)^{2} &=&
-\TRS\bigg[\bigg(\Big[\hat{T}_{0}^{-1}\;d\hat{T}_{0}\;\boldsymbol{,}\;\delta\hat{\Lambda}\Big]_{\boldsymbol{-}}+
d(\delta\hat{\Lambda})\bigg)^{2}\bigg]  =
-2\trs\big[d(\delta\lambda_{s})\;d(\delta\lambda_{s})\big] +   \\  \no  &+& \sum_{a=1,2}
\trs\Big[\big(\hat{T}_{0}^{-1}\;d\hat{T}_{0}\big)_{ss\ppr}^{aa}\;
\big(\hat{T}_{0}^{-1}\;d\hat{T}_{0}\big)_{s\ppr s}^{aa}\;\;
\big(\delta\lambda_{s\ppr}-\delta\lambda_{s}\big)^{2}\Big] +    \\   \no  &+& \sum_{a,b=1,2}
\trs\Big[\big(\hat{T}_{0}^{-1}\;d\hat{T}_{0}\big)_{ss\ppr}^{a\neq b}\;
\big(\hat{T}_{0}^{-1}\;d\hat{T}_{0}\big)_{s\ppr s}^{b\neq a}\;\;
\big(\delta\lambda_{s\ppr}+\delta\lambda_{s}\big)^{2}\Big] =  \\  \no  &=&
-2\trs\big[d(\delta\lambda_{s})\;d(\delta\lambda_{s})\big] + 2\;
\trs\Big[\big(\hat{T}_{0}^{-1}\;d\hat{T}_{0}\big)_{ss\ppr}^{11}\;
\big(\hat{T}_{0}^{-1}\;d\hat{T}_{0}\big)_{s\ppr s}^{11}\;\;
\big(\delta\lambda_{s\ppr}-\delta\lambda_{s}\big)^{2}\Big] + \\  \no &+&2\;
\trs\Big[\big(\hat{T}_{0}^{-1}\;d\hat{T}_{0}\big)_{ss\ppr}^{12}\;
\big(\hat{T}_{0}^{-1}\;d\hat{T}_{0}\big)_{s\ppr s}^{21}\;\;
\big(\delta\lambda_{s\ppr}+\delta\lambda_{s}\big)^{2}\Big]\;.
\eeq
The infinitesimal variation \(\hat{T}^{-1}\;d\hat{T}\) of coset matrices consists of off-diagonal
\(\mbox{so}(4)\,/\,\mbox{u}(2)\) coset generators for pair condensates and block diagonal
\(\mbox{u}(2)\) sub-algebra generators for the density part of the self-energy; however,
the latter \(\mbox{u}(2)\) density and sub-algebra part within \(\hat{T}^{-1}\;d\hat{T}\)
can be absorbed into \(\hat{Q}^{-1}\;d\hat{Q}\) of \(\mbox{u}(2)\) sub-algebra generators
with the corresponding, independent density field variables of the self-energy
\be  \lb{sa_5}
\hat{T}_{0}^{-1}\;d\hat{T}_{0}=\hat{Q}\;\Big(\hat{T}^{-1}\;d\hat{T}-\hat{Q}^{-1}\;d\hat{Q}\Big)\;\hat{Q}^{-1}\;.
\ee
Therefore, we can reduce the block diagonal density summation with '11' (or '22') of
\((\hat{T}_{0}^{-1}\;d\hat{T}_{0})_{ss\ppr}^{11}\) to \((\hat{Q}^{-1}\;d\hat{Q})_{ss\ppr}^{11}\)
within following relation
\beq \lb{sa_6}
\big(ds_{\mbox{\scz SO}(4)}\big)^{2} &=&
-2\trs\big[d(\delta\lambda_{s})\;d(\delta\lambda_{s})\big] +
2\;\trs\Big[\big(\hat{Q}^{-1}\;d\hat{Q}\big)_{ss\ppr}^{11}\;\big(\hat{Q}^{-1}\;d\hat{Q}\big)_{s\ppr s}^{11}\;
\big(\delta\lambda_{s\ppr}-\delta\lambda_{s}\big)^{2}\Big] +  \\ \no &+&2\;
\trs\Big[\big(\hat{T}^{-1}\;d\hat{T}\big)_{ss\ppr}^{12}\;
\big(\hat{T}^{-1}\;d\hat{T}\big)_{s\ppr s}^{21}\;\;
\big(\delta\lambda_{s\ppr}+\delta\lambda_{s}\big)^{2}\Big]\;.
\eeq
As one applies the introduced, independent parameter fields of diagonalizing matrices
\(\hat{Q}_{ss\ppr}^{aa}(\vec{x},t_{p})\) (\ref{sa_7}-\ref{sa_10}) and coset matrices
\(\hat{T}(\vec{x},t_{p})\) (\ref{sa_11},\ref{sa_12}) (compare sections \ref{s2}), one
can eventually compute the invariant length \((ds_{\mbox{\scz SO}(4)})^{2}\) in terms
of density fields and pair condensate terms of the \(\mbox{so}(4)\) self-energy
\beq  \lb{sa_7}
\hat{Q}_{ss\ppr}^{11}(\vec{x},t_{p})   &=&
\left(\bea{cc} \cos\big(|\mscr{F}_{\uparrow\downarrow}(\vec{x},t_{p})|\big)  &
\im\;\frac{\sin\big(|\mscr{F}_{\uparrow\downarrow}(\vec{x},t_{p})|\big)}{|\mscr{F}_{\uparrow\downarrow}(\vec{x},t_{p})|}\;
\mscr{F}_{\uparrow\downarrow}(\vec{x},t_{p})    \\
\im\;\frac{\sin\big(|\mscr{F}_{\uparrow\downarrow}^{*}(\vec{x},t_{p})|\big)}{|\mscr{F}_{\uparrow\downarrow}(\vec{x},t_{p})|}\;
\mscr{F}_{\uparrow\downarrow}(\vec{x},t_{p})  &  \cos\big(|\mscr{F}_{\uparrow\downarrow}(\vec{x},t_{p})|\big) \eea\right)_{
ss\ppr}  \;;             \\ \lb{sa_8} \mscr{F}_{\uparrow\downarrow}(\vec{x},t_{p})  &=&
|\mscr{F}_{\uparrow\downarrow}(\vec{x},t_{p})|\;\;\exp\big\{\im\;\varphi(\vec{x},t_{p})\big\}
\;;  \\  \lb{sa_9}
\hat{Q}_{ss\ppr}^{11}(\vec{x},t_{p})   &=&
\left(\bea{cc} \cos\big(|\mscr{F}_{\uparrow\downarrow}(\vec{x},t_{p})|\big)  &
\im\;\sin\big(|\mscr{F}_{\uparrow\downarrow}(\vec{x},t_{p})|\big)\;\;\exp\big\{\im\;\varphi(\vec{x},t_{p})\big\}    \\
\im\;\sin\big(|\mscr{F}_{\uparrow\downarrow}^{*}(\vec{x},t_{p})|\big)\;\;\exp\big\{-\im\;\varphi(\vec{x},t_{p})\big\}  &
\cos\big(|\mscr{F}_{\uparrow\downarrow}(\vec{x},t_{p})|\big) \eea\right)_{ss\ppr}^{11}  ;       \\ \lb{sa_10}
\big(\hat{Q}^{-1}\;d\hat{Q}\big)_{ss\ppr}^{11} &=&
\left(\bea{cc}  -\im\;\sin^{2}\big(|\mscr{F}|\big) \;\;d\varphi  &
e^{\im\;\varphi} \Big(\im\;d\big(|\mscr{F}|\big) -\frac{\sin\big(2\,|\mscr{F}|\big)}{2}\;d\varphi\Big)  \\
e^{-\im\;\varphi} \Big(\im\;d\big(|\mscr{F}|\big) +\frac{\sin\big(2\,|\mscr{F}|\big)}{2}\;d\varphi\Big) &
\im\;\sin^{2}\big(|\mscr{F}|\big) \;\;d\varphi  \eea\right)_{ss\ppr}^{11} \;;  \\ \lb{sa_11}
\hat{T}(\vec{x},t_{p})  &=&
\left(\bea{cc}  \hat{1}_{ss\ppr}\;\cosh\big(|f(\vec{x},t_{p})|\big)   &
-\sinh\big(|f(\vec{x},t_{p})|\big)\;\big(\hat{\tau}_{2}\big)_{ss\ppr}\;e^{\im\;\phi(\vec{x},t_{p})}  \\
-\sinh\big(|f(\vec{x},t_{p})|\big)\;\big(\hat{\tau}_{2}\big)_{ss\ppr}\;e^{-\im\;\phi(\vec{x},t_{p})} &
\hat{1}_{ss\ppr}\;\cosh\big(|f(\vec{x},t_{p})|\big)   \eea\right)^{ab} \;;  \\ \lb{sa_12}
\big(\hat{T}^{-1}\;d\hat{T}\big)_{ss\ppr}^{ab}  &=&
\left(\bea{cc} \im\;\sinh^{2}\big(|f|\big)\;d\phi\;\hat{1}_{ss\ppr} & \hspace*{-0.5cm}
-e^{\im\;\phi}\;\big(\hat{\tau}_{2}\big)_{ss\ppr}\;\Big(d\big(|f|\big)+\im\;\frac{\sinh\big(2\,|f|\big)}{2}\;d\phi\Big)
\hspace*{-0.2cm}  \\    \hspace*{-0.2cm}
-e^{-\im\;\phi}\;\big(\hat{\tau}_{2}\big)_{ss\ppr}\;\Big(d\big(|f|\big)-\im\;\frac{\sinh\big(2\,|f|\big)}{2}\;d\phi\Big)
  &   -\im\;\sinh^{2}\big(|f|\big)\;d\phi\;\hat{1}_{ss\ppr} \eea\right)^{ab}_{\mbox{;}}  \\  \no
\big(ds_{\mbox{\scz SO}(4)}\big)^{2} &=&\hspace*{-0.3cm}
-2\Big[\big(d(\delta\lambda_{\uparrow})\big)^{2}+\big(d(\delta\lambda_{\downarrow})\big)^{2}\Big]  -
4\;\bigg[\big(d|\mscr{F}|\big)^{2}+\big(d\varphi\big)^{2}\;\sin^{2}\big(|\mscr{F}|\big)\;\cos^{2}\big(|\mscr{F}|\big)\bigg]\;
\big(\delta\lambda_{\downarrow}-\delta\lambda_{\uparrow}\big)^{2}  +   \\  \lb{sa_13} &+&
8\;\bigg[\big(d|f|\big)^{2}+\big(d\phi\big)^{2}\;\sinh^{2}\big(|f|\big)\;\cosh^{2}\big(|f|\big)\bigg]\;
\big(\delta\lambda_{\downarrow}+\delta\lambda_{\uparrow}\big)^{2}  \;.
\eeq
We shift the eigenvalue self-energy densities in terms of a polynomial
\(\mscr{P}(\delta\hat{\lambda}(\vec{x},t_{p})\,)\) (\ref{sa_14}) from the coset and pair condensate part
to the density part \(d[\delta\hat{\Sigma}_{D}(\vec{x},t_{p})]\), resulting from the \(\mbox{U}(2)\)
sub-group integration measure, and achieve the invariant integration measure of densities (\ref{sa_15}) and
pair condensates (\ref{sa_16}) by performing the square root of the determinant of the metric tensors
within \((ds_{\mbox{\scz SO}(4)}\,)^{2}\) (\ref{sa_13})
\beq \lb{sa_14}
\mscr{P}\big(\delta\hat{\lambda}(\vec{x},t_{p})\big)  &=&
\prod_{\{\vec{x},t_{p}\}} \big|\delta\lambda_{\uparrow}(\vec{x},t_{p})+\delta\lambda_{\downarrow}(\vec{x},t_{p})\big|^{2} \;; \\   \lb{sa_15}
d[\delta\hat{\Sigma}_{D}(\vec{x},t_{p})]\;\;\mscr{P}\big(\delta\hat{\lambda}(\vec{x},t_{p})\big) &=&
d\big[d\hat{Q}(\vec{x},t_{p})\;\hat{Q}^{-1}(\vec{x},t_{p});\delta\hat{\lambda}(\vec{x},t_{p})\big]\;\;
\mscr{P}\big(\delta\hat{\lambda}(\vec{x},t_{p})\big) =  \\ \no &=&
\prod_{\{\vec{x},t_{p}\}} \Bigg\{8\;d\big(\delta\lambda_{\uparrow}(\vec{x},t_{p})\big)\;
d\big(\delta\lambda_{\downarrow}(\vec{x},t_{p})\big)\;\;
\big|\delta\lambda_{\uparrow}^{2}(\vec{x},t_{p})-
\delta\lambda_{\downarrow}^{2}(\vec{x},t_{p})\big|^{\boldsymbol{2}}\;\times \\ \no &\times&
d\big(|\mscr{F}(\vec{x},t_{p})|\big)\;\sin\big(|\mscr{F}(\vec{x},t_{p})|\big)\;
\cos\big(|\mscr{F}(\vec{x},t_{p})|\big)\;\;d\varphi(\vec{x},t_{p}) \Bigg\} \;;  \\ \lb{sa_16}
d\big[\hat{T}^{-1}(\vec{x},t_{p})\;d\hat{T}(\vec{x},t_{p})\big] &=&
\prod_{\{\vec{x},t_{p}\}} \bigg\{8\;d\big(|f(\vec{x},t_{p})|\big)\;\sinh\big(|f(\vec{x},t_{p})|\big)\;
\cosh\big(|f(\vec{x},t_{p})|\big)\;\;d\phi(\vec{x},t_{p})\bigg\} .
\eeq
Furthermore, we remind that the source term
\(\exp\{\im\;\mscr{A}_{\hat{J}_{\psi\psi}}[\hat{T}]\}\) (\ref{s2_106}) contains the correction
\(\cosh^{\boldsymbol{-3}}(2|f(\vec{x},t_{p})|)\) for the integration measure (\ref{sa_16})
so that the total, complete integration measure of pair condensates is in fact given by
\beq \lb{sa_17}
\lefteqn{d\big[\hat{T}^{-1}(\vec{x},t_{p})\;d\hat{T}(\vec{x},t_{p})\big]\;\times\; \prod_{\{\vec{x},t_{p}\}}\bigg(
\frac{1}{\cosh^{\boldsymbol{3}}(2\,|f(\vec{x},t_{p})|)}\bigg)= } \\ \no &=&
\prod_{\{\vec{x},t_{p}\}} \bigg\{8\;d\big(|f(\vec{x},t_{p})|\big)\;\sinh\big(|f(\vec{x},t_{p})|\big)\;
\cosh\big(|f(\vec{x},t_{p})|\big)\;\;d\phi(\vec{x},t_{p})\bigg\}\;\times\; \prod_{\{\vec{x},t_{p}\}}\bigg(
\frac{1}{\cosh^{\boldsymbol{3}}(2\,|f(\vec{x},t_{p})|)}\bigg)  \\ \no &=&\prod_{\{\vec{x},t_{p}\}} \bigg\{4\;
d\big(|f(\vec{x},t_{p})|\big)\;\frac{\sinh\big(2\,|f(\vec{x},t_{p})|\big)}{\cosh^{\boldsymbol{3}}(2\,|f(\vec{x},t_{p})|)}
\;\;d\phi(\vec{x},t_{p})\bigg\} = \prod_{\{\vec{x},t_{p}\}} \bigg\{
-d\Big(\cosh^{\boldsymbol{-2}}\big(2\,|f(\vec{x},t_{p})|\big)\Big)\;\;d\phi(\vec{x},t_{p})\bigg\}  \\ \no &=&
\prod_{\{\vec{x},t_{p}\}} d\big[\tanh^{\boldsymbol{2}}(2|f(\vec{x},t_{p})|)-1\big]\;\;
d\big[\phi(\vec{x},t_{p})\big] = \prod_{\{\vec{x},t_{p}\}} d\big[\tanh^{\boldsymbol{2}}(2|f(\vec{x},t_{p})|)\big]\;\;
d\big[\phi(\vec{x},t_{p})\big]\;.
\eeq

\end{appendix}


\end{document}